\documentclass[12pt,a4paper]{article}
\usepackage[pdftex,colorlinks=true]{hyperref}
\usepackage[nopatch=footnote]{microtype}
\usepackage{amssymb,enumitem,booktabs,subfig,xcolor,setspace,paralist,fullpage, makecell, longtable, bm, orcidlink, dsfont}
\usepackage[top=0.9in, bottom=0.9in, left=1in, right=1in]{geometry}
\usepackage{natbib}
\usepackage{url}
\definecolor{darkblue}{rgb}{0,0,.6}
\hypersetup{citecolor=darkblue,linkcolor=darkblue,urlcolor=darkblue}
\usepackage{amsmath}
\usepackage{amsfonts}
\usepackage{epsfig}
\usepackage{graphics}
\usepackage{palatino,eulervm}
\usepackage{graphicx}
\usepackage{lscape}
\usepackage{rotating}
\usepackage[linewidth=1pt]{mdframed}
\allowdisplaybreaks[4]

\newsavebox\CBox

\usepackage{algorithm}
\usepackage{algpseudocode}
\usepackage{tikz}
\usetikzlibrary{decorations.pathreplacing}
\newcommand{\Rlogo}{\protect\includegraphics[height=1.8ex,keepaspectratio]{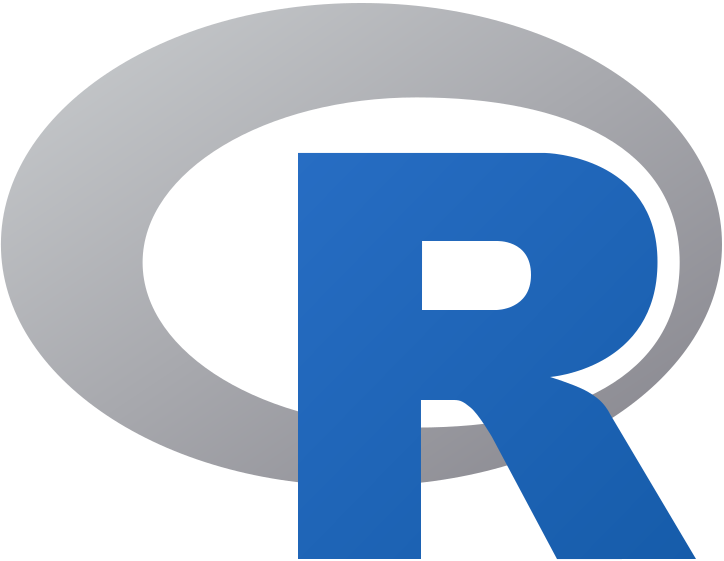}}
\usepackage[font=small]{caption}

\providecommand{\U}[1]{\protect\rule{.1in}{.1in}}

\graphicspath{{plots/}}

\usepackage{amsthm,thmtools}
\usepackage{mathrsfs}

\makeatletter
\def\th@newremark{\th@remark\thm@headfont{\bfseries}}
\makeatletter
\theoremstyle{newremark}

\declaretheoremstyle[
  spaceabove=6pt, spacebelow=6pt,
  headfont=\bfseries,
  notefont=\mdseries, notebraces={(}{)},
bodyfont=\normalfont,
  postheadspace=0.5em]{mystyle}

\begin{document}

\title{\Large\bf Visualizing and forecasting subnational life-table death counts: \mbox{Gap forecasting methods}}
\author{\normalsize Han Lin Shang \orcidlink{0000-0003-1769-6430}\thanks{Corresponding address: Department of Actuarial Studies and Business Analytics, Macquarie University, Sydney, NSW 2068, Australia; Telephone number: +61(2) 9850 4689; Email: hanlin.shang@mq.edu.au}\\ \normalsize Department of Actuarial Studies and Business Analytics \\ \normalsize Macquarie University \\
\\
\normalsize Andrea Nigri \orcidlink{0000-0002-2707-3678}\\
\normalsize Department of Social Sciences \\
\normalsize University of Foggia
}
\date{\normalsize \today}

\maketitle

\begin{abstract}

Subnational life-table death counts are highly correlated across time and space and differ by gender. While these associations are helpful in improving forecasts through joint modeling, less attention has been paid to identifying and understanding mortality disparities in gender and regional gaps. We propose a forecasting framework to model and forecast female life-table death counts at the national or subnational level, and to model and forecast the associated gender gap. For either females or males, we could forecast national life-table death counts and the regional gap relative to national data. By combining gender and regional gaps, we also explore the double gap by prioritizing national data and female data. Life-table death counts are unique due to their non-negativity and summability constraints. To address the constraints, we apply a one-to-one transformation, termed cumulative distribution function transformation, to obtain one- to 15-step-ahead forecasts. Using Japanese age-specific life-table death counts between ages 0 and 110+ from 1947 to 2023, we evaluate and compare point and interval forecast accuracy across gender, region, and double gaps. By focusing on these gaps, we can deepen our understanding of the possible factors driving gender and regional mortality variations.

\vspace{.04in}

\end{abstract}

\newpage

\setstretch{1.53}

\section{Introduction}\label{sec:1}

Mortality evolution is crucial for demographers and actuaries looking to develop more accurate forecasting models, with applications to government policymaking, pension planning, and annuity pricing. The majority of forecasting models used by demographers and actuaries tend to predict future longevity for specific countries separately for males and females. However, females tend to have a different mortality profile than males; they live longer than males at all ages in almost all countries \citep[see, e.g.,][]{Austad06, BMG+21}. In particular, differences in life expectancy can be observed between the very old, among centenarians, and super-centenarians ($\geq 110$), when women outnumber men by more than 9 to 1 \citep{PF98}. This gender gap can be attributed to biological and social factors. 

Because of their unique mortality profiles by sex, forecasts are often produced separately, although some efforts have been made toward multi-population modeling and forecasting. For instance, \cite{LL05} proposed a method for both sexes that is expected to converge, based on an augmented common factor model. \cite{Li13} introduced a Poisson common factor model to forecast female and male mortality jointly. \cite{HBY13} proposed a method to model and forecast products and ratios among multiple populations. The product-ratio method models the geometric mean of subpopulation rates and the ratio of subpopulation rates to product rates. \cite{SSB+16} considered a multilevel functional time-series model to identify common and population-specific trends among multiple populations. Through a two-way analysis of variance, \cite{JSS24} decomposed multiple populations into a grand mean, a row mean, a column mean, and residuals. \cite{TSY+25} considered a functional factor model to capture cross-sectional and temporal dependences on subnational age-specific mortality rates.

Targeting the gender gap in mortality, \cite{LN26} developed an age-period-cohort framework specifically among young and early adults. Their approach models the sex ratio of age-specific death rates, which yields coherent inference on the relative male-female mortality disadvantage and is less sensitive to overall mortality levels. For forecasting life expectancy, \cite{RCG+13, RLG14} introduced a two-sex model and developed a Bayesian paradigm to forecast life expectancy for both sexes; the female life expectancy is first forecast, and then the gap between female and male life expectancies is modeled, accounting for their correlation. 

\cite{TV12} was based on the idea that future human longevity is driven by a general trend. Targeting the gap between the record and the current life expectancy of a specific population, their model forecasts the world's record life expectancy and then assuming a tendency towards convergence with the predicted record level. \cite{PCV18} proposed a double-gap life expectancy model: the first gap is between a country's female life expectancy and the world record for female life expectancy, and the second gap is the difference in life expectancy between males and females in a country.

Due to the availability of subnational mortality data, we contribute to the literature by presenting a gap modeling framework, namely gender gap and regional heterogeneity, to visualize, model, and forecast subnational life-table death counts. We consider a new mortality instrument introduced in \cite{SH25}, developed from the cumulative sum of the normalized life-table death counts $d_x$ by age $x=0,1,\dots,109,110+$. Termed cumulative relative life-table death counts, we normalize life-table death counts to a probability by dividing the radix $l_0 = 10^5$ in year~$t$. It shares some similarity to the cumulative probability of death in \cite{SMI20}. Using the cumulative sum, we transform the probability density function (PDF) into a cumulative distribution function (CDF). With an additional benefit of monotonicity, the CDF endures one constraint less than the life-table death counts, namely, summability. Moreover, it provides a scale-free measure for pairwise comparison of any two populations, such as female and male data within a country or subnational and national data for a given gender. Comparison of female and male data reveals gender differences, while comparison of subnational and national data reveals regional heterogeneity. 

The outline of this paper is given as follows: In Section~\ref{sec:2}, we present the subnational life-table death count in Japan, which has the largest number of centenarians in the world \citep{SYR12}. To demonstrate the scale-free property of cumulative relative life-table death counts, we present several visualization tools in Section~\ref {sec:3} to reveal the gender gap and regional heterogeneity. In Section~\ref{sec:4}, we present a forecasting method to model gender, regional, and double gaps, and compare their point and interval forecast accuracy in Section~\ref{sec:5}. The conclusion is presented in Section~\ref{sec:6}, along with ideas for further extending the methodology.

\section{Subnational life-table death counts in Japan}\label{sec:2}

In many developed countries, such as Japan, the increase in human life expectancy and an aging population have led governments to raise concerns about the sustainability of the pension, health and aged care systems \citep[see, e.g.,][]{Coulmas07}. These concerns have resulted in a surge of interest among government policymakers in accurately modeling and forecasting mortality by \textit{age}, \textit{sex}, and \textit{region}. Subnational forecasts of age-specific mortality rates help inform local policy, and improvements in forecast accuracy are beneficial for allocating current and future resources at the national and subnational levels.

\begin{figure}[!htb]
\centering
{\includegraphics[width=8.6cm]{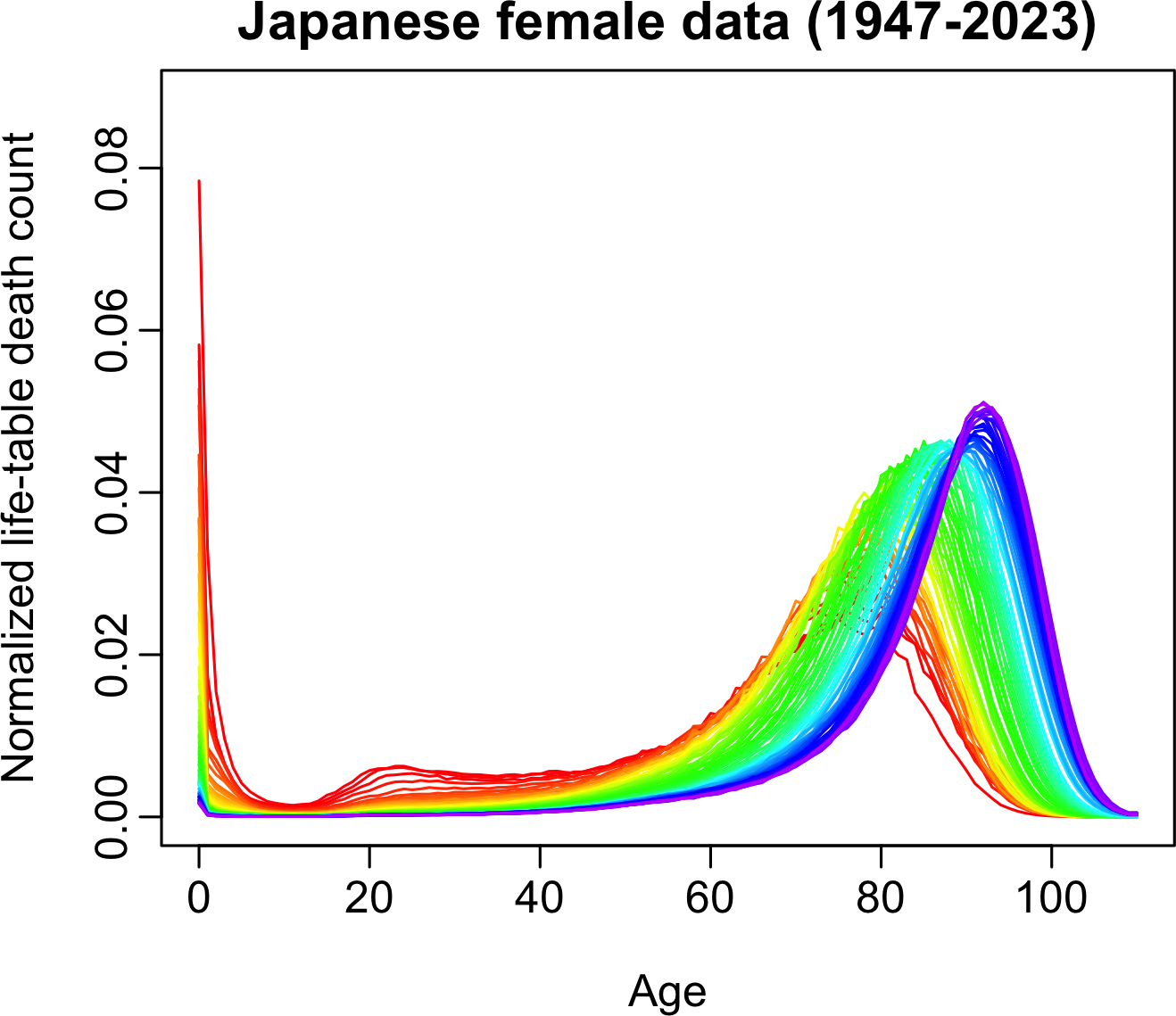}\label{fig:1a}}
\quad
{\includegraphics[width=8.6cm]{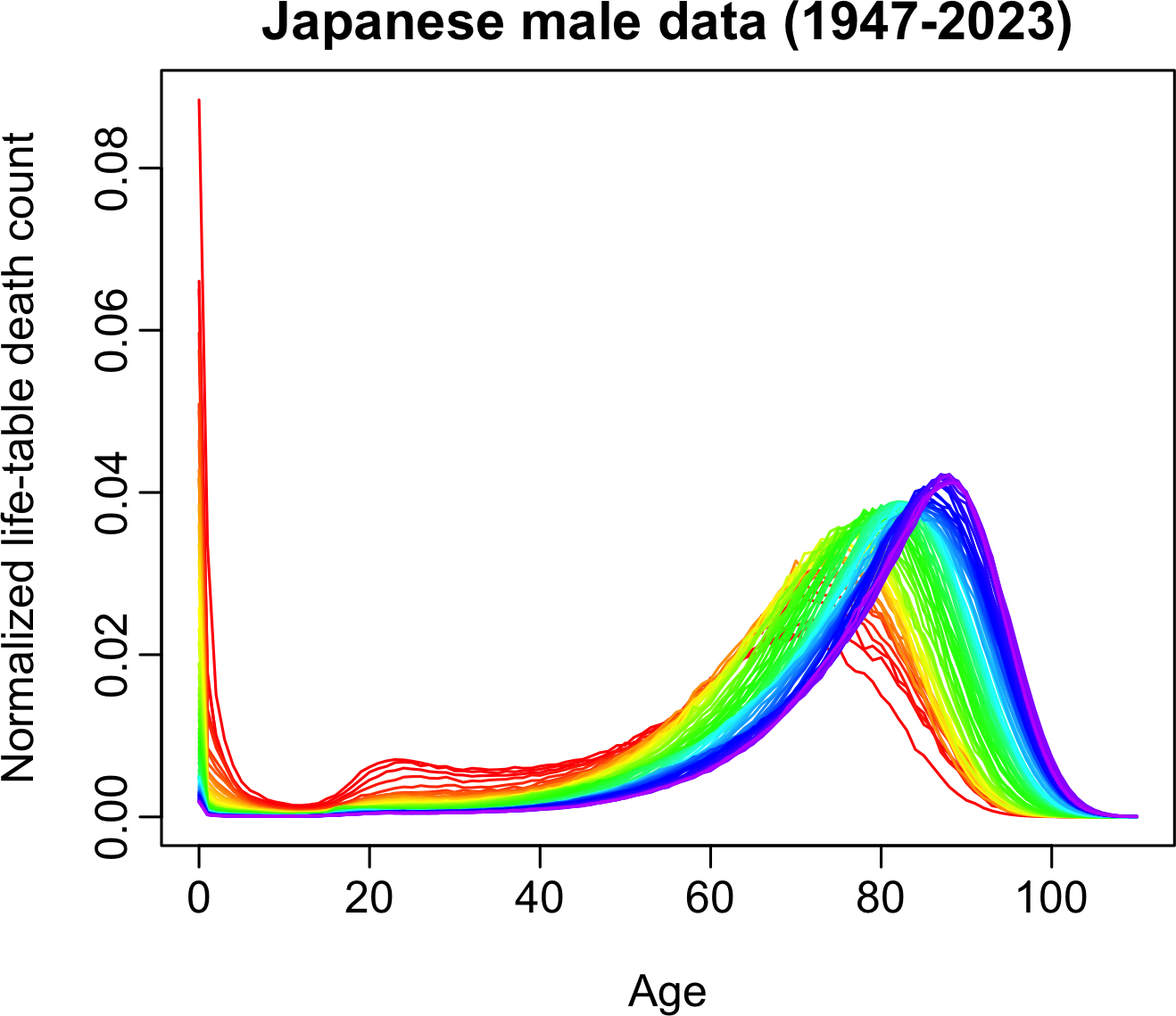}\label{fig:1b}}
\caption{\small Rainbow plots of age-specific normalized life-table death counts, denoted by $\tilde{d}_{t,x}^{s,g}$, from 1947 to 2023 in a single-year group, where $\sum_{x=0}^{110+}\tilde{d}_{t,x}^{s,g}=1$. The oldest years are shown in red, with the most recent years in violet. Curves are ordered chronologically according to the colors of the rainbow.}\label{fig:1}
\end{figure}

Sourced from the \cite{JMD25}, we consider Japanese period life-table death counts by age, sex, and region. For a given calendar year $t$, we observe life-table deaths, denoted by $d_{t,x}^{s,g}$, for region $s$, gender $g$, and age $x$; it is defined as the number of deaths occurring between two successive ages in a period life table. Life-table death counts are constrained by non-negativity and summability (i.e., the radix of $10^5$), and thus resemble a PDF after normalization, as shown in Figure~\ref{fig:1}. 

\begin{figure}[!htb]
\centering
\subfloat[Japanese females]
{\includegraphics[width=8.6cm]{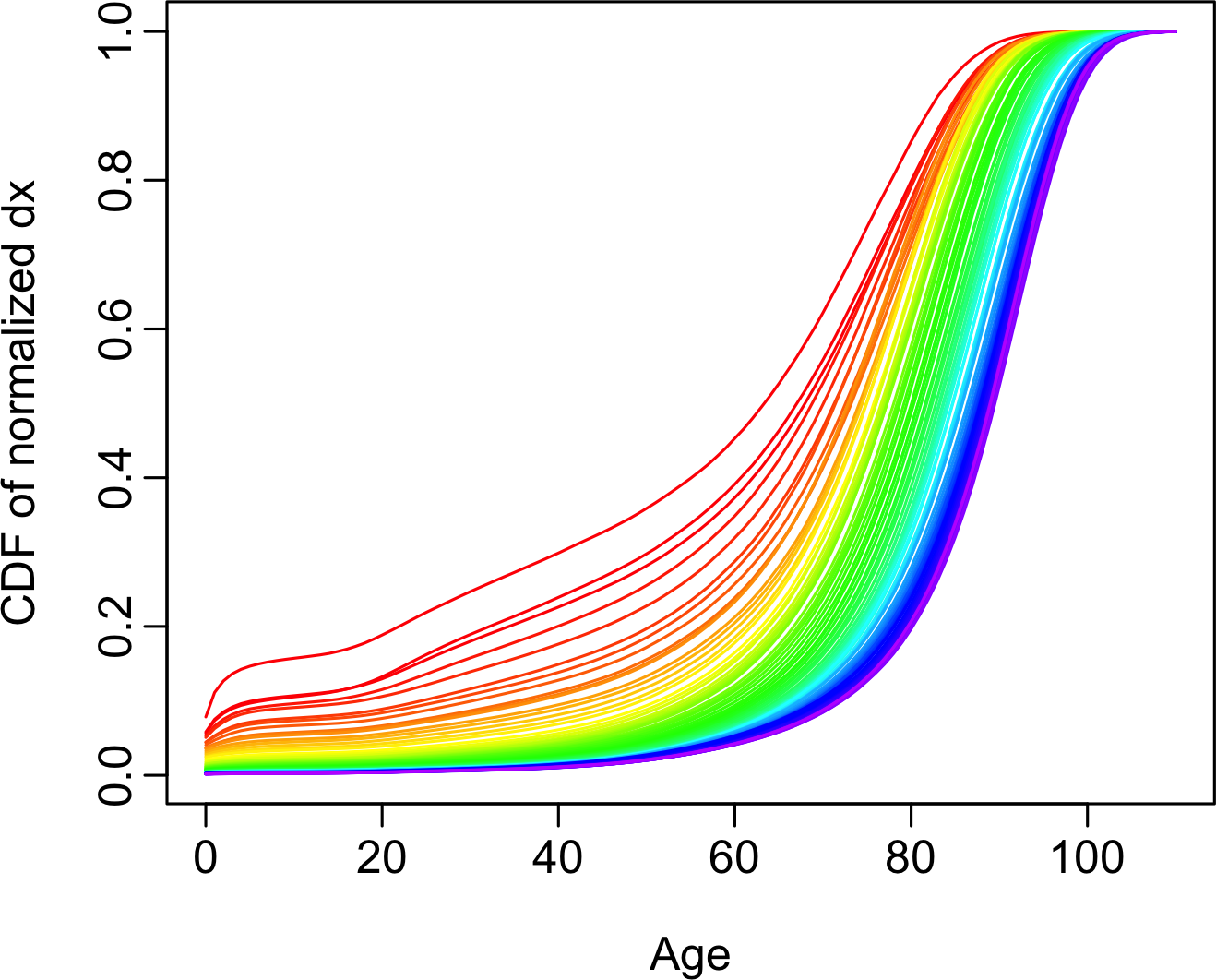}\label{fig:2a}}
\quad
\subfloat[Japanese males]
{\includegraphics[width=8.6cm]{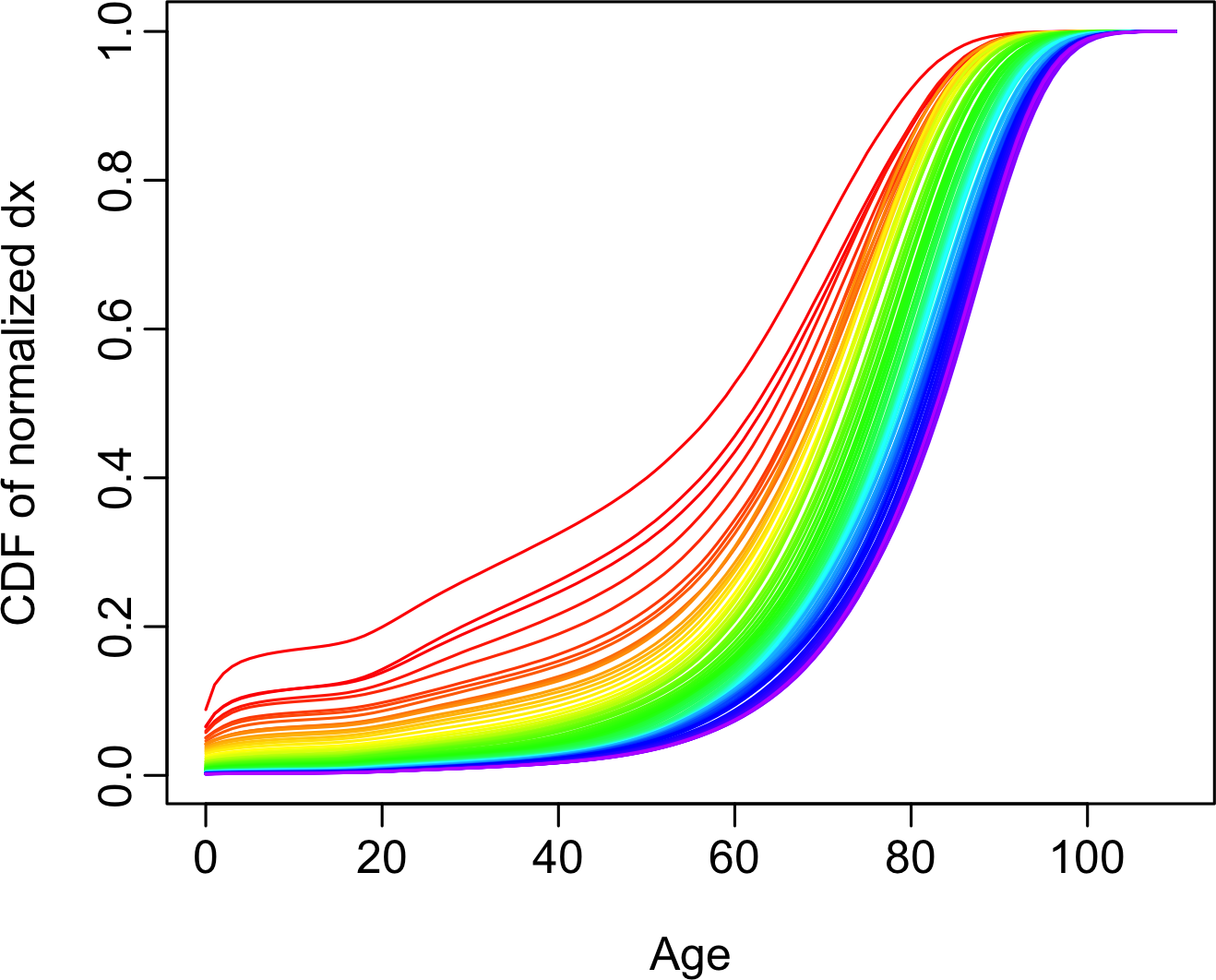}\label{fig:2b}}
\caption{\small We present their empirical CDFs, denoted by $D_{t,x}^g$, of age-specific life-table death counts in Japan.}\label{fig:2}
\end{figure}

From the normalized~$\tilde{d}_{t,x}^{s,g}$, we implement a cumulative sum to obtain its CDF,
\begin{equation*}
D_{t,x}^{s,g} = \sum^{x}_{i=1}\tilde{d}_{t,i}^{s,g},\qquad t=1, 2,\dots,n,
\end{equation*}
where $n$ denotes the number of years in a data set, $x=1, 2,\dots,111$ represents ages from 0 to 109 in a single year of age, and the last age group of 110+, and $D_{t,111}^{s,g}=1$. In Figure~\ref{fig:2}, we display a time series of the empirical CDFs of the Japanese national female and male data from 1947 to 2023.

\section{Visualization}\label{sec:3}

\subsection{Gender difference}\label{sec:3.1}

Gender differences in mortality are a long-standing and widely documented feature of population health, reflecting biological, behavioral, and social factors \citep[see, e.g.,][]{LAO+13}. 

One of the most relevant and extensively studied patterns is the systematic survival advantage of females over males, observed both globally \citep{luy2014women} and historically across European populations \citep[see, e.g.,][]{glei2007, zarulli2021, zarulli2021death}. Although the average sex gap in life expectancy in Europe amounts to several years, this summary measure conceals substantial heterogeneity between countries and regions. Evidence suggests that this longevity gap reflects the combined influence of biological susceptibility and gendered behavioral patterns, particularly differential exposure to health-related risks. Cardiovascular disease, smoking, alcohol consumption, and external causes of death, especially during working-age adulthood, have been identified as key contributors to excess male mortality and, consequently, to the observed sex gap in life expectancy \citep[see, e.g.,][]{mccartney2011contribution, janssen15, beltran2015}.

\begin{figure}[!htb]
\centering
\subfloat[$G_{t,x}^{\text{N}, \text{M-F}} = D_{t,x}^{\text{N}, \text{M}} - D_{t,x}^{\text{N}, \text{F}}$]
{\includegraphics[width=8.6cm]{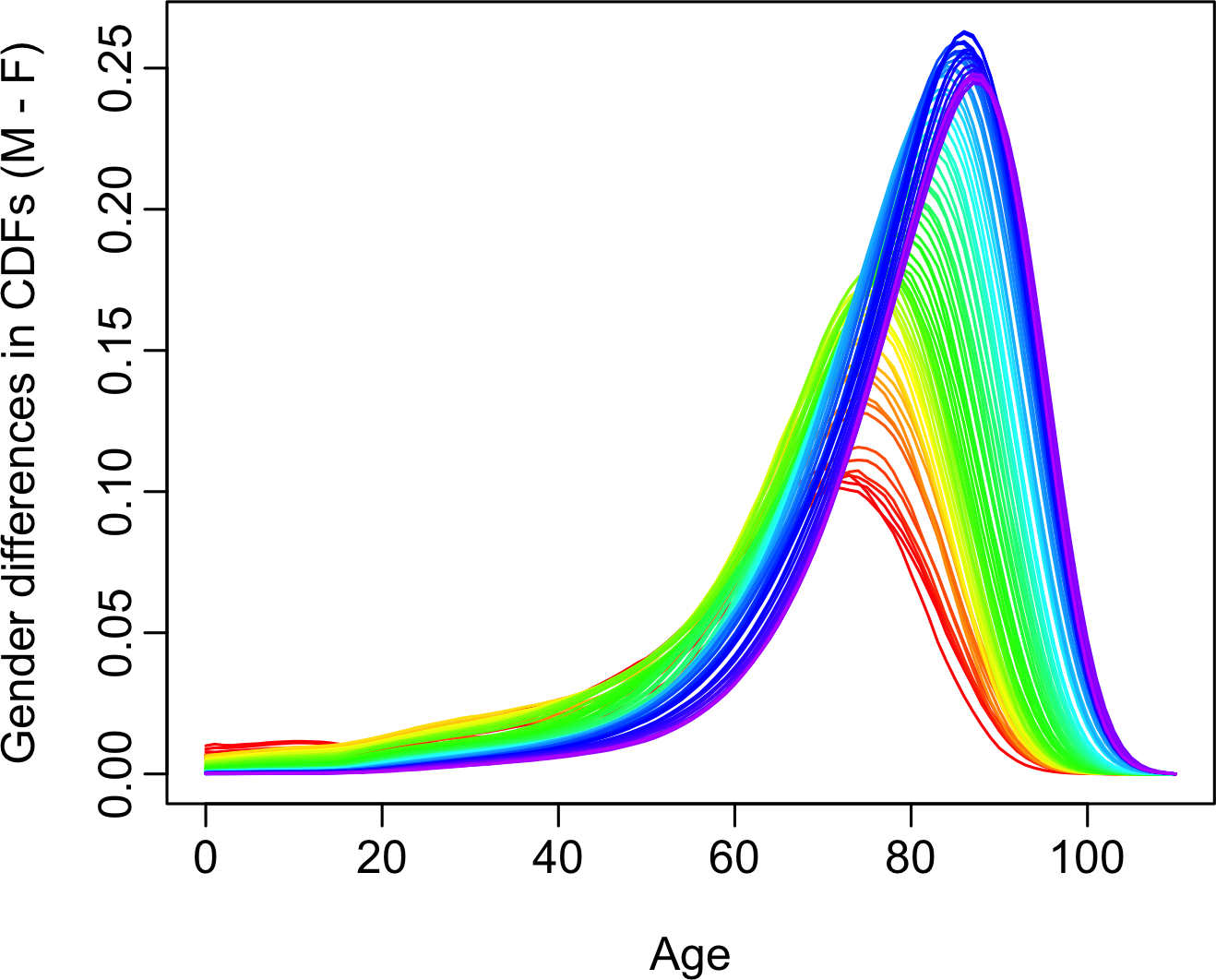}\label{fig:3a}}
\quad
\subfloat[$S_t^{\text{N}, \text{M-F}} = \sum_{x=1}^{111} (D_{t,x}^{\text{N}, \text{M}} - D_{t,x}^{\text{N}, \text{F}})$]
{\includegraphics[width=8.6cm]{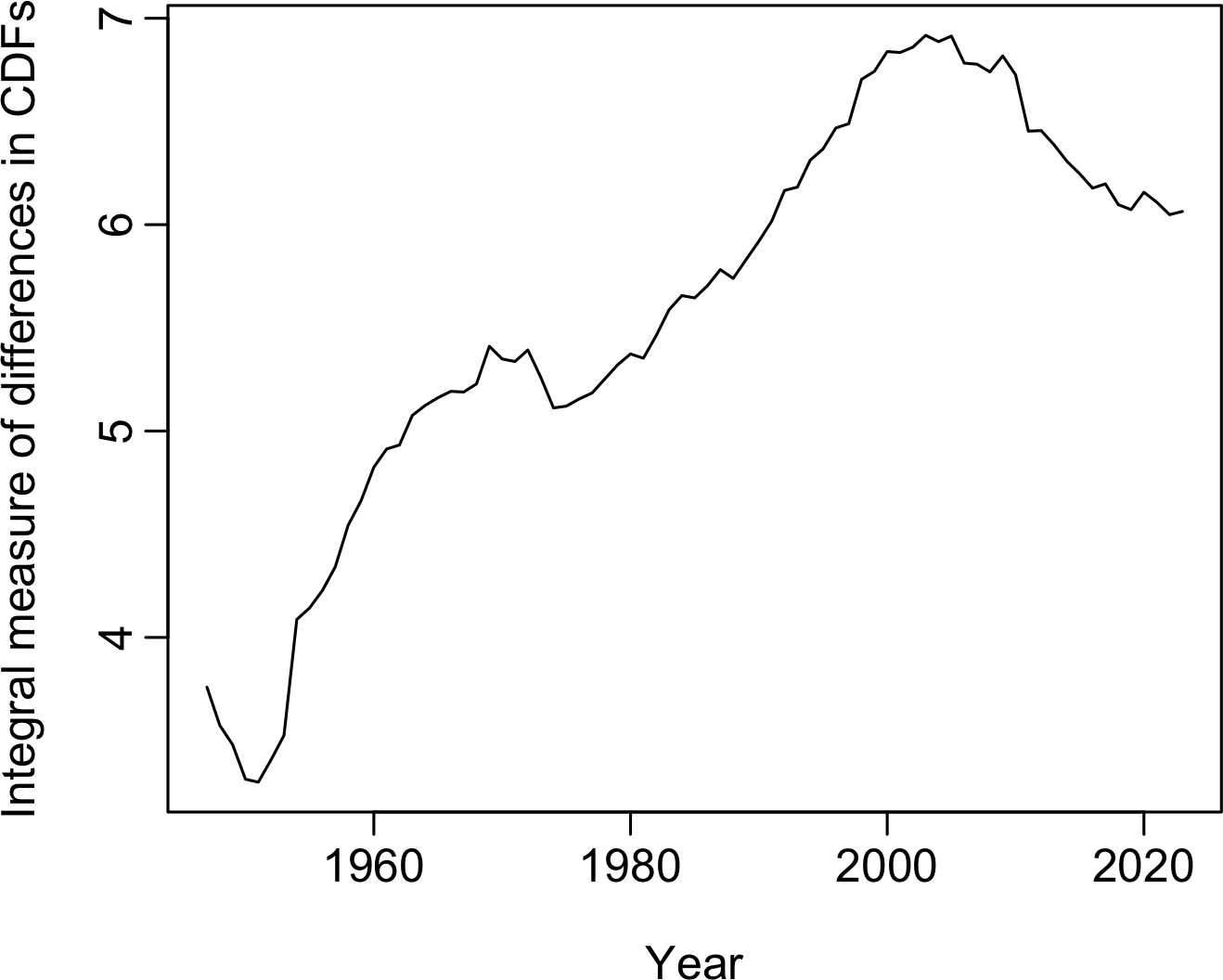}\label{fig:3b}}
\caption{\small Age-specific and integral measure of gender differences in the cumulative relative life-table death counts between Japanese males and females from 1947 to 2023. The superscript $\textsuperscript{N}$ represents national data.}\label{fig:3}
\end{figure}

In Japan, across all years, males experience higher cumulative relative life-table deaths than females. Among centenarians, the ratio of females to males is about 7:1. Based on the national cumulative relative life-table death counts, we compute the gender gap as
\begin{equation*}
G_{t,x}^{\text{N}, \text{M-F}} = D_{t,x}^{\text{N}, \text{M}} - D_{t,x}^{\text{N}, \text{F}},
\end{equation*}
which can be visualized in Figure~\ref{fig:3a}; the superscript $\textsuperscript{N}$ represents national data. By adding the gender gap across all ages, we compute an integral measure and visualize its trend in Figure~\ref{fig:3b}.

The range of the gender difference is $(-0.00003,  0.26298)$. In almost all ages and years, males in Japan experience a higher cumulative relative life-table death count than their female counterparts. In Figure~\ref{fig:3b}, the integral measure of the gender gap, $S_t^{\text{N}, \text{M-F}}$, reveals a smaller gender gap in previous years and a larger gap in recent years. The largest gender gap occurred around 2005. 

Instead of the integral measure, we can also consider the Wasserstein distance of order~1.
\begin{equation*}
W_{t,1}^{s,\text{M-F}}  = \frac{1}{111}\sum_{x=1}^{111} \left|(\text{D}^{s,\text{M}}_{t,x})^{-1} - (\text{D}^{s,\text{F}}_{t,x})^{-1}\right|,
\end{equation*}
where $(\cdot)^{-1}$ represents a chosen quantile based on the empirical CDF. The absolute Wasserstein distance is called Earth Mover's distance and represents the minimum average absolute distance by which one must move the mass to transform $\text{D}^{s,\text{M}}_{t,x}$ to $\text{D}^{s,\text{F}}_{t,x}$.

\begin{figure}[!htb]
\centering
\includegraphics[width=8.7cm]{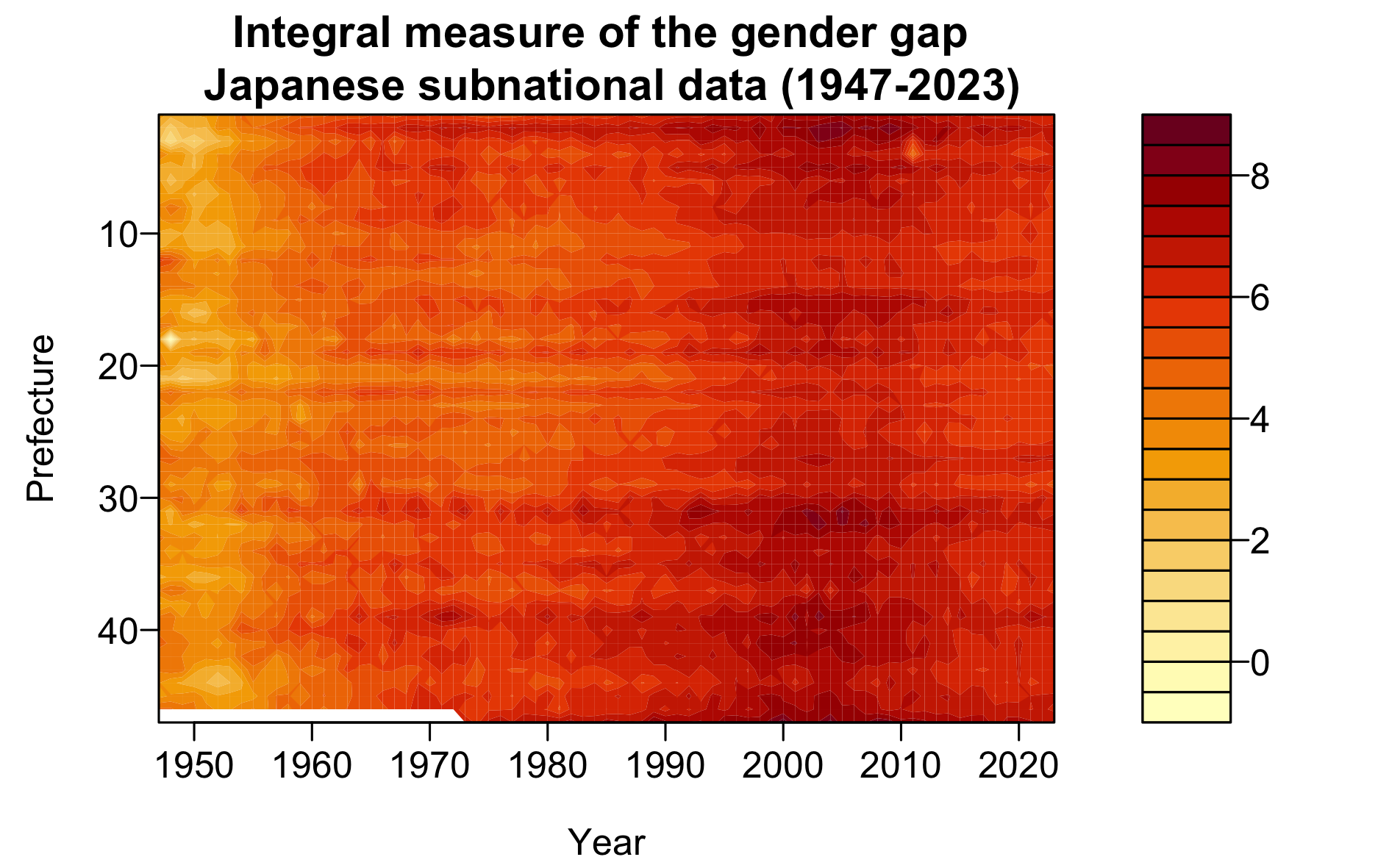}
\quad
\includegraphics[width=8.7cm]{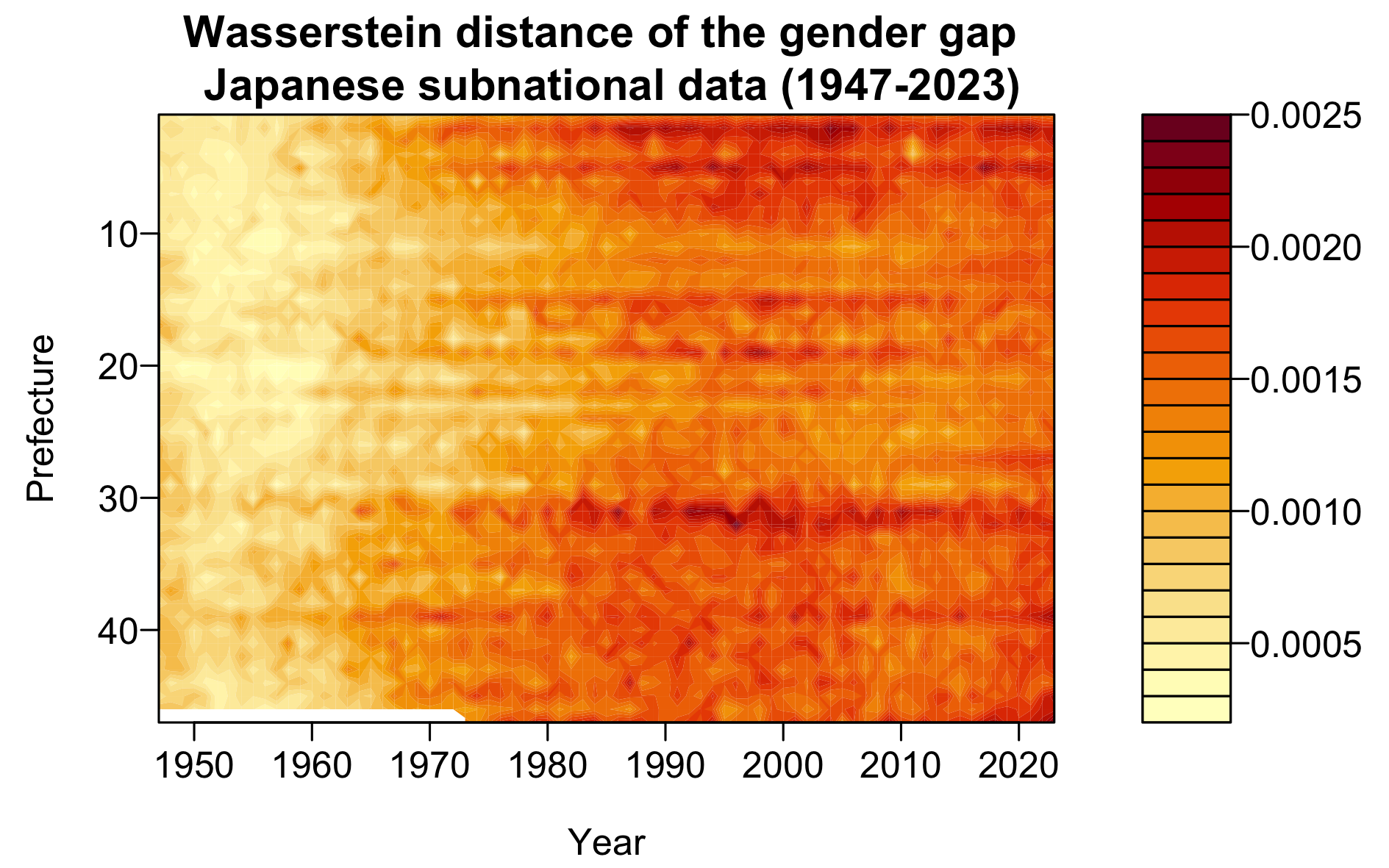}
\caption{\small Integral measure and Wasserstein distance of the gender gap in Japanese subnational cumulative relative life-table death counts from 1947 to 2023. Over time, the gender gap widens, particularly between 2000 and 2010, with the rate of increase varying across prefectures. Prefectures are ordered geographically from the northern region, Hokkaido, to the southern region, Okinawa. In Okinawa, data are observed from 1973 to 2023.}\label{fig:4}
\end{figure}

At the subnational level, we compute the integral measure of the gender gap for each of the 47 prefectures relative to the national data. For some prefectures, the gender gaps are greater than in others. Ordered geographically from North to South, we display an image plot of $S_t^{s, \text{M-F}}$ and $W_{t,1}^{s, \text{M-F}}$ in Figure~\ref{fig:4}.

\subsection{Regional heterogeneity}\label{sec:}

In addition to the gender gap, there is a regional gap, with some prefectures, such as Okinawa, known to have more centenarians than others \citep{WWH+08, PH24}. This more in-depth geographic consideration offers a nuanced understanding of mortality differentials. It equips researchers and policymakers with valuable information on how regional distinctions, such as health policies, education attainment, socioeconomic status, and other factors, can influence mortality between regions \citep[see, e.g.,][]{FD20, BPB23}. Recall that $D_{t,x}^{s,g}$ denotes the cumulative relative life-table death count in year $t$, age $x$, region $s$, and gender $g$, and the difference between the subnational and national data can be represented as
\begin{equation*}
U_{t,x}^{s, g} = D_{t,x}^{s, g} - D_{t,x}^{\text{N}, g}, \quad g\in \{\text{F}, \text{M}\}
\end{equation*}

\begin{figure}[!htb]
\centering
\includegraphics[width=8.7cm]{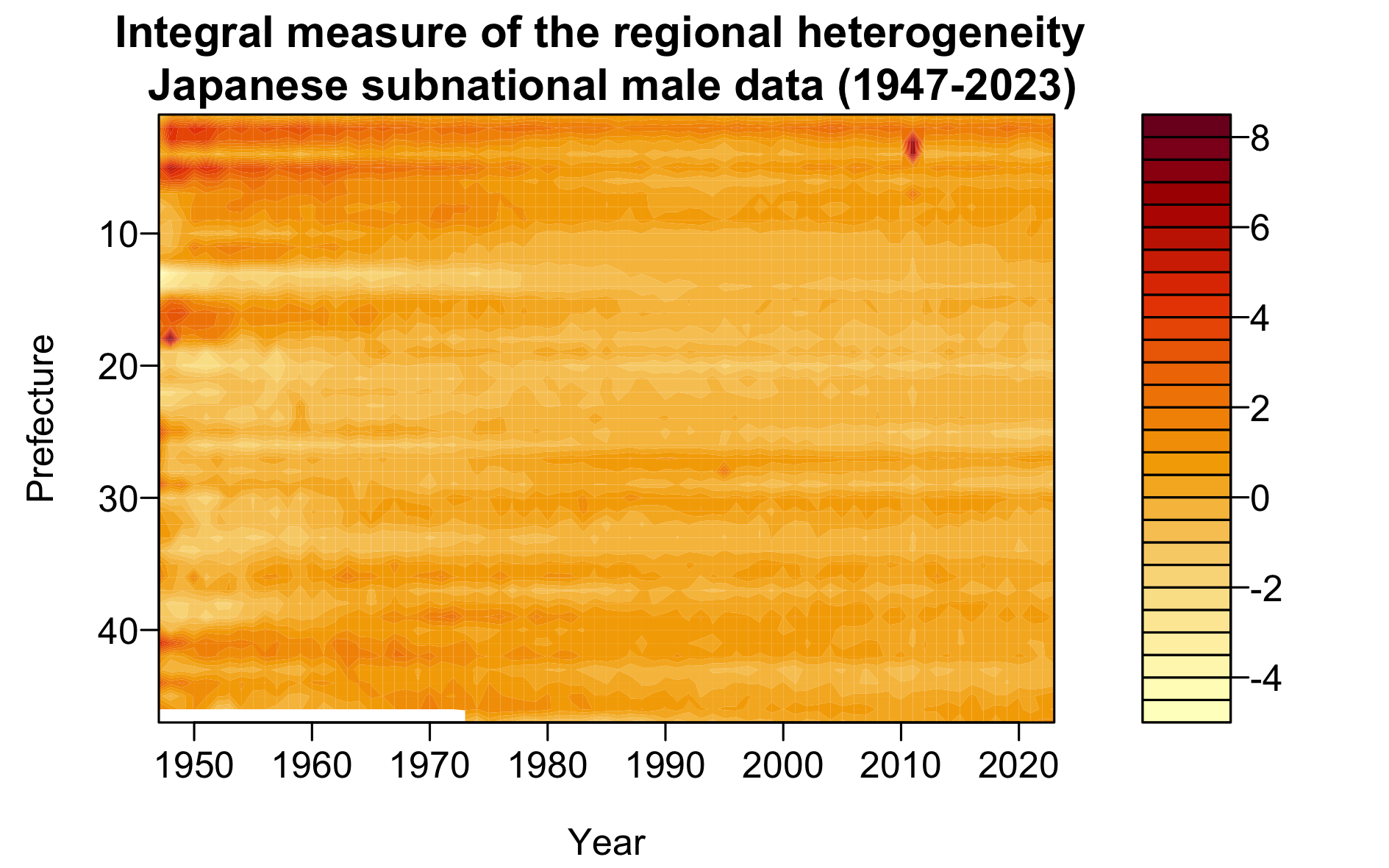}
\quad
\includegraphics[width=8.7cm]{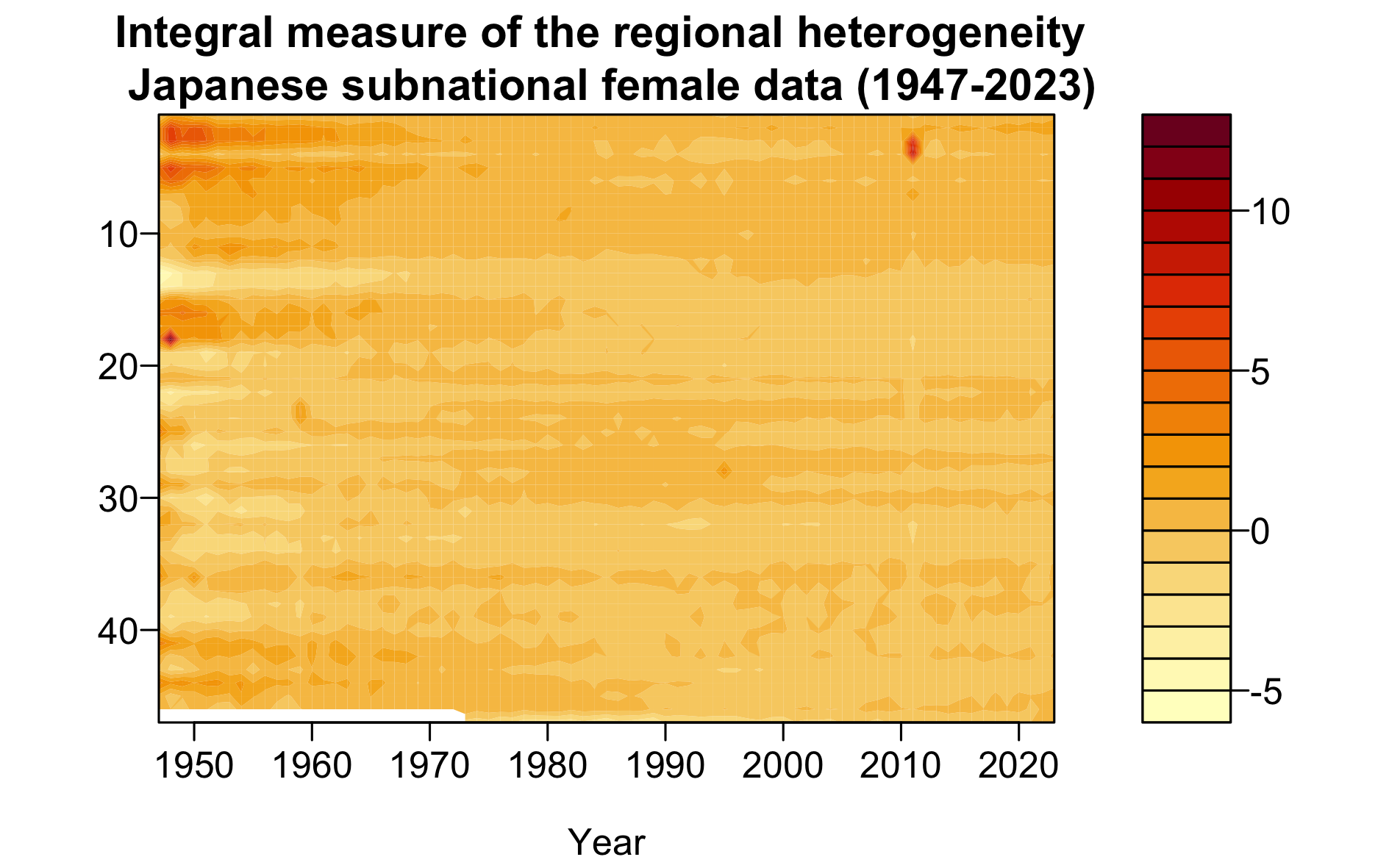}
\caption{\small Integral measure of the regional gap in Japanese subnational cumulative relative life-table death counts from 1947 to 2023. In Okinawa, data are observed from 1973 to 2023. In contrast to the gender gap, regional heterogeneity has narrowed over time.}\label{fig:5}
\end{figure}

By summing up the regional gap across all ages, we obtain an integral measure
\begin{align*}
V_{t}^{s, \text{F}} &= \sum_{x=1}^{111} (D_{t,x}^{s, \text{F}} - D_{t,x}^{\text{N}, \text{F}}), \\
V_{t}^{s, \text{M}} &= \sum_{x=1}^{111} (D_{t,x}^{s, \text{M}} - D_{t,x}^{\text{N}, \text{M}}),
\end{align*}
which are visualized in Figure~\ref{fig:5}. Okinawa has data from 1973 to 2023, while all other prefectures have data from 1947 to 2023.

\section{Modeling and Forecasting}\label{sec:4}

\subsection{Gender gaps in the Japanese national and subnational data}\label{sec:4.1}

Since the gap in the cumulative relative life-table death counts often lies between $-1$ and $1$ behaved like correlation, we take Fisher's $Z$ transformation \citep[see, e.g.,][]{VA23} as
\begin{equation*}
F_{t,x}^{\text{N}, \text{M-F}}=\frac{1}{2}\ln\left[\frac{1+G_{t,x}^{\text{N}, \text{M-F}}}{1-G_{t,x}^{\text{N}, \text{M-F}}}\right]=artanh(G_{t,x}^{\text{N}, \text{M-F}}),
\end{equation*}
which is a \textit{one-to-one} mapping. By the inverse Fisher $Z$ transformation, we obtain the differences in the two CDFs
\begin{align}
G_{t,x}^{\text{N}, \text{M-F}} &= \frac{\exp(2F_{t,x}^{\text{N}, \text{M-F}})-1}{\exp(2F_{t,x}^{\text{N}, \text{M-F}})+1}\notag\\
&=tanh(F_{t,x}^{\text{N}, \text{M-F}}) \label{eq:Fisher_Z_inv}.
\end{align}

Denote $\bm{F}_t^{\text{N}, \text{M-F}}=\big(F_{t,1}^{\text{N}, \text{M-F}}, F_{t,2}^{\text{N}, \text{M-F}}, \dots, F_{t,111}^{\text{N}, \text{M-F}}\big)^{\top}$ as a vector of gender gap among 111 ages, where $^{\top}$ symbolizes vector transpose. From a time series of $\bm{F}^{\text{N}, \text{M-F}} = \big(\bm{F}_1^{\text{N}, \text{M-F}},\bm{F}_{2}^{\text{N}, \text{M-F}}, \dots, \bm{F}_n^{\text{N}, \text{M-F}}\big)$, we implement a principal component analysis to decompose the matrix of the gender gap into a set of principal components and their associated principal component scores, expressed as
\begin{equation*}
F_{t,x}^{\text{N}, \text{M-F}} = \mu^{\text{N}, \text{M-F}}_{x} + \sum^K_{k=1}\beta^{\text{N}, \text{M-F}}_{t,k}\phi^{\text{N}, \text{M-F}}_{k,x} + \epsilon^{\text{N}, \text{M-F}}_{t,x},
\end{equation*}
where $\mu^{\text{N}, \text{M-F}}_x$ denotes the mean term, approximated by $\widehat{\mu}^{\text{N}, \text{M-F}}_x=\frac{1}{n}\sum_{t=1}^{n}F_{t,x}^{\text{N}, \text{M-F}}$; $\phi^{\text{N}, \text{M-F}}_{k,x}$ denotes the $k$\textsuperscript{th} principal component of age $x$; $\beta^{\text{N}, \text{M-F}}_{t,k}$ denotes the $k$\textsuperscript{th} estimated principal component score for year $t$; and $\epsilon^{\text{N}, \text{M-F}}_{t,x}$ denotes the error for year $t$ and age $x$. The principal components are commonly estimated via eigen-analysis of sample covariance. For forecasting purposes, overestimating $K$ leads to a smaller loss in accuracy than underestimating it. Following \cite{HBY13}, we set the number of components, $K=6$.

Since the temporal dependence is manifested in the principal component scores, we apply a univariate time-series forecasting method, such as autoregressive integrated moving average (ARIMA) or exponential smoothing (ETS), to obtain $h$-step-ahead forecasts of the scores, denoted by $\widehat{\beta}^{\text{N},\text{M-F}}_{n+h|n,k}$. For stationary principal component scores, a vector autoregression can also be used. Conditional on the observed data, the estimated principal components $\bm{\Phi}^{\text{N}, \text{M-F}}_x = \big(\widehat{\phi}^{\text{N}, \text{M-F}}_{1,x},\widehat{\phi}^{\text{N}, \text{M-F}}_{2,x},\dots, \widehat{\phi}^{\text{N}, \text{M-F}}_{K,x}\big)$, and the mean term, we obtain $h$-step-ahead forecasts of the gender gap, expressed as
\begin{align*}
\widehat{F}_{n+h|n,x}^{\text{N}, \text{M-F}} &= E\left[F_{n+h,x}^{\text{N}, \text{M-F}}|\bm{F}^{\text{N}, \text{M-F}}, \bm{\Phi}^{\text{N}, \text{M-F}}_x, \widehat{\mu}^{\text{N}, \text{M-F}}_x\right] \\
&= \widehat{\mu}^{\text{N}, \text{M-F}}_x + \sum^K_{k=1}\widehat{\beta}^{\text{N}, \text{M-F}}_{n+h|n,k}\widehat{\phi}_{k,x}^{\text{N}, \text{M-F}}.
\end{align*}

\begin{figure}[!htb]
\centering
\subfloat[Gender gap forecasts]
{\includegraphics[width=8.7cm]{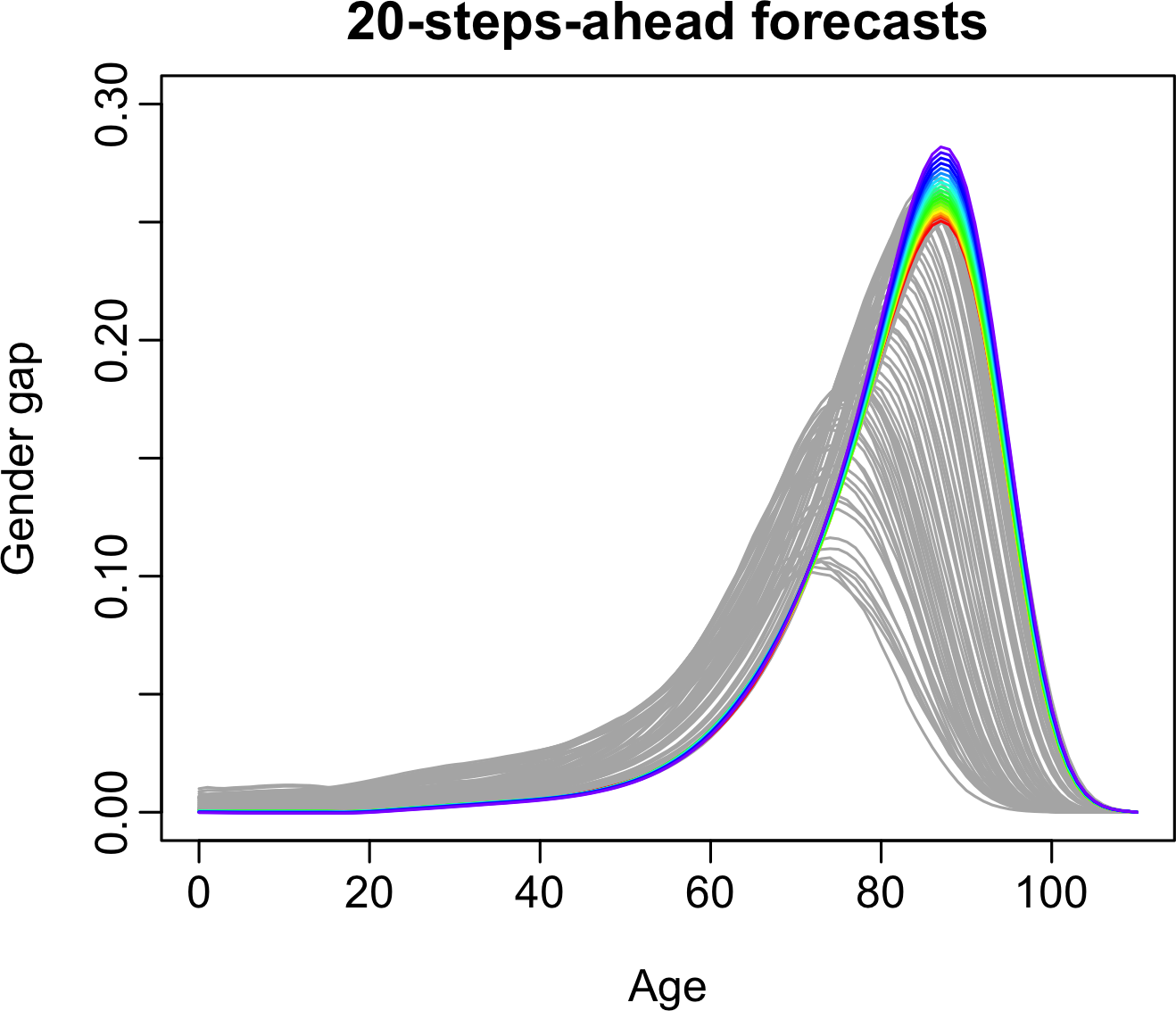}\label{fig:6a}}
\quad
\subfloat[CDF forecasts]
{\includegraphics[width=8.7cm]{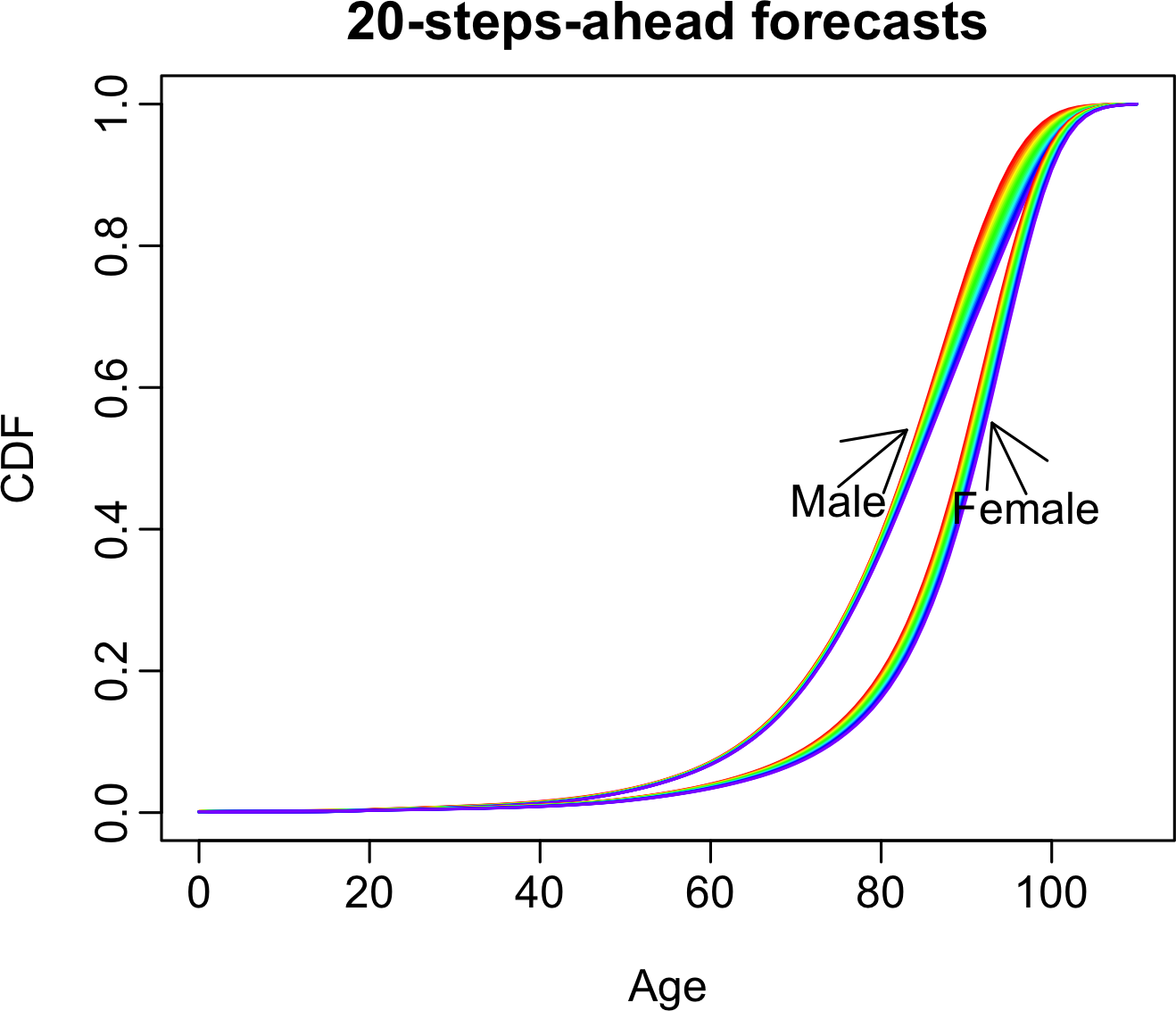}\label{fig:6b}} 
\\
\subfloat[Life-table death count forecasts]
{\includegraphics[width=8.7cm]{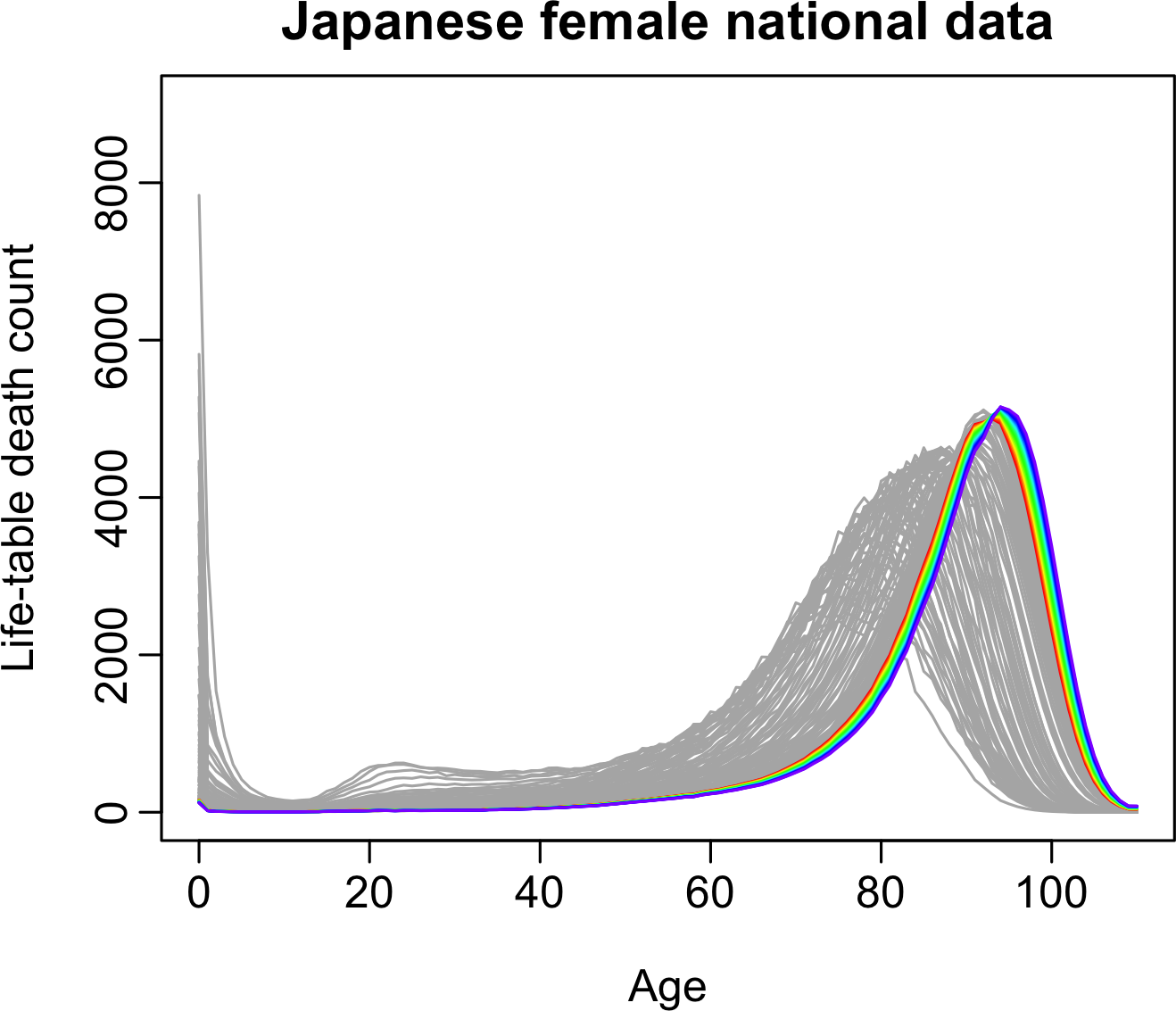}\label{fig:6c}}
\quad
\subfloat[Life-table death count forecasts]
{\includegraphics[width=8.7cm]{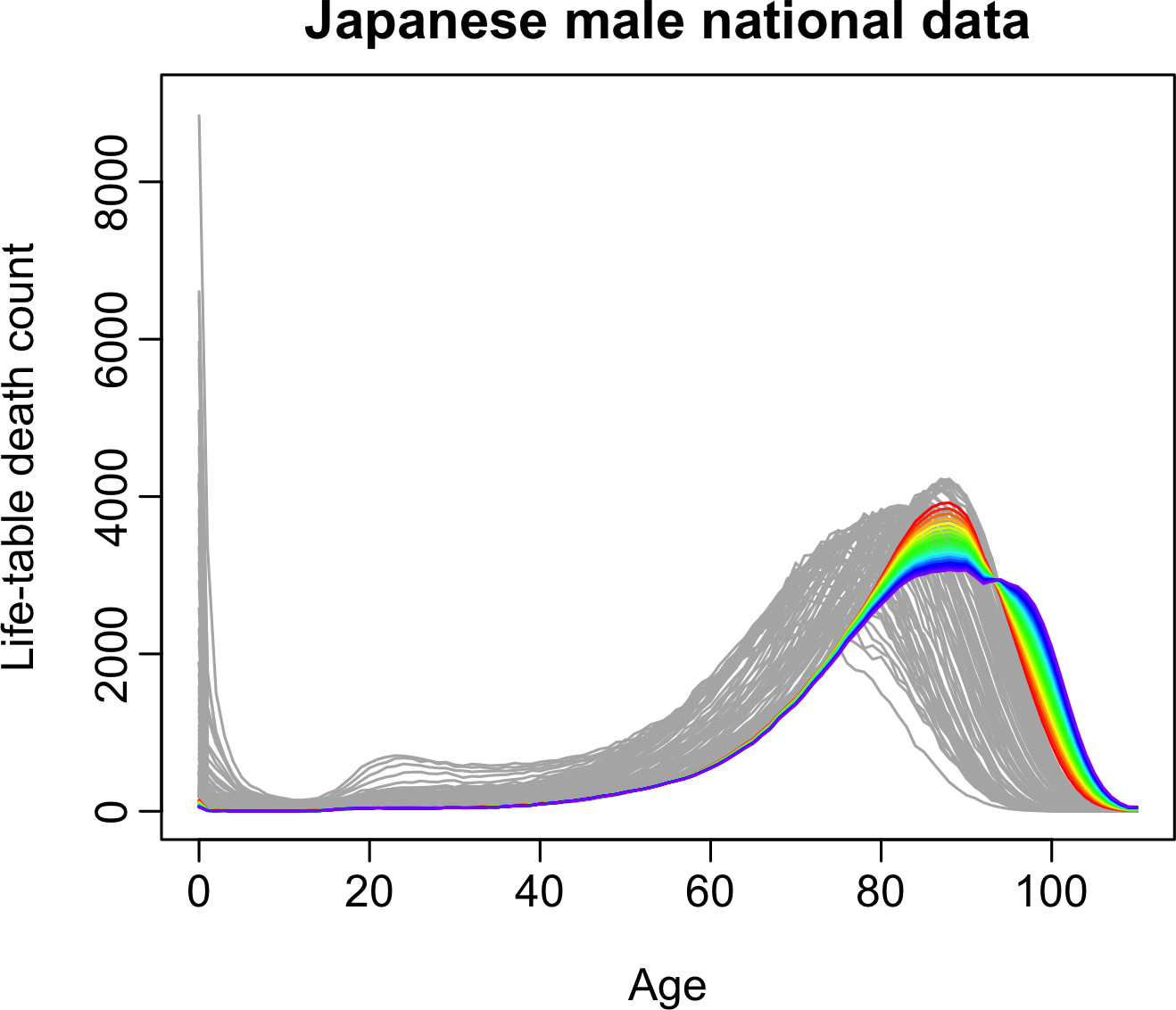}\label{fig:6d}} 
\caption{\small Using the ETS univariate time-series forecasting method, the 20-years-ahead forecasts of the gender gap, 20-years-ahead forecasts of the female and male cumulative distribution functions, and 20-years-ahead forecasts of the female and male life-table death counts (radix of $10^5$) in Japan.}\label{fig:6}
\end{figure}

Since the female series generally has a higher data quality than the male series for most developed countries, including Japan. Using the same forecasting method, we obtain the 20-years-ahead forecasts of the female cumulative relative life-table death counts, denoted by $\widehat{D}_{n+h|n,x}^{\text{N}, \text{F}}$. In Figure~\ref{fig:6a}, we present the 20-years-ahead forecasts of the gender gap, denoted by $\widehat{G}_{n+h|n,x}^{\text{N}, \text{M-F}}$, after the inverse Fisher $Z$ transformation in~\eqref{eq:Fisher_Z_inv}. By adding the forecasts of the gender gap and female cumulative relative life-table death counts, we obtain the 20-years-ahead forecasts of the male cumulative relative life-table death counts, $\widehat{D}_{n+h|n,x}^{\text{N}, \text{M}} = \widehat{D}_{n+h|n,x}^{\text{N}, \text{F}} + \widehat{F}_{n+h|n,x}^{\text{N}, \text{M-F}}$, in Figure~\ref{fig:6b}.

By taking the first-order differencing, we obtain
\begin{equation*}
\widehat{d}_{n+h|n,x}^{\text{N}, g} = \Delta^x_{i=1}\widehat{D}_{n+h|n,i}^{\text{N}, g}, \quad g\in \{\text{F}, \text{M}\},
\end{equation*}
where $\Delta$ represents the first-order differencing. In Figures~\ref{fig:6c} and~\ref{fig:6d}, we display the 20-years-ahead forecasts of the life-table death counts for Japanese females and males.

\subsection{Regional gaps in the Japanese female and male data}\label{sec:4.2}

Denote $\bm{F}_t^{\text{s-N},g}=\big(F_{t,1}^{\text{s-N},g},F_{t,2}^{\text{s-N},g},\dots,F_{t,111}^{\text{s-N},g}\big)^{\top}$ as a vector of regional gap among 111 ages. From a time series of $\bm{F}^{\text{s-N},g} = \big(\bm{F}_1^{\text{s-N},g},\bm{F}_{2}^{\text{s-N},g},\dots,\bm{F}_n^{\text{s-N},g}\big)$, we again perform a principal component analysis to decompose the matrix of the regional gap into a set of principal components and their associated principal component scores, expressed as
\begin{equation*}
F_{t,x}^{\text{s-N}, g} = \nu^{\text{s-N},\text{g}}_x + \sum^J_{j=1}\gamma^{\text{s-N},g}_{t,j}\psi_{j,x}^{\text{s - N}, g} + \varepsilon^{\text{s-N},\text{g}}_{t,x},
\end{equation*}
where $\nu^{\text{s-N},\text{g}}_x$ denotes the mean term, approximated by $\widehat{\nu}^{\text{s-N},\text{g}}_x = \frac{1}{n}\sum^n_{t=1}F_{t,x}^{\text{s-N}, g}$; $\psi_{x,j}^{\text{s - N}, g}$ denotes the $j$\textsuperscript{th} principal component for age $x$ and gender~$g$; $\gamma^{\text{s-N},g}_{t,j}$ denotes the $j$\textsuperscript{th} principal component score in year $t$ and gender~$g$; and $\varepsilon^{\text{s-N},\text{g}}_{t,x}$ represents the error term. The principal components are commonly estimated via eigen-analysis of sample covariance. The number of components $J$ is set again at six, as in \cite{HBY13}.

Since the temporal dependence is shown in the scores, we apply the ETS to obtain $h$-step-ahead forecasts of the scores, denoted by $\widehat{\gamma}^{\text{s-N},\text{g}}_{n+h|n,j}$. Conditional on the observed data, the estimated principal components $\bm{\Psi}^{\text{s-N},g}_x = \big(\widehat{\psi}^{\text{s-N},g}_{1,x},\widehat{\psi}^{\text{s-N},g}_{2,x},\dots,\widehat{\psi}^{\text{s-N},g}_{J, x}\big)$, and the mean term, we obtain $h$-step-ahead forecasts of the regional gap, expressed as
\begin{equation*}
\widehat{F}^{\text{s - N}, g}_{n+h|n,x} = \text{E}\left[F_{n+h, x}^{\text{s - N},\text{g}}|\bm{F}^{\text{s - N}, g}, \bm{\Psi}^{\text{s-N},g}_x, \nu^{\text{s-N},g}_x\right] 
=\widehat{\nu}^{\text{s-N},\text{g}}_{x} + \sum^J_{j=1}\widehat{\gamma}^{\text{s-N},g}_{n+h|n,j}\widehat{\psi}_{j,x}^{\text{s-N},g}.
\end{equation*}

\begin{figure}[!htb]
\centering
\subfloat[Regional gap forecasts]
{\includegraphics[width=8.78cm]{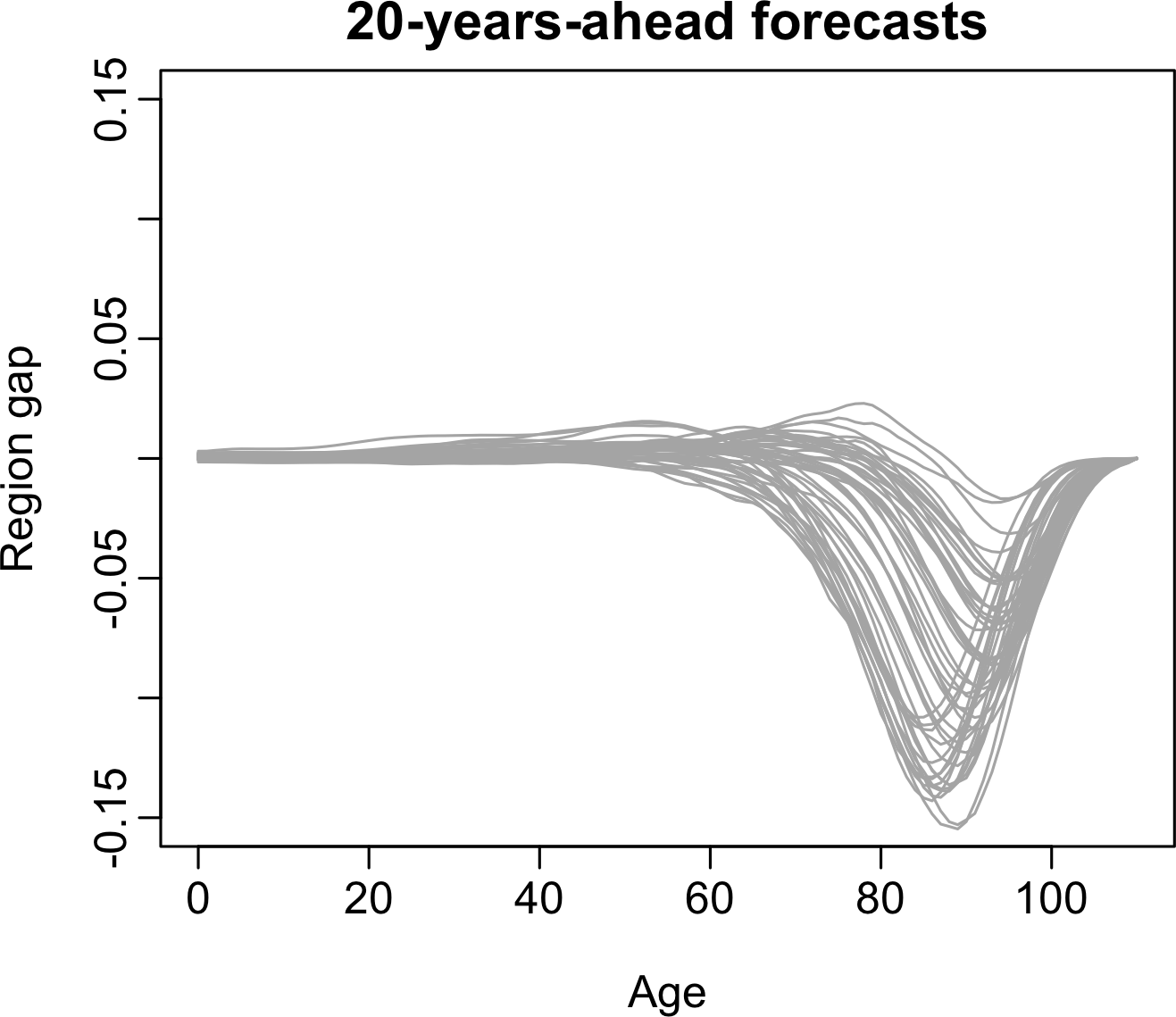}\label{fig:7a}}
\quad
\subfloat[CDF forecasts]
{\includegraphics[width=8.78cm]{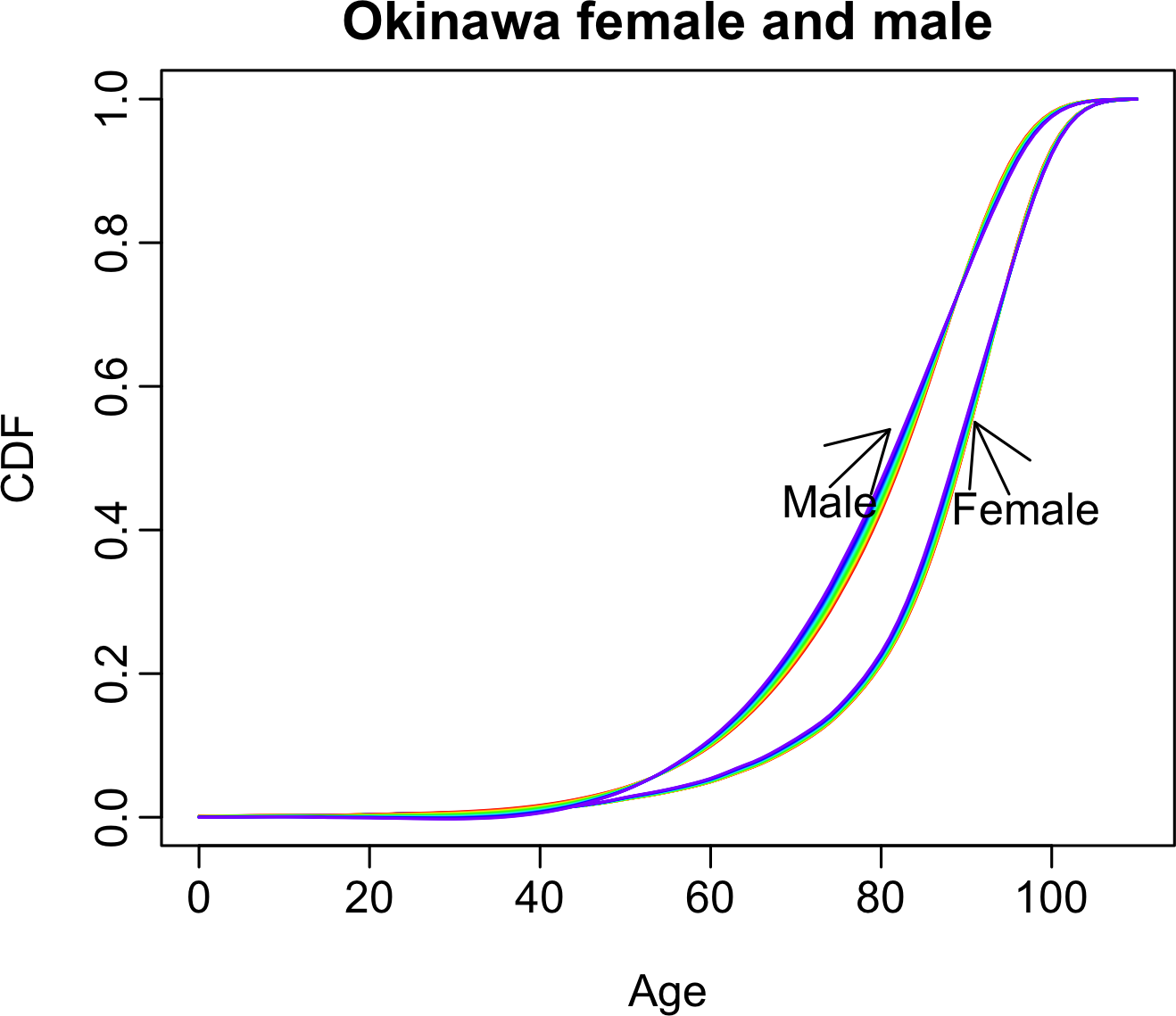}\label{fig:7b}}
\\
\subfloat[Life-table death count forecasts]
{\includegraphics[width=8.78cm]{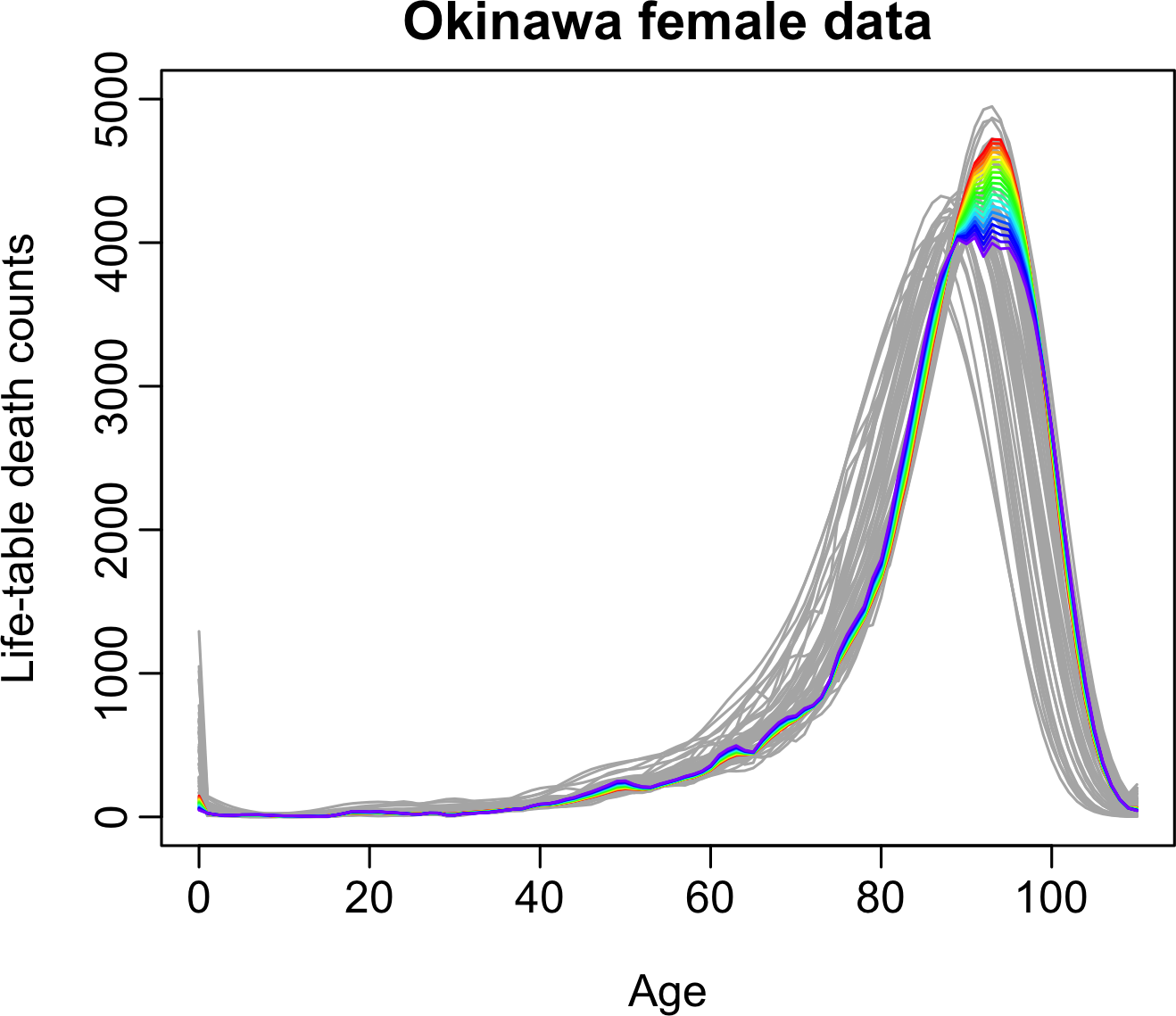}\label{fig:7c}}
\quad
\subfloat[Life-table death count forecasts]
{\includegraphics[width=8.78cm]{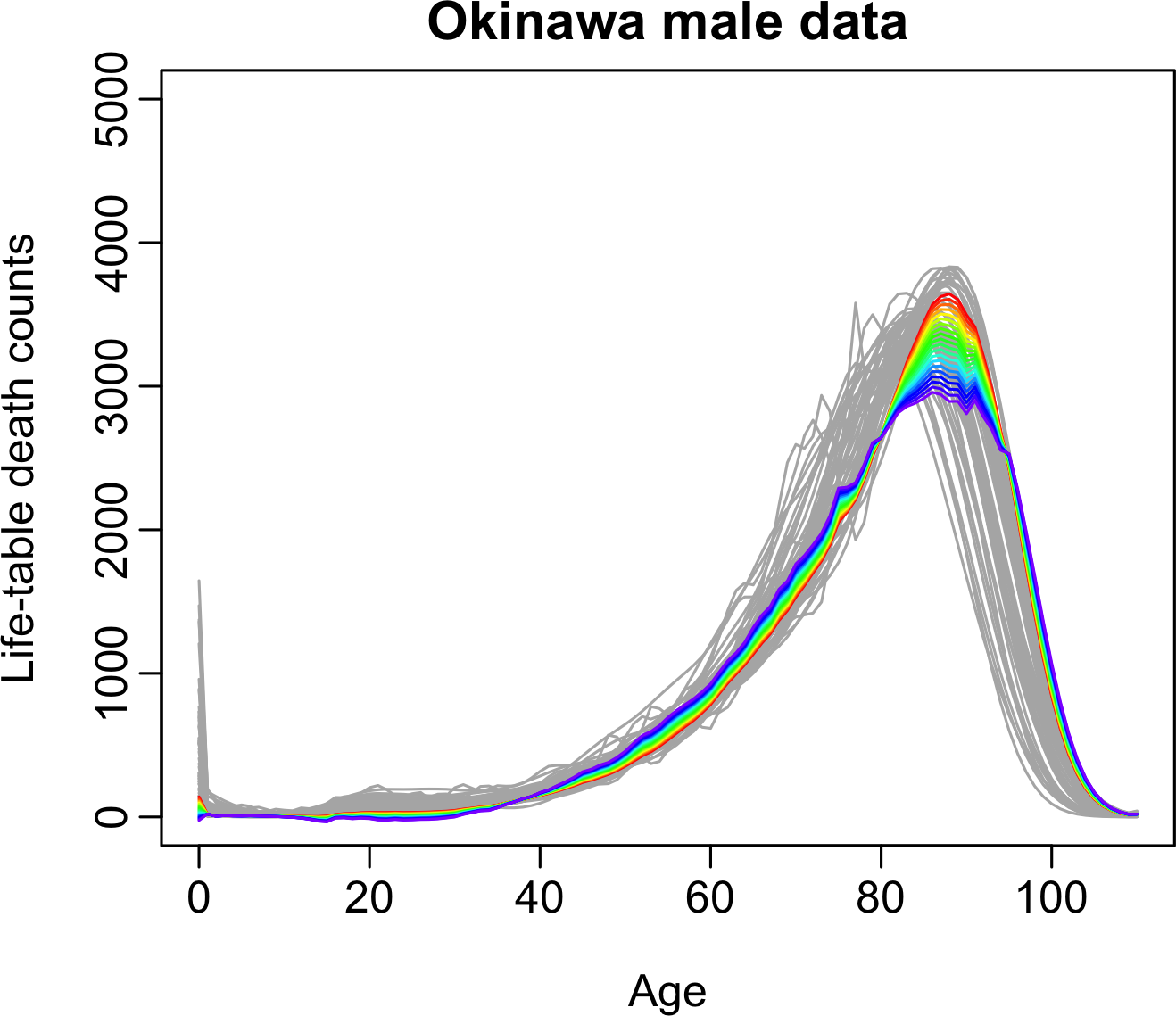}\label{fig:7d}}
\caption{\small Using the ETS univariate time-series forecasting method, we produce the 20-years-ahead forecasts of the regional gap and 20-years-ahead forecasts of the Okinawa female and male cumulative life-table death counts. By taking the first-order difference, we obtain female and male life-table death counts (radix of $10^5$) in Okinawa.\label{fig:7_add}}
\end{figure}

Since the national series has a higher data quality than the subnational series for most developed countries, using the same forecasting method, we obtain the 20-years-ahead forecasts of the national cumulative relative life-table death counts, denoted by $\widehat{D}_{n+h|n,x}^{\text{N}, g}$. In Figure~\ref{fig:7a}, we present the 20-years-ahead forecasts of the regional gap between Okinawa and national data, denoted by $\widehat{G}_{n+h|n,x}^{\text{Okinawa-N}, g}$ after taking the inverse Fisher $Z$ transformation. By adding the forecasts of the regional gap between national cumulative relative life-table death counts, we obtain the 20-years-ahead forecasts of the Okinawa cumulative relative life-table death counts, denoted by $\widehat{D}_{n+h|n,x}^{\text{Okinawa},g} = \widehat{D}_{n+h|n,x}^{\text{N}, g}+\widehat{G}_{n+h|n,x}^{\text{Okinawa-N},g}$, as shown in Figure~\ref{fig:7b}.

By taking the first-order differencing, we obtain
\begin{equation}
\widehat{d}_{n+h|n,x}^{s,g} = \Delta^x_{i=1}\widehat{D}_{n+h|n,i}^{s,g},\label{eq:2}
\end{equation}
where $s=1, 2, \dots,47$ and $g=\{\text{F},\text{M}\}$. In Figures~\ref{fig:7c} and~\ref{fig:7d}, we display the 20-years-ahead forecasts of the life-table death counts for females and males in Okinawa. Recall that Okinawa is the only prefecture with 51 years of data; our modeling framework can handle uneven data sample sizes.

\subsection{Double gap in the Japanese subnational female and male data}\label{sec:4.3}

\begin{figure}[!htb]
\centering
\subfloat[Japanese national female forecasts]
{\includegraphics[width=8.78cm]{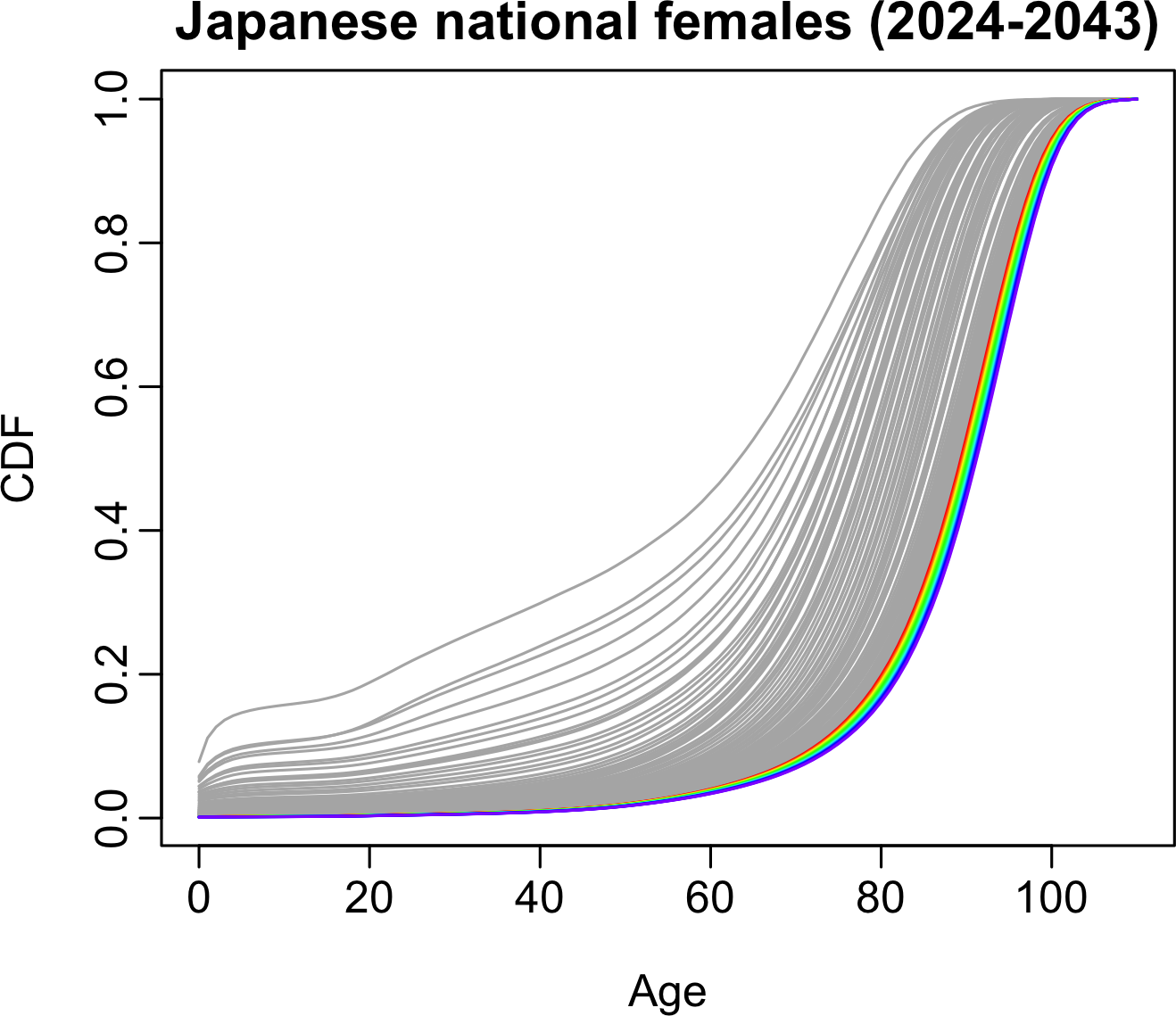}\label{fig:9a}}
\quad
\subfloat[Okinawa female forecasts]
{\includegraphics[width=8.78cm]{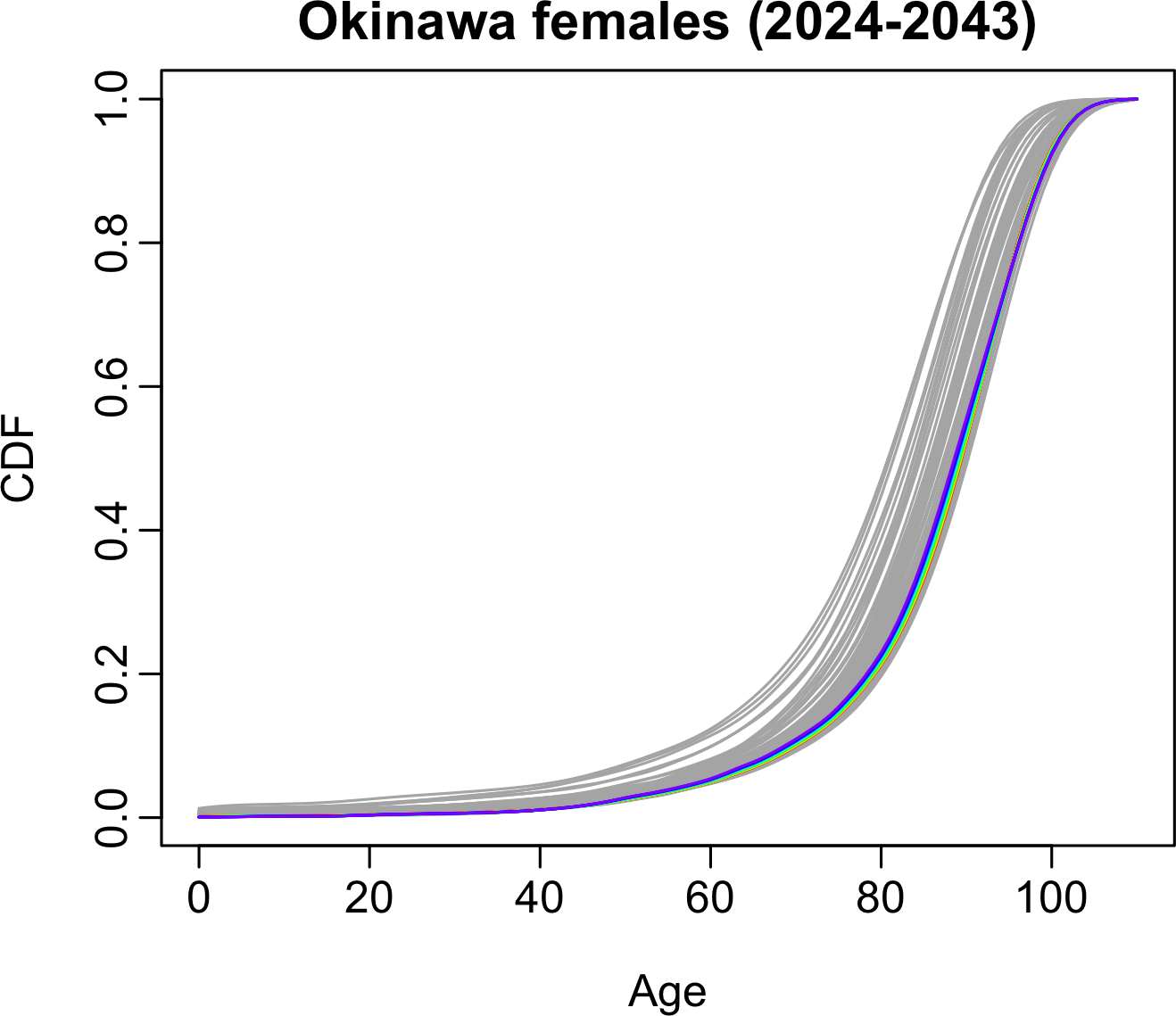}\label{fig:9b}}
\caption{\small The 20-years-ahead forecasts of the cumulative relative life-table death counts for Japanese national female data and for females in Okinawa.}\label{fig:9}
\end{figure}

The double gap is an example of a situation in which we consider both gender and regional gaps. We first produce 20-years-ahead CDF forecasts for Japanese national female data from 2024 to 2043 in Figure~\ref{fig:9a}. By modeling and forecasting the regional gap, we then obtain 20-years-ahead CDF forecasts for Okinawa females, as shown in Figure~\ref{fig:9b}. The region gap uses high-quality national data to guide the modeling of relatively lower-quality subnational data.

\begin{figure}[!htb]
\centering
\subfloat[Gender gap forecasts in Okinawa]
{\includegraphics[width=8.78cm]{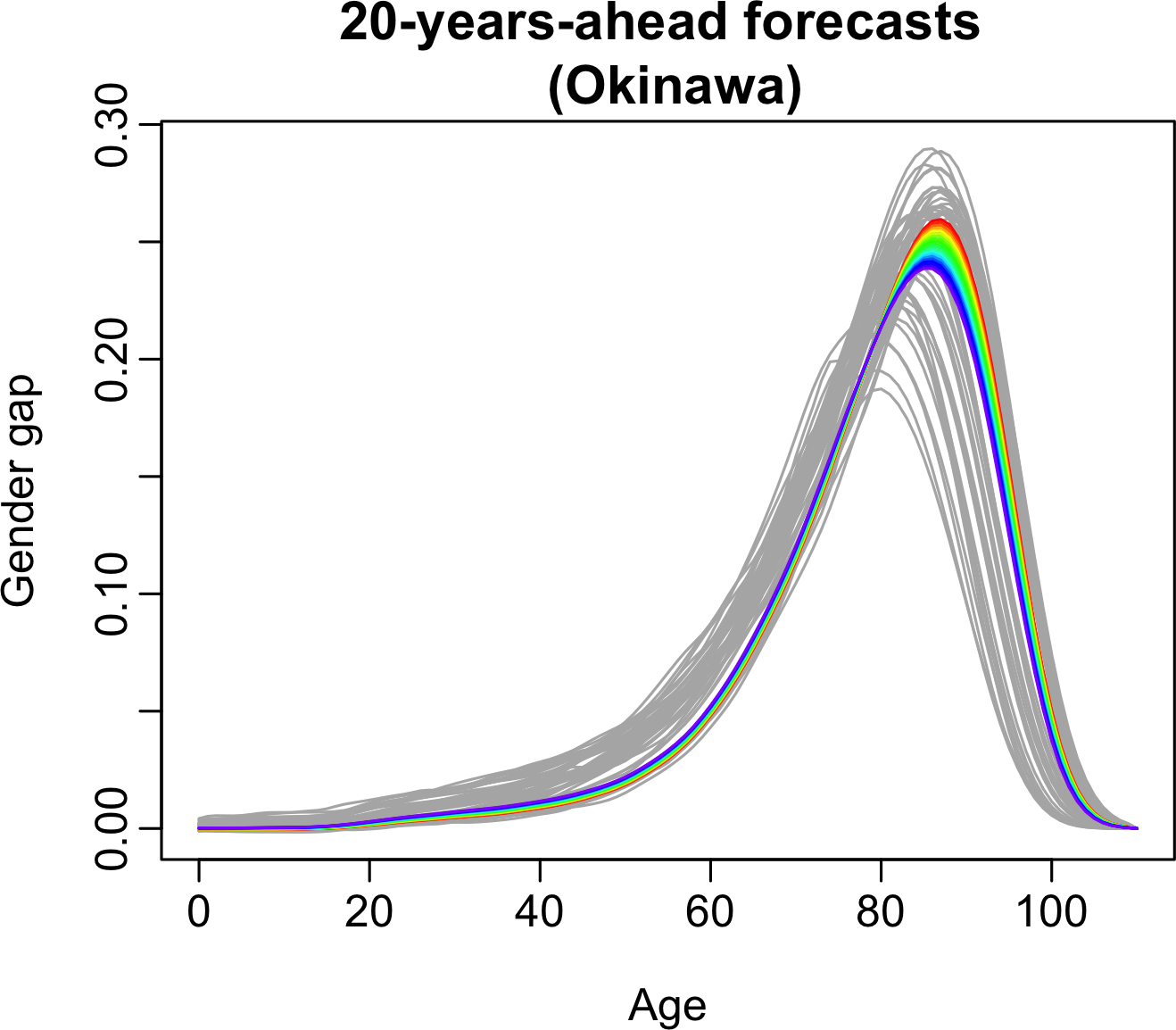}\label{fig:10a}}
\quad
\subfloat[Okinawa male forecasts]
{\includegraphics[width=8.78cm]{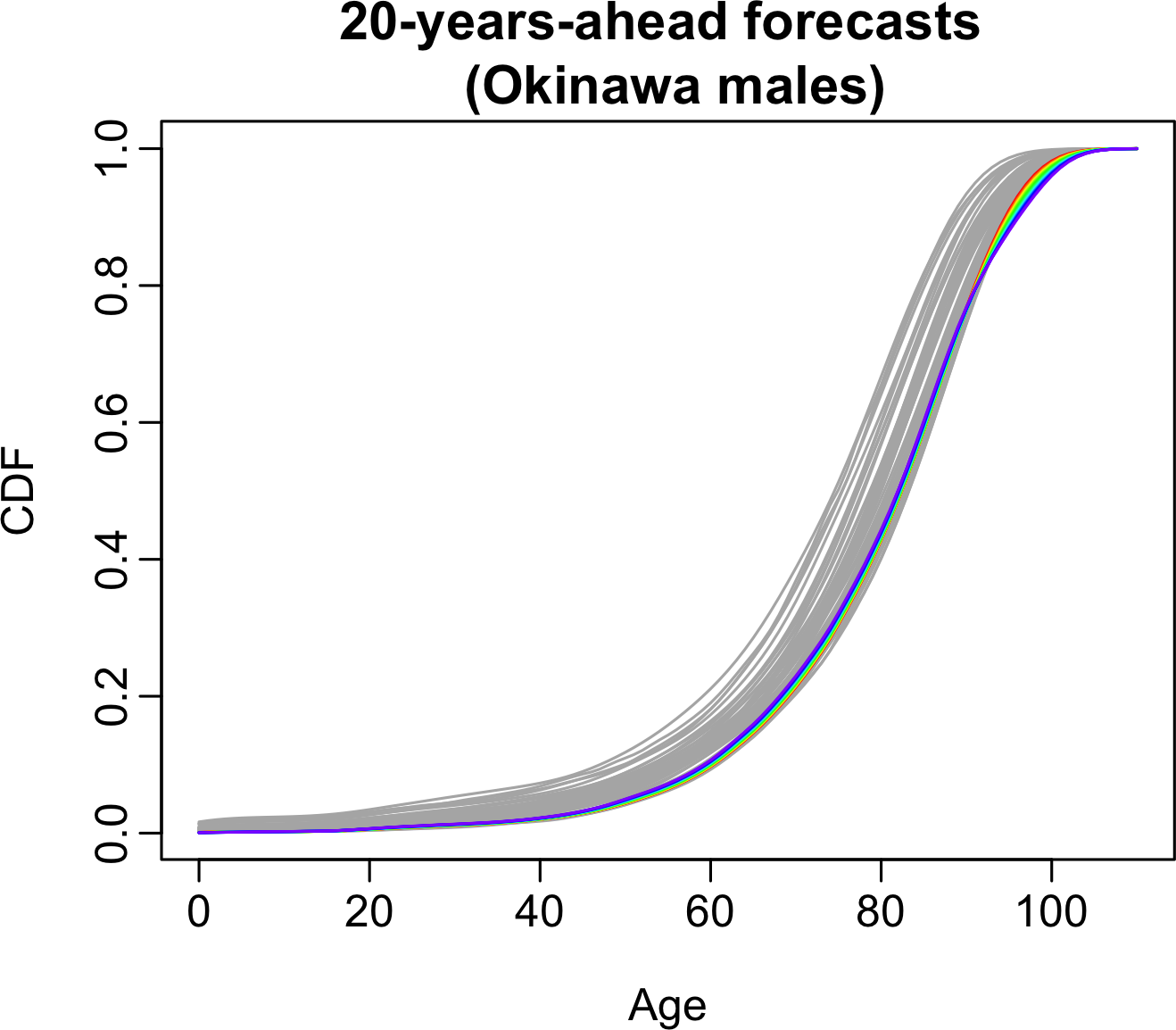}\label{fig:10b}}
\caption{\small By modeling and forecasting the gender gap in Okinawa, we obtain the 20-years-ahead CDF forecasts for males in Okinawa.}\label{fig:10}
\end{figure}

By exploring the gender gap between males and females in Okinawa, we forecast the 20-years-ahead gender gap in Figure~\ref{fig:10a}, from which we obtain the 20-years-ahead male CDF forecasts in Okinawa in Figure~\ref{fig:10b}.

\begin{figure}[!htb]
\centering
\subfloat[Female data]
{\includegraphics[width=8.79cm]{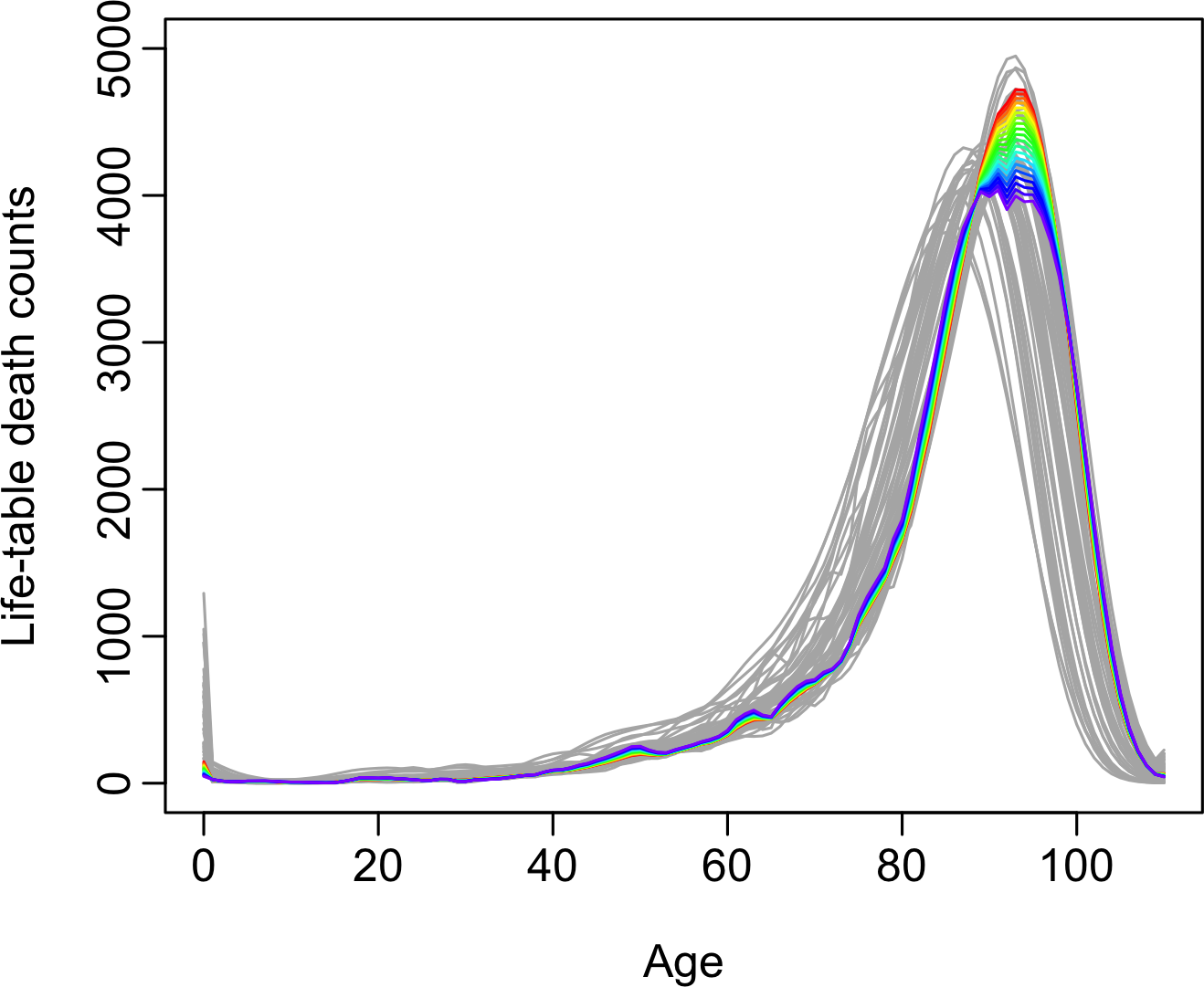}\label{fig:11a}}
\quad
\subfloat[Male data]
{\includegraphics[width=8.78cm]{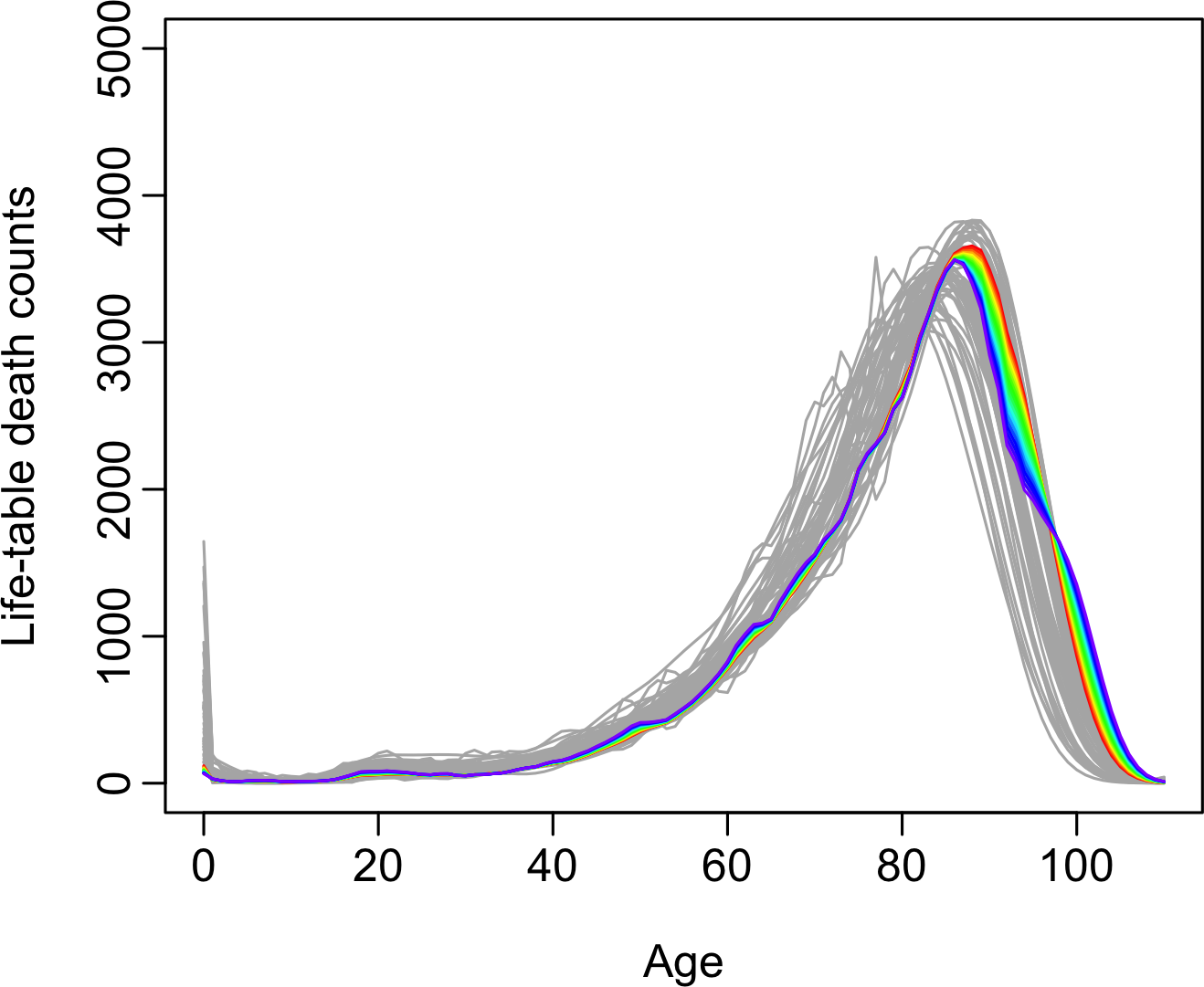}\label{fig:11b}}
\caption{\small The 20-years-ahead forecasts of the female and male life-table death counts in Okinawa.}\label{fig:11}
\end{figure}

From~\eqref{eq:2}, we obtain the 20-years-ahead life-table death counts for females and males in each prefecture, an example of which is Okinawa shown in Figure~\ref{fig:11}. The forecasts for both females and males indicate a tendency toward higher mortality, a surprising pattern that may be attributed to the recent COVID-19 pandemic.

As a sensitivity analysis, we present point and interval forecast errors using the age-period model in Appendix~\ref{sec:Appendix_B}.

\subsection{Sequential conformal prediction}\label{sec:4.4}

In statistics and machine learning, conformal prediction of \cite{SV08} and \cite{FZV23} is often used to construct probabilistic forecasts calibrated on out-of-sample errors. Compared with Bayesian or bootstrap methods, conformal prediction is faster computationally. Since its introduction in \cite{GVV98}, there have been many methodological advances in the field, resulting in a variety of conformal prediction extensions, including time series forecasting  \citep{ACT23}. Among various conformal prediction methods, we consider the sequential conformal prediction proposed by \cite{XX23}. Without calibrating parameters in a validation set, sequential conformal prediction can automatically adjust the predictive quantiles of absolute residuals as new data arrive.
\begin{center}
\begin{tikzpicture}[>=stealth, thick]
\coordinate (a) at (1,0);
\coordinate (b) at (6,0);
\coordinate (c) at (10,0);
\coordinate (d) at (14,0);
\draw[very thick] (a) -- (b);
\draw[decorate, decoration={brace, amplitude=6pt}]([yshift=2mm]a) -- ([yshift=2mm]b)
  node[midway, yshift=12pt] {Initial training};
\draw[very thick, dashed] (b) -- (c);
\draw[decorate, decoration={brace, amplitude=6pt}]([yshift=2mm]b) -- ([yshift=2mm]c)
  node[midway, yshift=12pt] {Initial validation};
\draw[very thick] (c) -- (d);
\draw[decorate, decoration={brace, amplitude=6pt}]([yshift=2mm]c) -- ([yshift=2mm]d)
  node[midway, yshift=12pt] {Test};
\node[below] at (a) {$6$};
\node[below] at (b) {$m+\ell-h$};
\node[below] at (c) {$m+\ell+h$};
\node[below] at (d) {$n$};
\end{tikzpicture}
\end{center}

Using the last 15 years as the test set, we compute absolute residuals using the other years. Absolute residuals are a common choice for a nonconformity score \citep{LGR+18}. At the quantile of $1-\alpha$, we fit a quantile regression on lagged residuals, where the order of the autoregression AR($p$) is determined by an information criterion, such as the Akaike information criterion (AIC). Conditional on the most recent $p$ number of absolute residuals as input, we produce a $h$-step-ahead forecast quantile, denoted as $\widehat{q}_{\alpha, (m+\ell)+h,x}^{s,g}$, where $\ell = 1, 2,\dots, (16-h)$. The prediction intervals of $d_{(m+\ell)+h,x}^{s,g}$ are then given by $\widehat{d}_{(m+\ell)+h,x}^{s,g}\pm \widehat{q}_{\alpha, (m+\ell)+h,x}^{s,g}$. Once $d_{(m+\ell)+h,x}^{s,g}$ arrives, we can update the absolute residuals and refit. We summarize the sequential conformal prediction with Algorithm~1.
\begin{algorithm}
\caption{Sequential conformal prediction}
\begin{algorithmic}[1] 
\Require National and subnational life-table death counts, forecast horizon $h=1, 2, \dots,15$; let $\alpha$ be a pre-specified level of significance
\For{$\ell = 1$ to $(16 - h)$}
    \State $n$ = number of years in a data set, $m$ = $n - 16$
    \State From initial training set $\zeta = 6:(m + \ell - h)$, produce $h$-step-ahead forecast $\widehat{d}_{\zeta+h,x}^{s,g}$ and compute the absolute residuals, $|\widehat{d}_{\zeta+h,x}^{s,g} - d_{\zeta+h,x}^{s,g}|$, where $\zeta$ is a jump-off year
    \For{$x= 1$ to $111$}
    \State Fit a quantile regression of the absolute residual for each $x$ at the $(1-\alpha)$ quantile
    \State Fit an AR$(p)$ model where order $p$ is determined by the AIC
    \State From the most recent $p$ absolute residuals for a given age, predict quantile $q^{s,g}_{\alpha, (m+\ell)+h, x}$
    \EndFor
    \State Using observations $1:(m + \ell)$, produce $h$-step-ahead forecast $\widehat{d}_{(m+\ell)+h,x}^{s,g}$
    \State The lower and upper bounds are $\widehat{d}_{(m+\ell)+h,x}^{s,g} \pm q^{s,g}_{\alpha, (m+\ell)+h, x}$, respectively
\EndFor 
\\
\Return For holdout data $(d^{s,g}_{(m+\ell)+h,x},\dots,d^{s,g}_{n,x})$, we evaluate its empirical coverage probability (ECP) and mean interval score
\end{algorithmic}
\end{algorithm}

\vspace{-.2in}

\section{Comparison of forecast accuracy}\label{sec:5}

\subsection{Expanding-window forecast scheme}\label{sec:5.1}

\begin{figure}[!htb]
\begin{center}
\begin{tikzpicture}
\draw[->] (0,0) -- (10,0) node[right] {Time};
    
\draw[fill=blue!20] (0,-0.5) rectangle (3,0.5) node[midway] {Train};
\draw[fill=red!20] (3,-0.5) rectangle (3.5,0.5) node[midway] {F};
    
\draw[fill=blue!20] (0,-1.5) rectangle (5,-0.5) node[midway] {Train};
\draw[fill=red!20] (5,-1.5) rectangle (5.5,-0.5) node[midway] {F};
    
\draw[fill=blue!20] (0,-2.5) rectangle (7,-1.5) node[midway] {Train};
\draw[fill=red!20] (7,-2.5) rectangle (7.5,-1.5) node[midway] {F};
    
\draw[fill=blue!20] (0,-3.5) rectangle (9,-2.5) node[midway] {Train};
\draw[fill=red!20] (9,-3.5) rectangle (9.5,-2.5) node[midway] {F};
    
\node[left] at (0,0) {1947:2008};
\node[left] at (0,-1) {1947:2009};
\node[left] at (0,-2) {\hspace{-0.8in}{$\vdots$}};
\node[left] at (0,-3) {1947:2022};
    
\draw[fill=blue!20] (6.5,1) rectangle (7,1.5);
\node[right] at (7,1.25) {Training Window};
\draw[fill=red!20] (6.5,0.5) rectangle (7,1);
\node[right] at (7,0.75) {Forecast (F) when $h=1$};
\end{tikzpicture}
\end{center}
\caption{\small A diagram of the expanding-window forecast scheme. The numbers on the left show the year indices for the training set. The initial training samples span 1947-2008, and we sequentially compute $h$-step-ahead forecasts with the last training samples spanning 1947-2022.}\label{fig:tikz_9}
\end{figure}
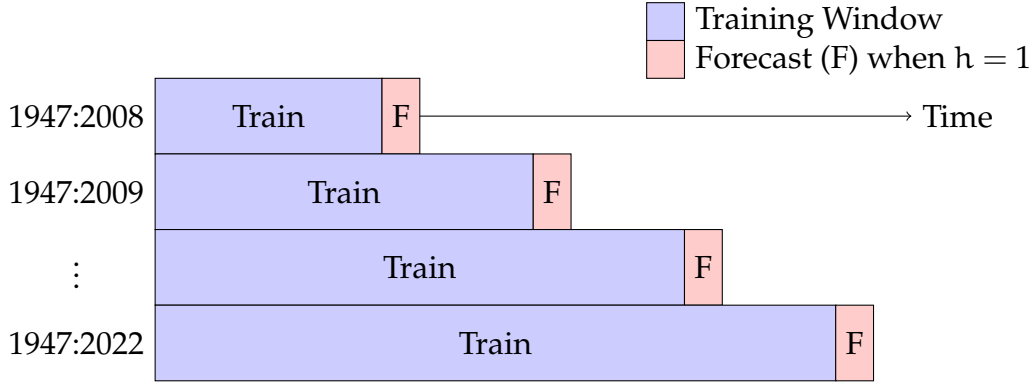

To assess the accuracy of our models, we examine differences between holdout test data and their forecasts through a back-testing exercise in the spirit of \cite{BHT+06} and \cite{JK11}. From 1947 to 2023, there are 77 years of data for Japan across 46 of 47 prefectures, with the exception of Okinawa, where the data period covers 1973 to 2023. For all series, we divide their data into a training set and a testing set. The testing set contains the most recent 15 years of data, while the initial training set can vary depending on the data length. Using the forecasting method, we produce one- to 15-step-ahead forecasts. Using an expanding window in Figure~\ref{fig:tikz_9}, we increase the training set by one year and then produce one- to 14-step-ahead forecasts. We iterate this process until reaching the end of the data period. This produces 15 one-step-ahead forecasts, 14 two-step-ahead forecasts, $\dots$, one 15-step-ahead forecast.

\subsection{Point forecast accuracy metrics}\label{sec:5.2}

Since the life-table death counts can be treated as a probability density function, we consider two density-evaluation metrics. The metrics are the Kullback-Leibler divergence (KLD) \citep{KL51} and Jensen-Shannon divergence (JSD) \citep{Shannon48}. Both metrics are non-negative; thus, the best forecasting method tends to have the smallest divergence measures.

The KLD measures information loss by approximating an unknown density with its forecast. For two probability density functions, denoted by $d_{m+\xi,x}^{s,g}$ and $\widehat{d}_{m+\xi,x}^{s,g}$, the discrete KLD is given as
\begin{align*}
\text{KLD}(h) 	= \ & D_{\text{KL}}\big(d_{m+\xi,x}^{s,g}||\widehat{d}_{m+\xi,x}^{s,g}\big) + D_{\text{KL}}\big(\widehat{d}_{m+\xi,x}^{s,g}||d_{m+\xi,x}^{s,g}\big) \\
			= \ & \frac{1}{111\times (16-h)}\sum^{15}_{\xi=h}\sum_{x=1}^{111}\big[d_{m+\xi,x}^{s,g}\big(\ln d_{m+\xi,x}^{s,g} - \ln \widehat{d}_{m+\xi,x}^{s,g}\big) + \widehat{d}_{m+\xi,x}^{s,g} \big(\ln \widehat{d}_{m+\xi,x}^{s,g} - \ln d_{m+\xi,x}^{s,g}\big)\big],
\end{align*}
where $\xi$ denotes the years in the forecasting period, and $m$ denotes the end of the training data, known as the jump-off year in demography.

Although we have holdout samples in the forecasting period, they may not reflect the actual densities. An alternative is the JSD, which can be viewed as a symmetric, smoothed version of the KLD. The JSD is defined by
\begin{equation*}
\text{JSD}(h) = \frac{1}{2}D_{\text{KL}}\left[d_{m+\xi,x}^{s,g}||\delta_{m+\xi,x}^{s,g}\right] + \frac{1}{2}D_{\text{KL}}\left[\widehat{d}_{m+\xi,x}^{s,g}||\delta_{m+\xi,x}^{s,g}\right],
\end{equation*}
where $\delta_{m+\xi,x}^{s,g}$ measures a common quantity between $d_{m+\xi,x}^{s,g}$ and $\widehat{d}_{m+\xi,x}^{s,g}$. An example of $\delta_{m+\xi,x}^{s,g}$ can be its geometric mean $\delta_{m+\xi,x}^{s,g}=\sqrt{d_{m+\xi,x}^{s,g}\widehat{d}_{m+\xi,x}^{s,g}}$.

\subsection{Comparison of point forecast accuracy}\label{sec:5.3}

\begin{figure}[!htb]
\centering
\includegraphics[width=8.65cm]{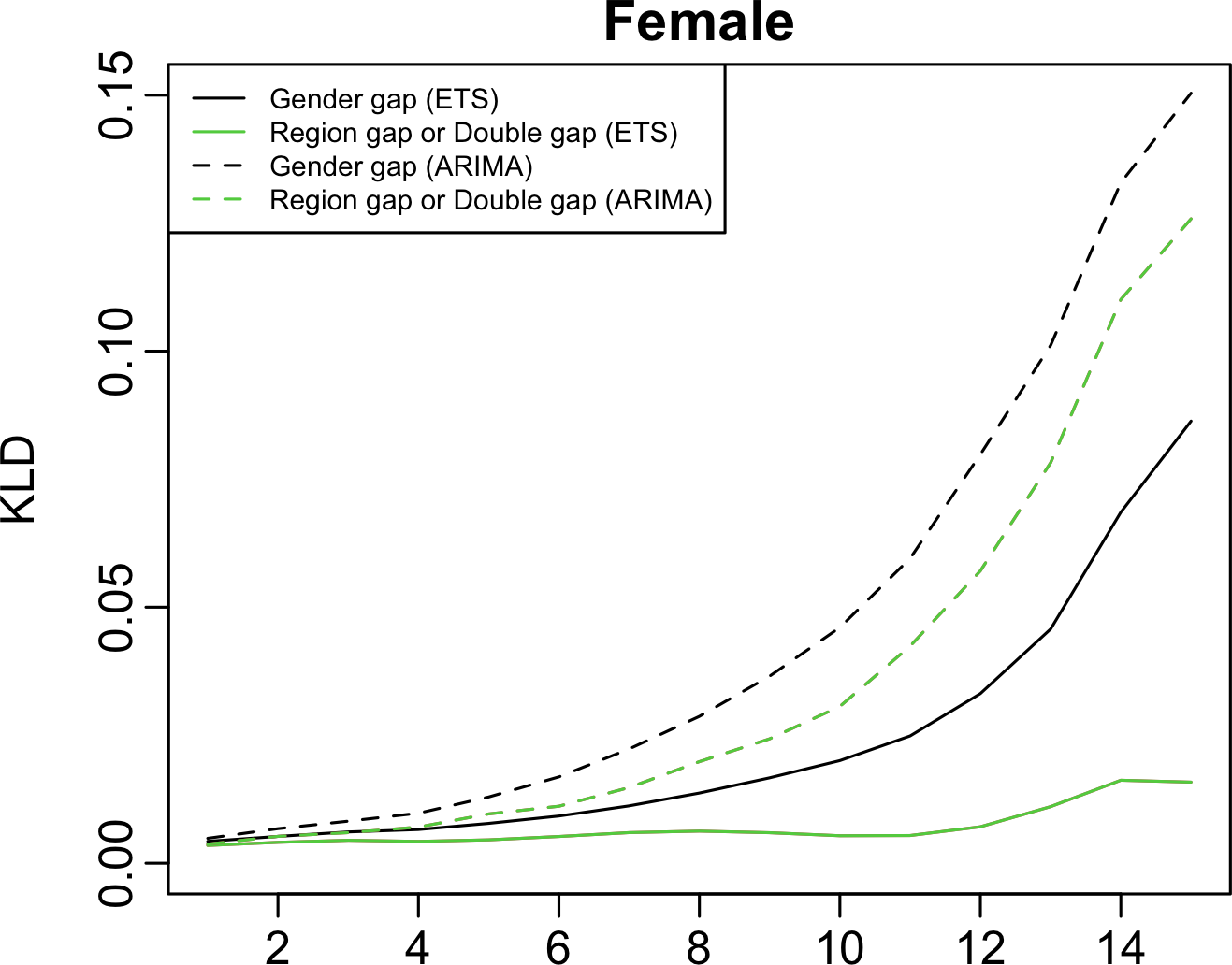}
\quad
\includegraphics[width=8.65cm]{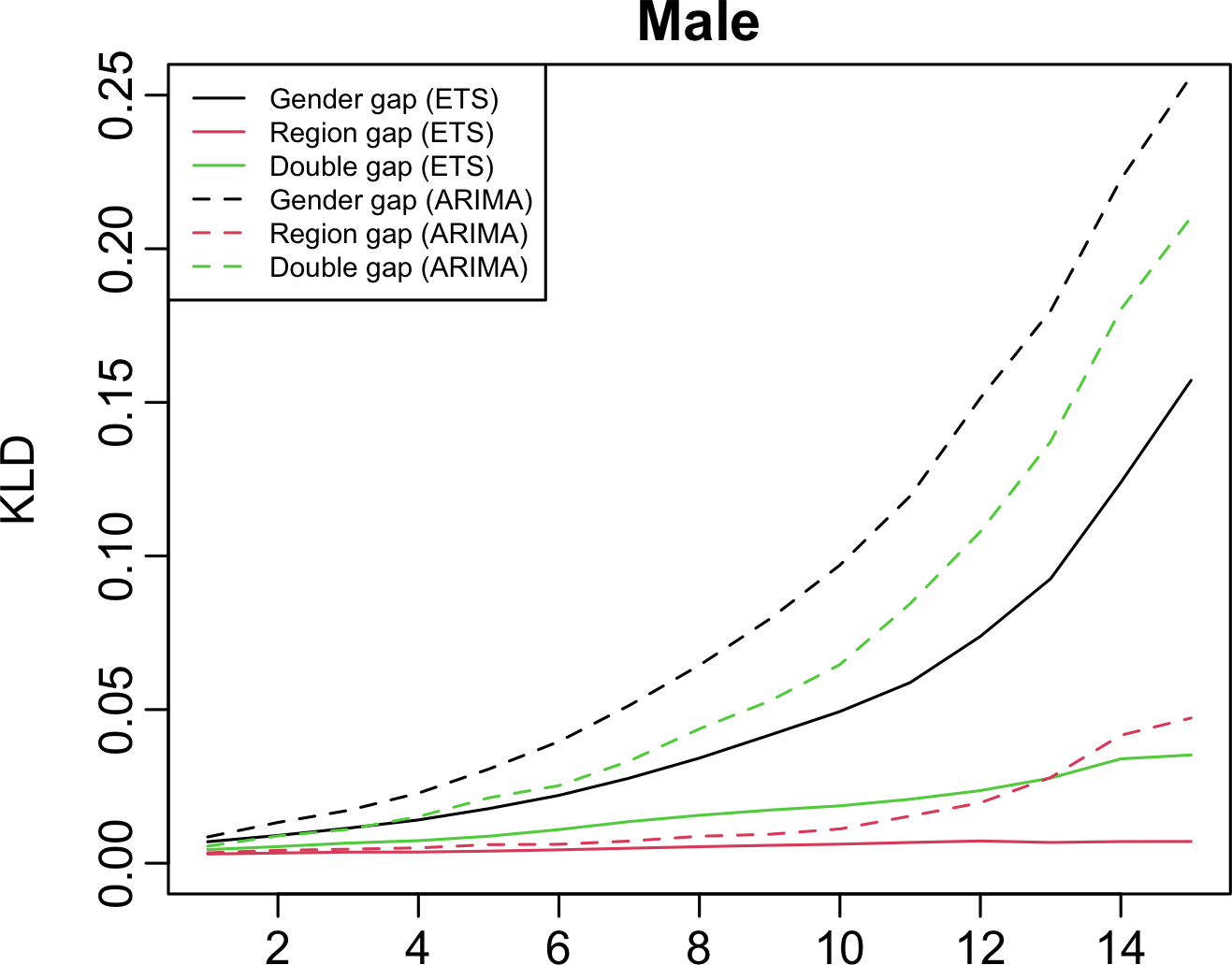}
\\
\includegraphics[width=8.65cm]{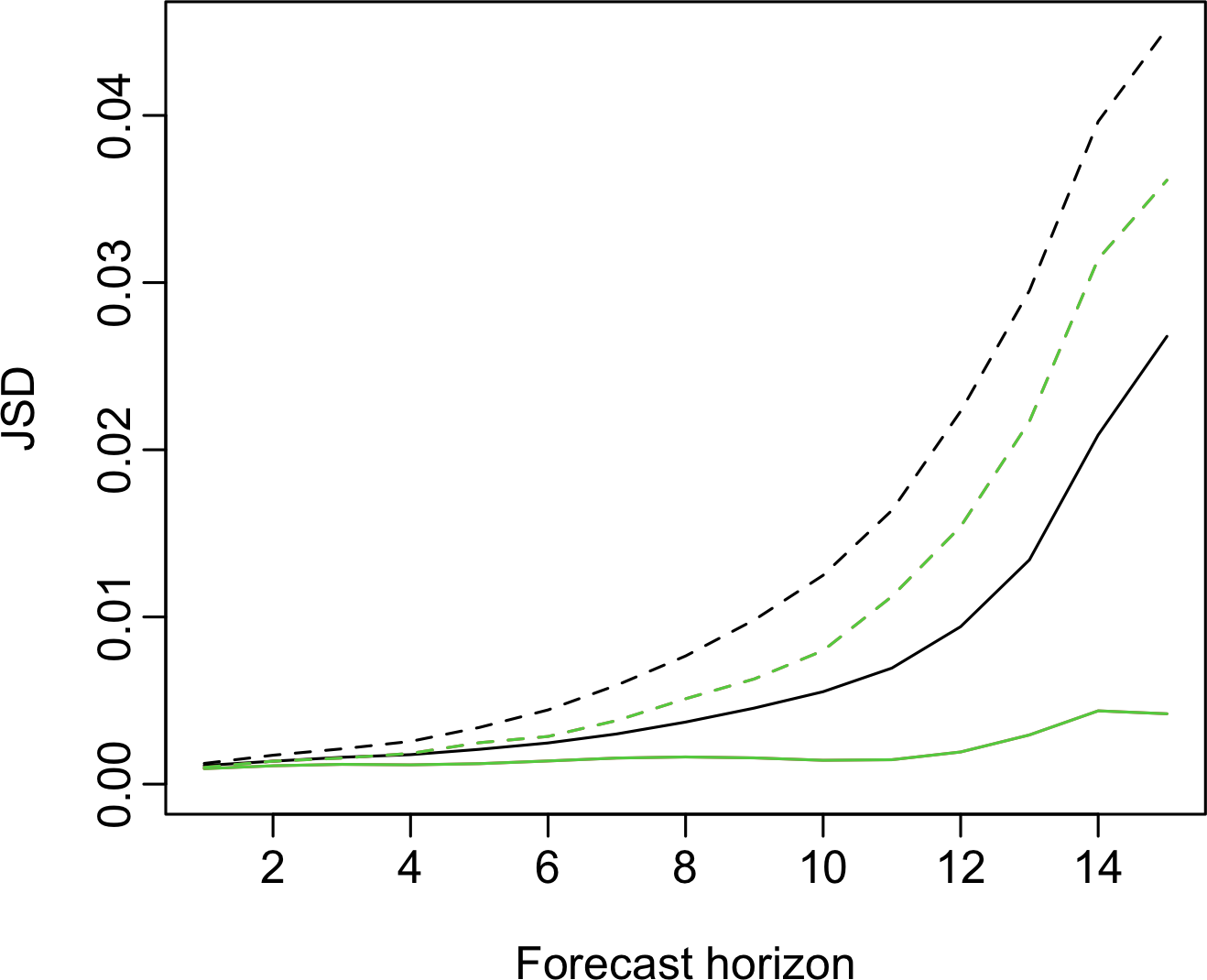}
\quad
\includegraphics[width=8.65cm]{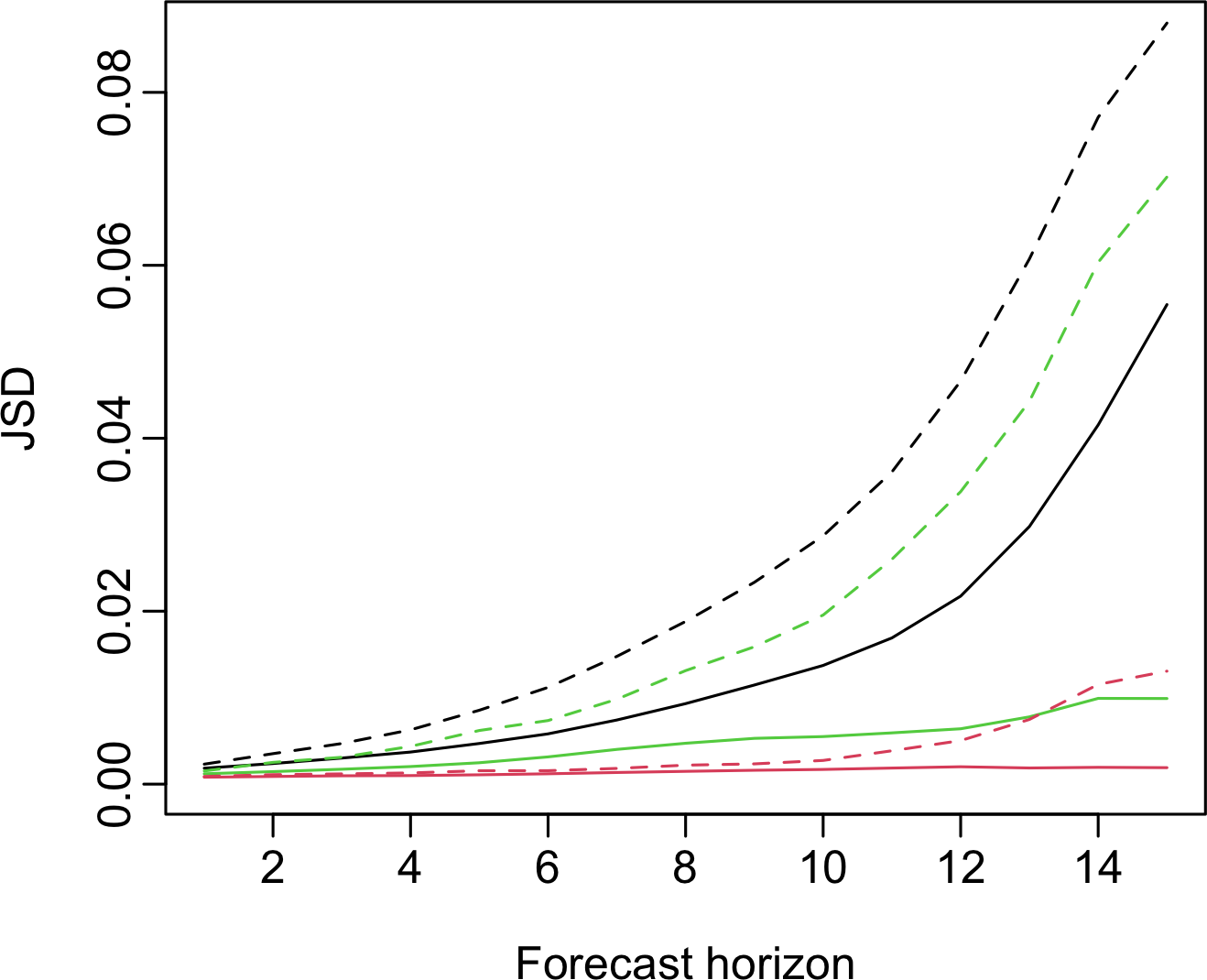}
\caption{\small Averaged over 47 prefectures, the one-to-15-step-ahead point forecast accuracy comparison using the functional time-series forecasting method with the gender gap, regional gap, and double gap. For the female data, the regional gap and double gap produce the same forecasts, due to the method design.}\label{fig:12}
\end{figure}

In Figure~\ref{fig:12}, we display one-to-15-step-ahead point forecast errors, as measured by the KLD and JSD. For the female data, the functional time-series forecasting method produces the same errors for the region and double gaps. For modeling male data, the functional time-series forecasting method produces the smallest errors using the regional gap, followed by the double gap.

In Table~\ref{tab:2}, we display horizon-specific point forecast errors, measured by the KLD and JSD, using different gap models. The region gap is recommended as the one with the smallest overall KLD and JSD. Between ARIMA and ETS, it is advantageous to use ETS for producing point forecasts, as it not only produces smaller errors but is also computationally faster. Compared to ARIMA, ETS can handle non-stationarity by explicitly modeling trend and seasonality. Computationally, the orders of the ARIMA and ETS are automatically determined using corrected AIC in the \textit{forecast} package \citep{HAB26} in \Rlogo \ \citep{Team25}.

\begin{table}[!htb]
\centering
\tabcolsep 0.04in
\renewcommand{\arraystretch}{1.03}
\caption{Averaged over 47 prefectures, we compare the one- to 15-step-ahead point forecast accuracy, as measured by the KLD and JSD, using the functional time-series forecasting method with the gender gap, regional gap, and double gap.}\label{tab:2}
\begin{small}
\begin{tabular}{@{}lrrrrrrrrrrrr@{}}
\toprule
    &  \multicolumn{6}{c}{KLD} &  \multicolumn{6}{c}{JSD} \\
    \cmidrule(lr){2-7}\cmidrule(lr){8-13}
    & \multicolumn{3}{c}{Female} & \multicolumn{3}{c}{Male} & \multicolumn{3}{c}{Female} & \multicolumn{3}{c}{Male} \\
    \cmidrule(lr){2-4}\cmidrule(lr){5-7}\cmidrule(lr){8-10}\cmidrule(lr){11-13}
$h$ & Gender & Region & Double & Gender & Region & Double & Gender & Region & Double & Gender & Region & Double \\ 
\midrule
\multicolumn{2}{l}{\hspace{-.08in} \underline{ETS}} & \\
1 & 0.0043 & 0.0035 & 0.0035 & 0.0070 & 0.0030 & 0.0044 & 0.0011 & 0.0009 & 0.0009 & 0.0018 & 0.0008 & 0.0012 \\ 
  2 & 0.0052 & 0.0041 & 0.0041 & 0.0090 & 0.0033 & 0.0054 & 0.0014 & 0.0011 & 0.0011 & 0.0024 & 0.0009 & 0.0014 \\ 
  3 & 0.0061 & 0.0045 & 0.0045 & 0.0114 & 0.0036 & 0.0066 & 0.0016 & 0.0012 & 0.0012 & 0.0030 & 0.0010 & 0.0017 \\ 
  4 & 0.0066 & 0.0043 & 0.0043 & 0.0141 & 0.0036 & 0.0073 & 0.0018 & 0.0012 & 0.0012 & 0.0037 & 0.0010 & 0.0020 \\ 
  5 & 0.0078 & 0.0046 & 0.0046 & 0.0178 & 0.0039 & 0.0088 & 0.0021 & 0.0012 & 0.0012 & 0.0047 & 0.0011 & 0.0025 \\ 
  6 & 0.0092 & 0.0052 & 0.0052 & 0.0221 & 0.0044 & 0.0109 & 0.0025 & 0.0014 & 0.0014 & 0.0058 & 0.0012 & 0.0032 \\ 
  7 & 0.0112 & 0.0060 & 0.0060 & 0.0277 & 0.0049 & 0.0136 & 0.0030 & 0.0016 & 0.0016 & 0.0074 & 0.0013 & 0.0040 \\ 
  8 & 0.0137 & 0.0063 & 0.0063 & 0.0342 & 0.0054 & 0.0156 & 0.0037 & 0.0016 & 0.0016 & 0.0093 & 0.0015 & 0.0047 \\ 
  9 & 0.0167 & 0.0060 & 0.0060 & 0.0417 & 0.0058 & 0.0173 & 0.0046 & 0.0016 & 0.0016 & 0.0115 & 0.0016 & 0.0053 \\ 
  10 & 0.0200 & 0.0054 & 0.0054 & 0.0494 & 0.0062 & 0.0187 & 0.0055 & 0.0014 & 0.0014 & 0.0137 & 0.0017 & 0.0055 \\ 
  11 & 0.0249 & 0.0054 & 0.0054 & 0.0588 & 0.0068 & 0.0208 & 0.0069 & 0.0015 & 0.0015 & 0.0169 & 0.0019 & 0.0059 \\ 
  12 & 0.0331 & 0.0071 & 0.0071 & 0.0738 & 0.0072 & 0.0236 & 0.0094 & 0.0019 & 0.0019 & 0.0217 & 0.0020 & 0.0064 \\ 
  13 & 0.0457 & 0.0110 & 0.0110 & 0.0926 & 0.0068 & 0.0277 & 0.0134 & 0.0029 & 0.0029 & 0.0298 & 0.0019 & 0.0078 \\ 
  14 & 0.0685 & 0.0162 & 0.0162 & 0.1239 & 0.0070 & 0.0340 & 0.0209 & 0.0044 & 0.0044 & 0.0415 & 0.0019 & 0.0099 \\ 
  15 & 0.0863 & 0.0158 & 0.0158 & 0.1572 & 0.0071 & 0.0352 & 0.0268 & 0.0042 & 0.0042 & 0.0555 & 0.0019 & 0.0099 \\ 
  \midrule
  Mean & 0.0240 & 0.0070 & 0.0070 & 0.0494 & 0.0053 & 0.0167 & 0.0070 & 0.0019 & 0.0019 & 0.0153 & 0.0014 & 0.0048 \\ 
\midrule  
\multicolumn{2}{l}{\hspace{-.08in} \underline{ARIMA}} & \\
   1 & 0.0049 & 0.0037 & 0.0037 & 0.0085 & 0.0034 & 0.0055 & 0.0012 & 0.0010 & 0.0010 & 0.0023 & 0.0009 & 0.0015 \\ 
  2 & 0.0067 & 0.0052 & 0.0052 & 0.0133 & 0.0042 & 0.0089 & 0.0017 & 0.0014 & 0.0014 & 0.0035 & 0.0011 & 0.0025 \\ 
  3 & 0.0082 & 0.0060 & 0.0060 & 0.0172 & 0.0046 & 0.0110 & 0.0021 & 0.0016 & 0.0016 & 0.0047 & 0.0012 & 0.0031 \\ 
  4 & 0.0098 & 0.0071 & 0.0071 & 0.0228 & 0.0050 & 0.0152 & 0.0026 & 0.0018 & 0.0018 & 0.0063 & 0.0013 & 0.0044 \\ 
  5 & 0.0129 & 0.0096 & 0.0096 & 0.0306 & 0.0060 & 0.0213 & 0.0034 & 0.0025 & 0.0025 & 0.0085 & 0.0015 & 0.0062 \\ 
  6 & 0.0169 & 0.0111 & 0.0111 & 0.0396 & 0.0062 & 0.0253 & 0.0044 & 0.0029 & 0.0029 & 0.0112 & 0.0016 & 0.0074 \\ 
  7 & 0.0223 & 0.0148 & 0.0148 & 0.0513 & 0.0072 & 0.0333 & 0.0059 & 0.0038 & 0.0038 & 0.0148 & 0.0018 & 0.0098 \\ 
  8 & 0.0287 & 0.0198 & 0.0198 & 0.0644 & 0.0088 & 0.0437 & 0.0077 & 0.0051 & 0.0051 & 0.0188 & 0.0022 & 0.0131 \\ 
  9 & 0.0365 & 0.0243 & 0.0243 & 0.0795 & 0.0094 & 0.0528 & 0.0098 & 0.0063 & 0.0063 & 0.0233 & 0.0023 & 0.0159 \\ 
  10 & 0.0460 & 0.0306 & 0.0306 & 0.0970 & 0.0112 & 0.0647 & 0.0125 & 0.0080 & 0.0080 & 0.0287 & 0.0027 & 0.0196 \\ 
  11 & 0.0595 & 0.0424 & 0.0424 & 0.1195 & 0.0153 & 0.0845 & 0.0164 & 0.0112 & 0.0112 & 0.0361 & 0.0039 & 0.0260 \\ 
  12 & 0.0798 & 0.0571 & 0.0571 & 0.1514 & 0.0197 & 0.1079 & 0.0223 & 0.0154 & 0.0154 & 0.0466 & 0.0050 & 0.0338 \\ 
  13 & 0.1011 & 0.0782 & 0.0782 & 0.1798 & 0.0279 & 0.1372 & 0.0295 & 0.0217 & 0.0217 & 0.0607 & 0.0075 & 0.0443 \\ 
  14 & 0.1329 & 0.1101 & 0.1101 & 0.2226 & 0.0417 & 0.1802 & 0.0396 & 0.0314 & 0.0314 & 0.0771 & 0.0115 & 0.0603 \\ 
  15 & 0.1504 & 0.1258 & 0.1258 & 0.2562 & 0.0473 & 0.2103 & 0.0452 & 0.0361 & 0.0361 & 0.0880 & 0.0131 & 0.0702 \\ 
  \midrule
  Mean & 0.0478 & 0.0364 & 0.0364 & 0.0902 & 0.0145 & 0.0668 & 0.0136 & 0.0100 & 0.0100 & 0.0287 & 0.0038 & 0.0212 \\ 
   \bottomrule
\end{tabular}
\end{small}
\end{table}

These results indicate that it is more accurate to model the regional gap than the gender gap. Regional gaps between subnational and national data tend to have a higher data quality than the gender gaps between female and male data within a given prefecture.

\subsection{Interval forecast accuracy metrics}\label{sec:5.4}

To evaluate interval forecast accuracy, we consider the coverage probability difference (CPD) between the empirical and nominal coverage probabilities and the mean interval score of \cite{GR07}. For each year in the forecasting period, the $h$-step-ahead prediction intervals are computed at the $100(1-\alpha)\%$ nominal coverage probability. We consider the common case of the symmetric $100(1-\alpha)\%$ prediction intervals, with lower and upper bounds that are predictive quantiles at $\alpha/2$ and $1-\alpha/2$, denoted $\widehat{d}_{m+\xi,x}^{{\text{lb}}, s, g}$ and $\widehat{d}_{m+\xi,x}^{{\text{ub}}, s, g}$. The CPD is defined as
\begin{equation*}
\text{CPD}_h^{s,g} = \frac{1}{111\times (16-h)}\sum^{15}_{\xi=h}\sum^{111}_{x=1}\left[\mathds{1}\big\{d_{m+\xi,x}^{s,g}>\widehat{d}_{m+\xi,x}^{{\text{ub}}, s, g}\big\} + \mathds{1}\big\{d_{m+\xi,x}^{s,g} < \widehat{d}_{m+\xi,x}^{{\text{lb}}, s, g}\big\}\right].
\end{equation*}
For different years in the forecasting period, the mean CPD is defined by 
\begin{equation*}
\overline{\text{CPD}}^{s,g} = \frac{1}{15}\sum^{15}_{h=1}\text{CPD}_h^{s,g}.
\end{equation*}

As defined by \cite{GR07}, a scoring rule for the prediction intervals at time point $d_{m+\xi,x}^{s,g}$ is
\begin{align*}
S_{\alpha,\xi}\big(\widehat{d}_{m+\xi,x}^{\text{lb},s,g}, \widehat{d}_{m+\xi,x}^{\text{ub},s,g}, d_{m+\xi,x}^{s,g}\big) = \big(\widehat{d}_{m+\xi,x}^{\text{ub},s,g}, d_{m+\xi,x}^{s,g} - \widehat{d}_{m+\xi,x}^{\text{lb},s,g}\big) &+ \frac{2}{\alpha}\big(\widehat{d}_{m+\xi,x}^{\text{lb},s,g} - d_{m+\xi,x}^{s,g}\big)\mathds{1}\big\{d_{m+\xi,x}^{s,g}<\widehat{d}_{m+\xi,x}^{\text{lb},s,g}\big\} \\
&+\frac{2}{\alpha}\big(d_{m+\xi,x}^{s,g} - \widehat{d}_{m+\xi,x}^{\text{ub},s,g}\big)\mathds{1}\big\{d_{m+\xi,x}^{s,g}>\widehat{d}_{m+\xi,x}^{\text{ub},s,g}\big\},
\end{align*}
where $\mathds{1}\{\cdot\}$ represents the binary indicator function and $\alpha$ denotes the level of significance, customarily $\alpha = 0.2$ or 0.05. Averaging over different \textit{ages} and \textit{years} in the forecasting period, the mean interval score $\overline{S}_{\alpha}\big(\widehat{d}^{\text{lb},s,g}, \widehat{d}^{\text{ub},s,g}, d^{s,g}\big)$ is given as
\begin{equation*}
\overline{S}_{\alpha}\big(\widehat{d}^{\text{lb},s,g}, \widehat{d}^{\text{ub},s,g}, d^{s,g}\big) = \frac{1}{111\times (16-h)}\sum^{15}_{\xi=h}\sum^{111}_{x=1}S_{\alpha,\xi}\big(\widehat{d}_{m+\xi,x}^{\text{lb},s,g}, \widehat{d}_{m+\xi,x}^{\text{ub},s,g}, d_{m+\xi,x}^{s,g}\big).
\end{equation*}
The mean interval score rewards a narrow prediction interval, if and only if the holdout actual observation lies within the prediction interval. The optimal interval score is achieved when $d_{m+\xi,x}^{s,g}$ lies between $\widehat{d}_{m+\xi,x}^{\text{lb},s,g}$ and $\widehat{d}_{m+\xi,x}^{\text{ub},s,g}$, and the distance between $\widehat{d}_{m+\xi,x}^{\text{lb},s,g}$ and $\widehat{d}_{m+\xi,x}^{\text{ub},s,g}$ is minimal.

\subsection{Comparison of interval forecast accuracy: ETS method}\label{sec:5.5}

\begin{figure}[!htb]
\centering
\includegraphics[width=8.7cm]{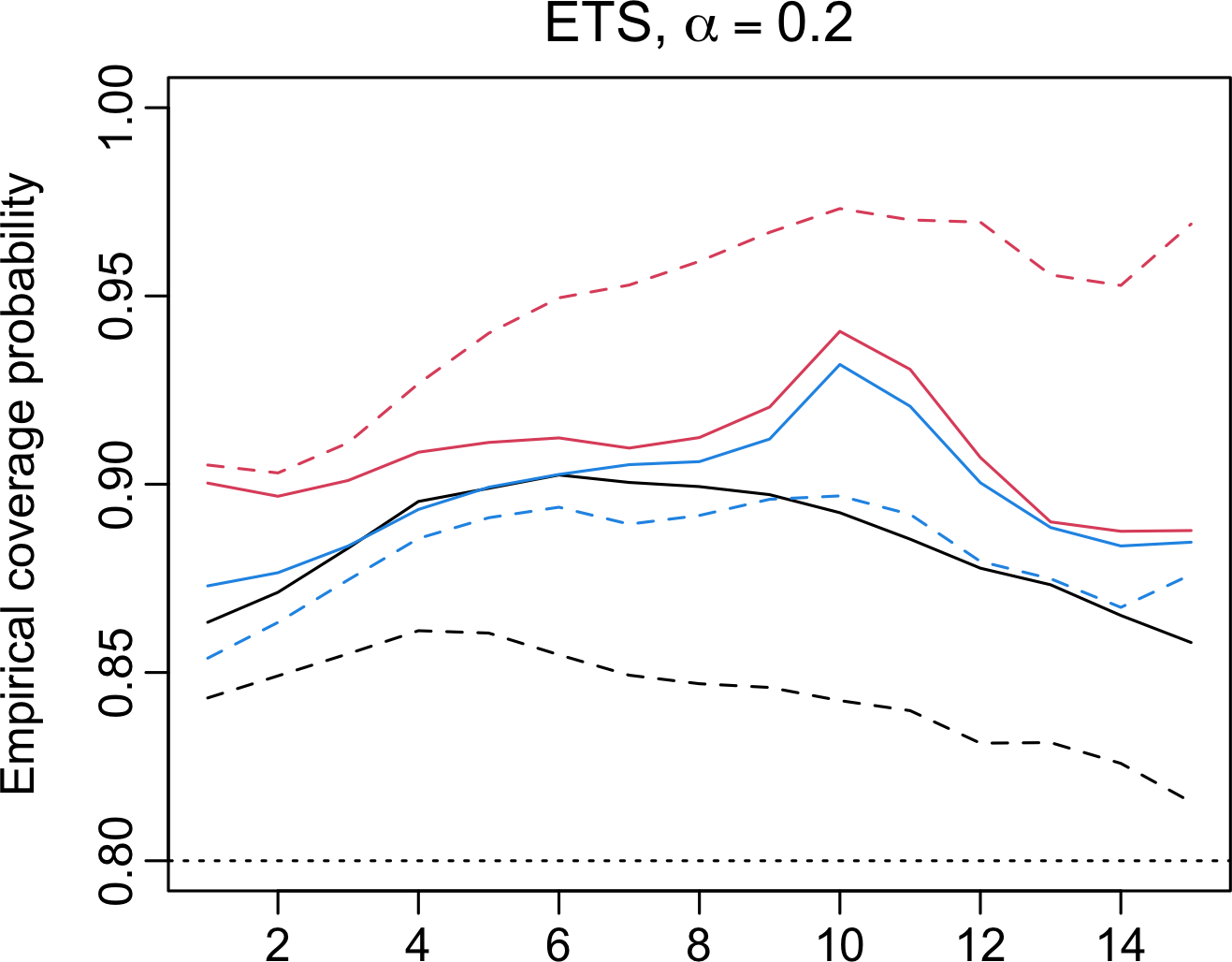}
\quad
\includegraphics[width=8.7cm]{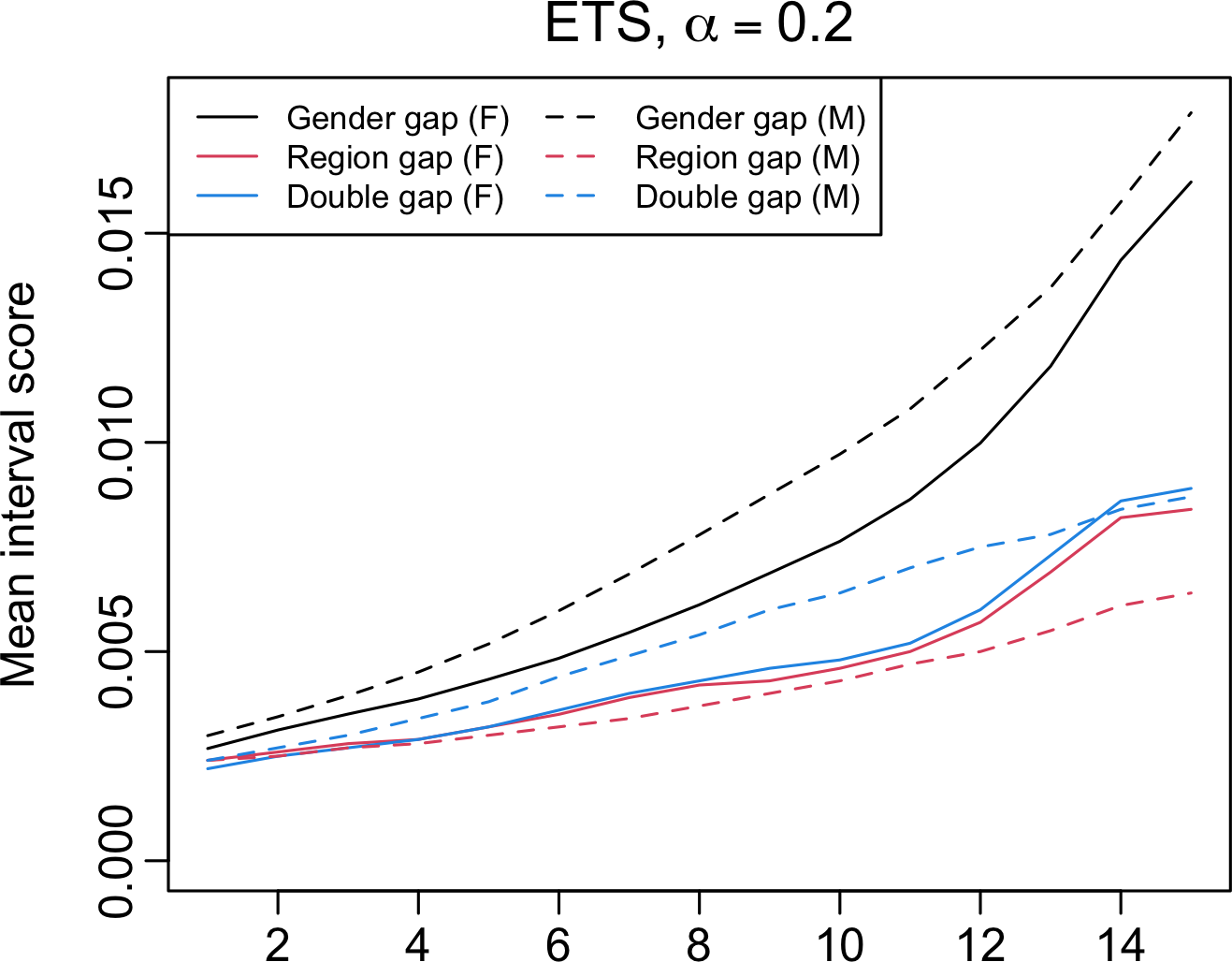}
\\
\vspace{.12in}
\includegraphics[width=8.7cm]{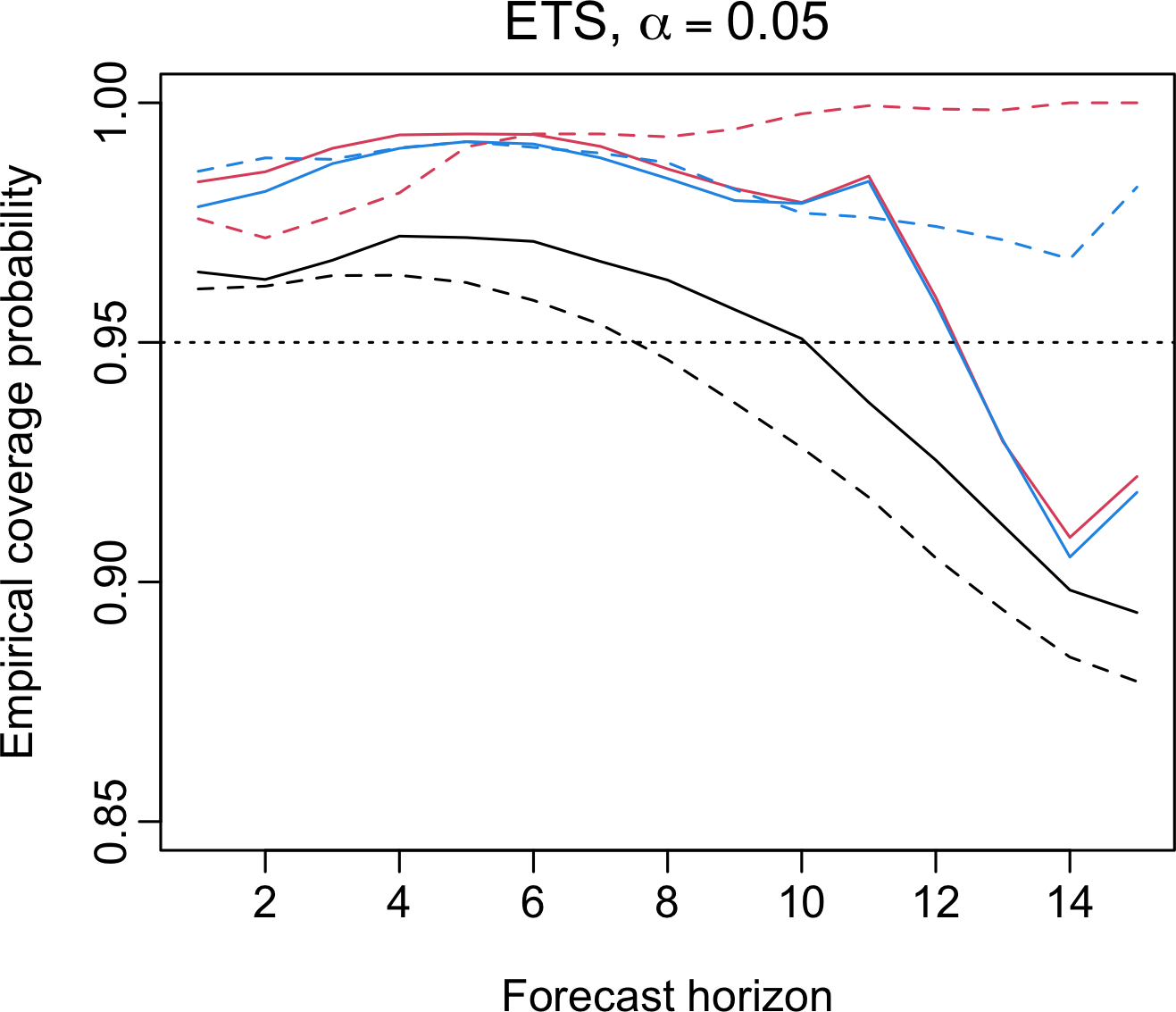}
\quad
\includegraphics[width=8.7cm]{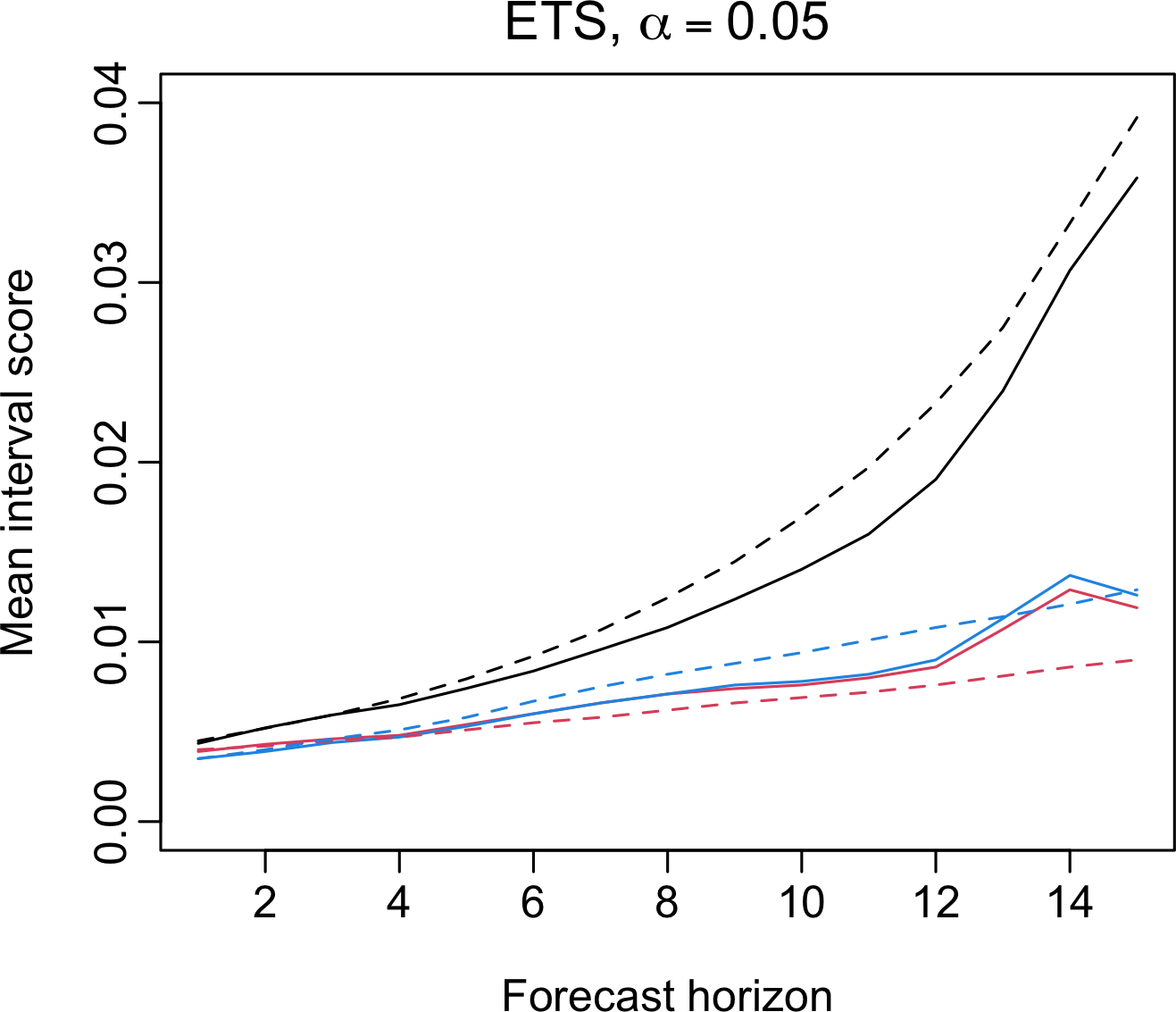}
\caption{\small Averaged over 47 prefectures, the one-to-15-step-ahead interval forecast accuracy comparison at the nominal coverage probabilities of 80\% and 95\%, using the combination of the functional time-series and the ETS forecasting method with the gender gap, regional gap, and double gap.}\label{fig:13}
\end{figure}

In Figure~\ref{fig:13}, we display one-to-15-step-ahead interval forecast errors, as measured by the ECP and mean interval score, using the functional time-series forecasting method. 

At the 80\% nominal level, the ECP obtained from the three gap approaches exceeds the nominal level. At the 95\% nominal coverage probability, the ECPs from the region and double gaps are above the nominal level for the male data and, for almost all horizons, for the female data. By examining the mean interval score, the regional gap approach is recommended as it produces the smallest mean interval scores across all forecast horizons for both the female and male data.

In Table~\ref{tab:1}, we present an overall comparison of the interval forecast accuracy between the ARIMA and ETS forecasting methods at two different levels of nominal coverage probability. While the ETS forecasting method tends to produce ECP higher than the nominal coverage probability, the ARIMA forecasting methods demonstrate a slight under-estimation, especially at the 95\% nominal coverage. Based on the mean interval score, the ETS forecasting method yields smaller forecast errors than the ARIMA method, highlighting the importance of achieving correct coverage. For completeness, we display the horizon-specific results using the ARIMA method in Appendix~\ref{sec:Appendix_A}.
\begin{table}[!htb]
\centering
\tabcolsep 0.21in
\caption{\small Averaged over 47 prefectures and 15 forecast horizons, we compute the overall interval forecast accuracy, as measured by the ECP and mean interval scores, between the ARIMA and ETS forecasting methods for two levels of significance, $\alpha = 0.2$ and 0.05, corresponding to the nominal coverage probabilities of 80\% and 95\%.}\label{tab:1}
\begin{tabular}{@{}lllrrrrrr@{}}
  \toprule
	  	&		&		& \multicolumn{3}{c}{Female} & \multicolumn{3}{c}{Male} \\
\cmidrule(lr){4-6}\cmidrule(lr){7-9}
Method 	& $\alpha$ & Gap 	& ECP 	& CPD 	& $\overline{S}_{\alpha}$ 	& ECP 	& CPD 	& $\overline{S}_{\alpha}$ \\ 
  \midrule
ARIMA 	& 0.2 	& Gender  & 0.811 & 0.054 & 0.011 & 0.773 & 0.049 & 0.013 \\ 
		& 		& Region  	& 0.826 & 0.062 & 0.009 & 0.833 & 0.059 & 0.006 \\ 
		&		& Double 	& 0.826 & 0.061 & 0.009 & 0.782 & 0.053 & 0.010 \\ 
\\
		& 0.05	&  Gender & 0.894 & 0.082 & 0.026 & 0.889 & 0.080 & 0.026 \\ 
		&		&  Region 	 & 0.910 & 0.072 & 0.021 & 0.896 & 0.084 & 0.012 \\ 
		&		&  Double  & 0.909 & 0.073 & 0.021 & 0.909 & 0.069 & 0.019 \\ 
\midrule
ETS 		& 0.2		&  Gender & 0.884 & 0.101 & 0.007 & 0.843 & 0.070 & 0.009 \\ 
		&		&  Region  & 0.908 & 0.110 & 0.005 & 0.947 & 0.149 & 0.004 \\ 
		&		&  Double  & 0.897 & 0.107 & 0.005 & 0.882 & 0.092 & 0.005 \\ 
\\
		& 0.05 	&  Gender & 0.948 & 0.056 & 0.014 & 0.935 & 0.059 & 0.016 \\ 
		&		&  Region  & 0.972 & 0.044 & 0.007 & 0.991 & 0.047 & 0.006 \\ 
		&		&  Double  & 0.970 & 0.045 & 0.007 & 0.983 & 0.040 & 0.008 \\ 
\bottomrule
\end{tabular}
\end{table}

\section{Conclusion}\label{sec:6}

We present a gap-modeling framework and demonstrate it with three variants: Our first approach to forecast male subnational life-table death counts combines separate forecasts from the female subnational forecasts and gender gap forecasts. Our second approach to forecasting subnational life-table death counts combines separate forecasts for each region and region-gap forecasts for females or males. By combining gender and regional gaps, we also consider a double-gap forecasting method. The current framework is not limited to a specific forecasting method, although we demonstrate the idea of gap modeling using a functional time-series method. 

Using the age-specific Japanese subnational life-table death counts, we evaluate and compare the finite-sample point and interval forecast accuracy. From point and interval forecast accuracy, the regional gap is recommended for modeling and forecasting subnational life-table death counts for Japanese females and males. To facilitate reproducibility, all \Rlogo\ code used in this study is collated and publicly available at \url{https://github.com/hanshang/Gap_modeling}.

There are several ways in which the methodology presented can be further extended, and we briefly mention a couple:
\begin{inparaenum}
\item[1)] The mortality gap could be the national mortality profile and the insurer's own mortality profile, and this gap may reveal the mortality difference between industry and academic practices.
\item[2)] The mortality gap could be divided by other factors, such as differences in socio-economics or household wealth \citep[see, e.g.,][]{VH14, CKR+19}, or causes-of-death \citep[see, e.g.,][]{BNS25}.
\end{inparaenum}

\section*{Acknowledgment}

We are grateful for the insightful comments provided by the participants at the International Symposia on Nonparametric Statistics in Thessaloniki, Greece in 2026 (ISNP2026). The first author is grateful for the financial support from an Australian Research Council Future Fellowship (FT240100338).

\newpage

\appendix
\section{Comparison of interval forecast accuracy: ARIMA method}\label{sec:Appendix_A}

\begin{figure}[!htb]
\centering
\includegraphics[width=8.7cm]{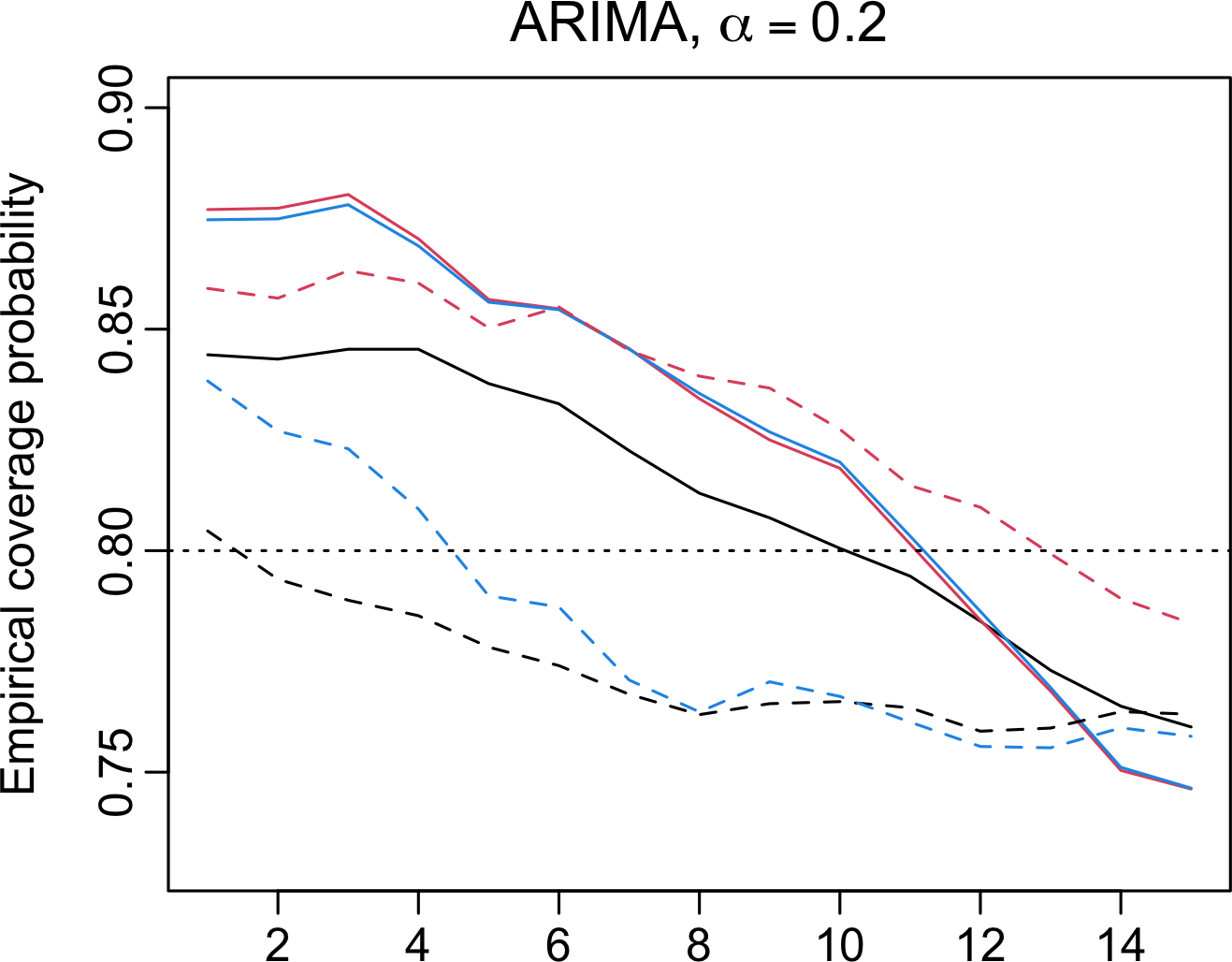}
\quad
\includegraphics[width=8.7cm]{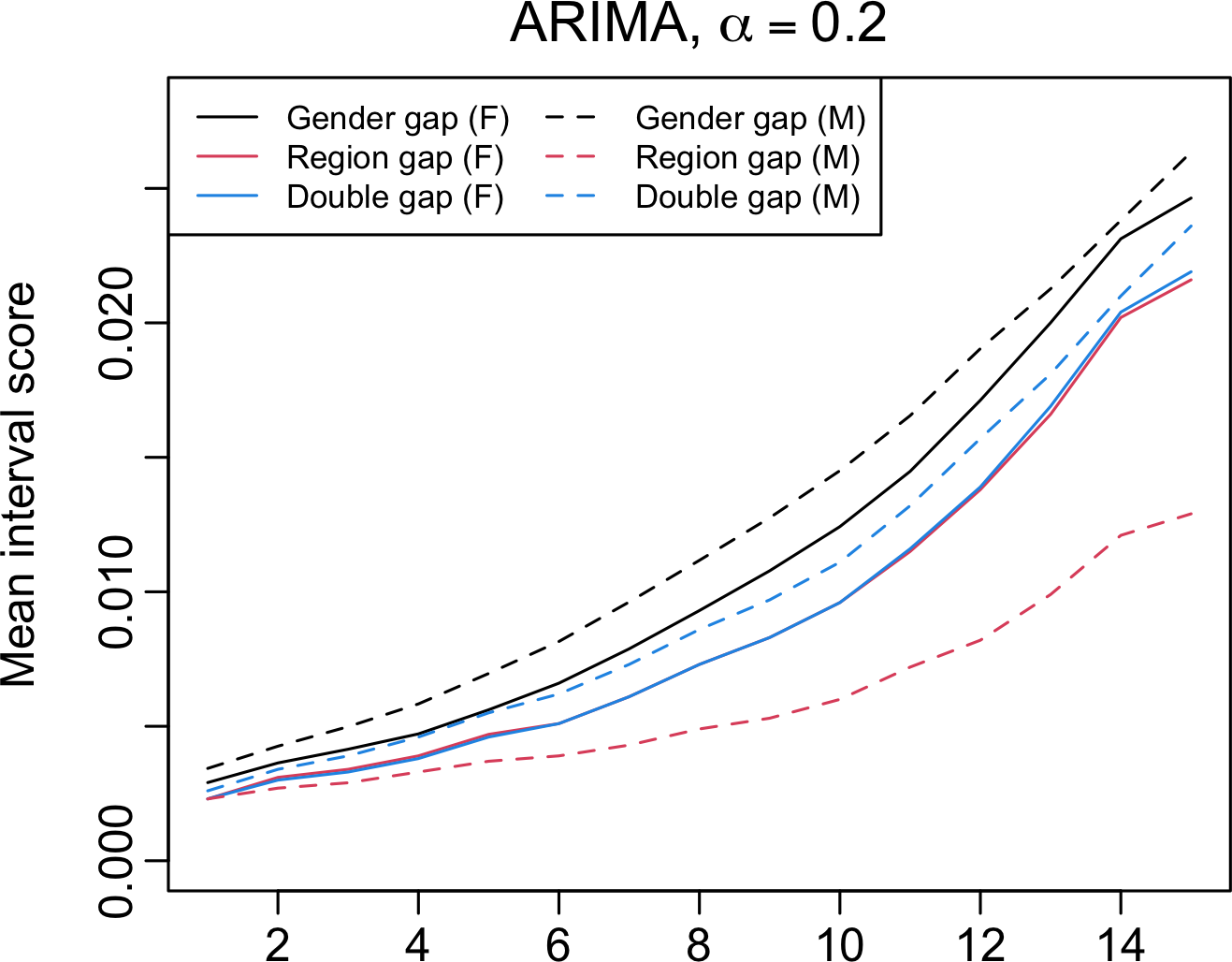}
\\
\vspace{.12in}
\includegraphics[width=8.7cm]{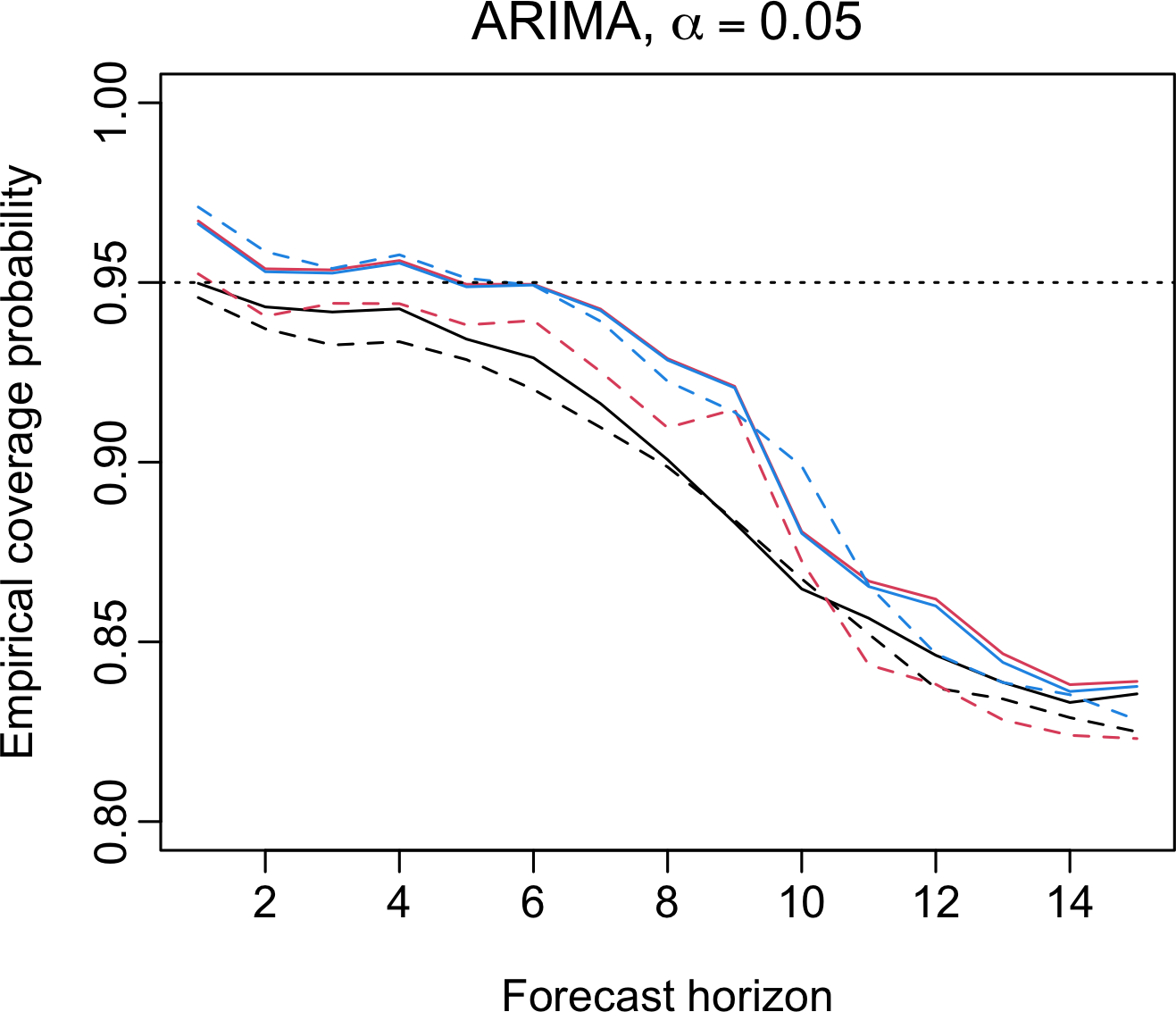}
\quad
\includegraphics[width=8.7cm]{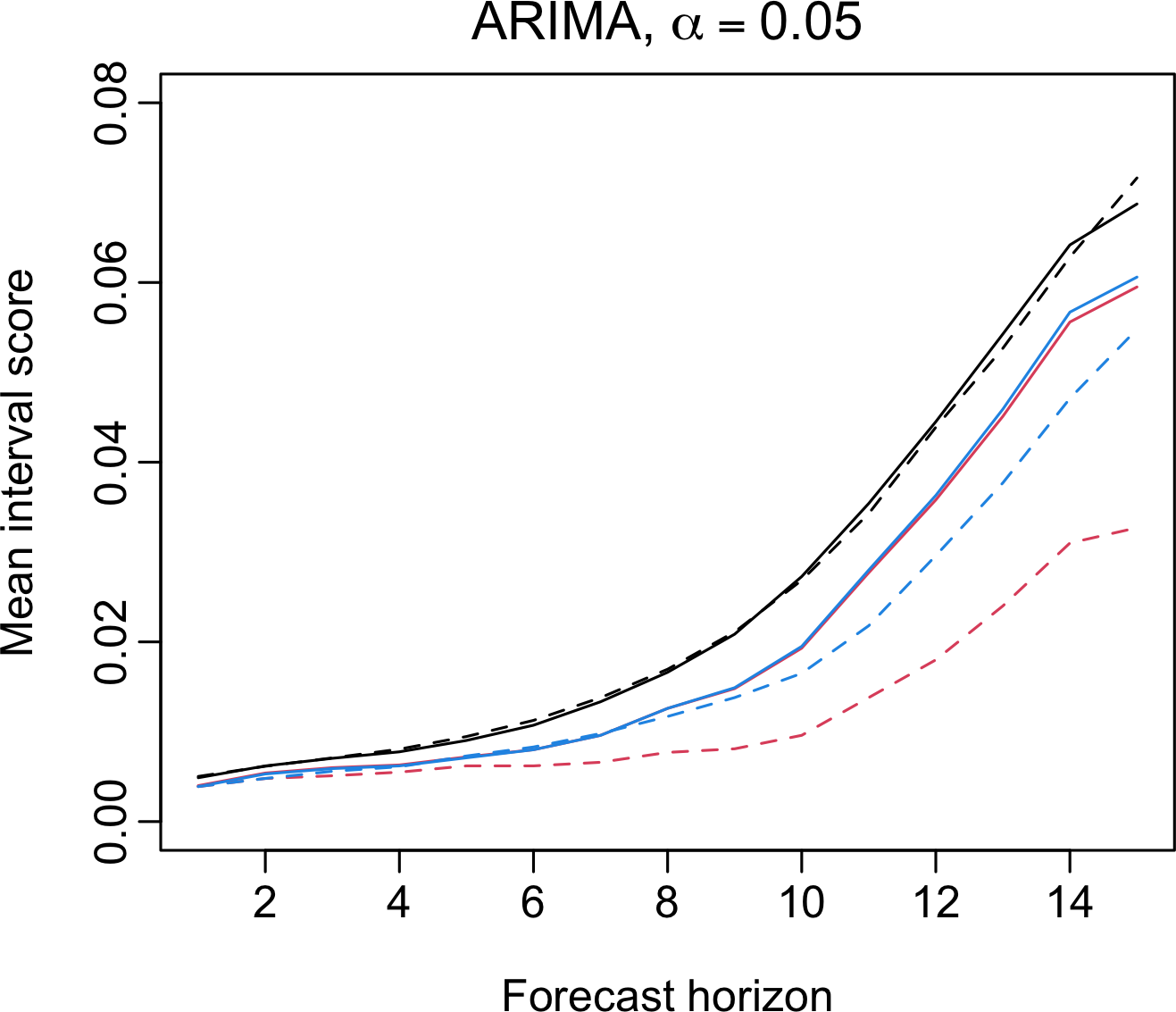}
\caption{\small Averaged over 47 prefectures, the one-to-15-step-ahead interval forecast accuracy comparison at the nominal coverage probabilities of 80\% and 95\%, using the combination of the functional time-series and the ARIMA forecasting method with the gender gap, regional gap, and double gap.}\label{fig:14}
\end{figure}

In Figure~\ref{fig:14}, we present the one-to-15-step-ahead interval forecast accuracy, as measured by the ECP, CPD, and the mean interval score, at the nominal coverage probabilities of 80\% and 95\%. The ARIMA method tends to produce ECPs that are lower than the nominal coverage probabilities for most forecast horizons. Compared to Figure~\ref{fig:13}, the mean interval scores obtained by the ARIMA method tend to be higher than those obtained by the ETS method.

\newpage
\section{Model sensitivity analysis: Age-period benchmark}\label{sec:Appendix_B}

In this appendix, we describe an age-period benchmark associated with the double-gap representation in Section~\ref{sec:4.3}. The age-period benchmark is applied to the same transformed and smoothed quantities used in the main text; therefore, we do not repeat the definitions of the CDF, the logit and Fisher-$Z$ transformations are implemented.

\subsection{Surfaces used in the age-period benchmark}

Let $x\in\{0, 1,\ldots,\omega\}$ denote age, $t\in\{1,\ldots,n\}$ calendar year, $g\in\{F,M\}$ gender, and $s\in\{1,\ldots,47\}$ prefecture. As in Section~4, we work with the following transformed (and already smoothed) surfaces:
\begin{inparaenum}
\item[(i)] the national female CDF on the logit scale, $\text{logit}\!\big(D^{N,F}_{t,x}\big)$ for $x=1,\ldots,\omega-1$ with a slight abuse of notation;
\item[(ii)] the regional female gap (subnational minus national CDF) on the Fisher--$Z$ scale, $F^{\,s-N,F}_{t,x}$ for $x=1,\ldots,\omega$;
\item[(iii)] the subnational gender gap (male minus female CDF) on the Fisher--$Z$ scale, $F^{\,s,M-F}_{t,x}$.
\end{inparaenum}

For notational convenience, we denote by $Y^{(i)}_{t,x}, Y^{(ii)}_{t,x}$ and $Y^{(iii)}_{t,x}$ any one of the above surfaces (with the appropriate index set in $x$), and we apply the same age-period specification separately to $\text{logit}(D^{N,F}_{t,x})$, $F^{\,s-N,F}_{t,x}$ and $F^{\,s,M-F}_{t,x}$ for all prefectures $s$.

\subsection{Age--period model and estimation}

For each surface $Y_{t,x}$ we use a very simple age--period decomposition:
\begin{equation}\label{eq:AP-model-simple}
Y_{t,x} \;=\; \mu \;+\; \lambda_x \;+\; \kappa_t \;+\; \eta_{t,x},
\end{equation}
where $\omega^\ast$ denotes the closing age, in our study 
$\omega^\ast=110$. Furthermore, $\mu$ is a global intercept, $\lambda_x$ is an age effect (common between years), $\kappa_t$ is a period effect (common between ages), and $\eta_{x,t}$ is a residual term with mean zero. To make the decomposition identifiable, we impose the usual constraints
\begin{equation*}
\sum_{x=0}^{\omega^\ast} \lambda_x = 0, \qquad \sum_{t=1}^{n} \kappa_t = 0.
\end{equation*}
In other words, $\mu$ collects the overall level, while $\lambda_x$ and $\kappa_t$ capture the deviations by age and period. The model in \eqref{eq:AP-model-simple} is estimated using ordinary least squares. 

\subsection{Period dynamics and forecasting}

The evolution of time is entirely driven by the period effect $\kappa_t$. For each surface, we treat the estimated period effect $\{\kappa_t\}_{t=t_1}^T$ as a univariate time series and fit either an ETS model or an ARIMA$(p,d,q)$ specification, with the choice guided by AIC on a small grid. In the main comparison, we focus on the ARIMA-based version of the benchmark.

Let $\kappa_{n+h\mid n}$ be the $h$--step--ahead forecast of $\kappa_{n+h}$, for $h=1,2,\ldots$.  The corresponding age-period forecast of the transformed surface is
\begin{equation}\label{eq:AP-forecast-surface}
Y_{n+h\mid n,x}=\mu + \lambda_x + \kappa_{n+h\mid n}, \qquad x=1,\ldots,\omega^{\ast}.
\end{equation}

We apply~\eqref{eq:AP-forecast-surface} to:
\begin{inparaenum}[(i)]
\item $Y^{(1)}_{t,x}$ (national female logit CDF),
\item $Y^{(2)}_{s,t,x}$ (regional female gap for each prefecture $s$),
\item $Y^{(3)}_{s,t,x}$ (subnational gender gap for each prefecture~$s$),
\end{inparaenum}
obtaining age-period forecasts for all components of the double-gap representation.

\subsection{Reconstruction and comparison}

The age-period benchmark is returned to the original scale exactly as in the double-gap model in Section~\ref{sec:4.3}.

\begin{asparaenum}[1)]
\item From $Y^{(1)}_{n+h\mid n,x}$ we obtain the forecast national female CDF $\widehat D^{N,F}_{n+h\mid n,x}$ by applying the inverse logit transform and enforcing $\widehat D^{N,F}_{n+h\mid n,\omega}=1$.
\item From $Y^{(2)}_{s,n+h\mid n,x}$ and $Y^{(3)}_{s,n+h\mid n,x}$ we obtain the forecast regional female and gender gaps $\widehat G^{\,s-N,F}_{n+h\mid n,x}$ and $\widehat G^{\,s,M-F}_{n+h\mid n,x}$ by applying the inverse Fisher-$Z$ transform.
\item For each prefecture~$s$, we reconstruct the subnational female and male CDFs via the double-gap decomposition:
\begin{align*}
\widehat D^{s,F}_{n+h\mid n,x} &= \widehat D^{N,F}_{n+h\mid n,x} + \widehat G^{\,s-N,F}_{n+h\mid n,x}, \\
\widehat D^{s,M}_{n+h\mid n,x} &= \widehat D^{N,F}_{n+h\mid n,x} + \widehat G^{\,s-N,F}_{n+h\mid n,x} + \widehat G^{\,s,M-F}_{n+h\mid n,x},
\end{align*}	
with the same truncation and monotonicity adjustments used in the main model.
\item For each $(s,g,h)$, we convert the CDF forecasts to life-table death counts by first-order differencing,
\begin{align*}
\widehat d^{s,g}_{n+h\mid n,1} &= \widehat D^{s,g}_{n+h\mid n,1}, \\
\widehat d^{s,g}_{n+h\mid n,x} &= \widehat D^{s,g}_{n+h\mid n,x} - \widehat D^{s,g}_{n+h\mid n,x-1}, \qquad x=2,3,\ldots,\omega.
\end{align*}
\end{asparaenum}

The forecasts are obtained through the expanding-window scheme to compare the age-period benchmark with the functional time-series model through the density-valued loss functions. From Tables~\ref{tab:2} and~\ref{tab:AP_table1}, the region gap method provides the smallest point forecast errors; between the functional time-series and age-period models, the former is recommended.

\begin{small}
\begin{center}
\tabcolsep 0.034in
\renewcommand{\arraystretch}{1.1}
\begin{longtable}{@{}lrrrrrrrrrrrr@{}}
\caption{Averaged over 47 prefectures, we compare the one- to 15-step-ahead point forecast accuracy, as measured by the KLD and JSD, using the age--period benchmark with the gender gap, regional gap, and double gap.}\label{tab:AP_table1}\\
\hline
    &  \multicolumn{6}{c}{KLD} &  \multicolumn{6}{c}{JSD} \\
    \cmidrule(lr){2-7}\cmidrule(lr){8-13}
    & \multicolumn{3}{c}{Female} & \multicolumn{3}{c}{Male} & \multicolumn{3}{c}{Female} & \multicolumn{3}{c}{Male} \\
    \cmidrule(lr){2-4}\cmidrule(lr){5-7}\cmidrule(lr){8-10}\cmidrule(lr){11-13}
$h$ & Gender & Region & Double & Gender & Region & Double & Gender & Region & Double & Gender & Region & Double \\ 
\hline
\endfirsthead
\hline
    &  \multicolumn{6}{c}{KLD} &  \multicolumn{6}{c}{JSD} \\
    \cmidrule(lr){2-7}\cmidrule(lr){8-13}
    & \multicolumn{3}{c}{Female} & \multicolumn{3}{c}{Male} & \multicolumn{3}{c}{Female} & \multicolumn{3}{c}{Male} \\
    \cmidrule(lr){2-4}\cmidrule(lr){5-7}\cmidrule(lr){8-10}\cmidrule(lr){11-13}
$h$ & Gender & Region & Double & Gender & Region & Double & Gender & Region & Double & Gender & Region & Double \\ 
\hline
\endhead
\midrule
\multicolumn{13}{r}{Continued on next page} \\
\endfoot
\hline
\endlastfoot
\multicolumn{2}{l}{\hspace{-.08in} \underline{ETS}} & \\
1  & 0.0330 & 0.0239 & 0.0239 & 0.0335 & 0.0229 & 0.0370 & 0.0077 & 0.0056 & 0.0056 & 0.0087 & 0.0054 & 0.0099 \\
2  & 0.0336 & 0.0242 & 0.0242 & 0.0346 & 0.0235 & 0.0385 & 0.0078 & 0.0057 & 0.0057 & 0.0090 & 0.0055 & 0.0104 \\
3  & 0.0347 & 0.0247 & 0.0247 & 0.0352 & 0.0240 & 0.0394 & 0.0080 & 0.0058 & 0.0058 & 0.0091 & 0.0057 & 0.0106 \\
4  & 0.0352 & 0.0245 & 0.0245 & 0.0351 & 0.0238 & 0.0397 & 0.0082 & 0.0058 & 0.0058 & 0.0091 & 0.0056 & 0.0107 \\
5  & 0.0370 & 0.0254 & 0.0254 & 0.0357 & 0.0244 & 0.0404 & 0.0086 & 0.0060 & 0.0060 & 0.0092 & 0.0058 & 0.0109 \\
6  & 0.0388 & 0.0262 & 0.0262 & 0.0366 & 0.0250 & 0.0412 & 0.0090 & 0.0062 & 0.0062 & 0.0094 & 0.0059 & 0.0111 \\
7  & 0.0404 & 0.0268 & 0.0268 & 0.0381 & 0.0257 & 0.0425 & 0.0093 & 0.0063 & 0.0063 & 0.0097 & 0.0061 & 0.0114 \\
8  & 0.0421 & 0.0274 & 0.0274 & 0.0399 & 0.0264 & 0.0440 & 0.0097 & 0.0064 & 0.0064 & 0.0101 & 0.0063 & 0.0118 \\
9  & 0.0438 & 0.0279 & 0.0279 & 0.0428 & 0.0272 & 0.0458 & 0.0100 & 0.0065 & 0.0065 & 0.0106 & 0.0064 & 0.0123 \\
10 & 0.0455 & 0.0287 & 0.0287 & 0.0457 & 0.0282 & 0.0480 & 0.0104 & 0.0066 & 0.0066 & 0.0112 & 0.0067 & 0.0129 \\
11 & 0.0467 & 0.0296 & 0.0296 & 0.0498 & 0.0294 & 0.0513 & 0.0107 & 0.0068 & 0.0068 & 0.0119 & 0.0069 & 0.0138 \\
12 & 0.0484 & 0.0305 & 0.0305 & 0.0562 & 0.0309 & 0.0554 & 0.0110 & 0.0069 & 0.0069 & 0.0133 & 0.0073 & 0.0148 \\
13 & 0.0396 & 0.0294 & 0.0294 & 0.0550 & 0.0319 & 0.0629 & 0.0091 & 0.0066 & 0.0066 & 0.0144 & 0.0074 & 0.0169 \\
14 & 0.0368 & 0.0285 & 0.0285 & 0.0640 & 0.0336 & 0.0715 & 0.0085 & 0.0063 & 0.0063 & 0.0167 & 0.0077 & 0.0192 \\
15 & 0.0401 & 0.0292 & 0.0292 & 0.0670 & 0.0353 & 0.0726 & 0.0092 & 0.0064 & 0.0064 & 0.0175 & 0.0081 & 0.0194 \\
\midrule
Mean & 0.0397 & 0.0271 & 0.0271 & 0.0446 & 0.0275 & 0.0487 & 0.0091 & 0.0063 & 0.0063 & 0.0113 & 0.0064 & 0.0131 \\
\midrule
\multicolumn{2}{l}{\hspace{-.08in} \underline{ARIMA}} & \\
1  & 0.0334 & 0.0244 & 0.0244 & 0.0381 & 0.0259 & 0.0346 & 0.0077 & 0.0058 & 0.0058 & 0.0099 & 0.0061 & 0.0093 \\
2  & 0.0339 & 0.0249 & 0.0249 & 0.0420 & 0.0282 & 0.0364 & 0.0078 & 0.0059 & 0.0059 & 0.0109 & 0.0066 & 0.0098 \\
3  & 0.0354 & 0.0256 & 0.0256 & 0.0457 & 0.0292 & 0.0376 & 0.0082 & 0.0060 & 0.0060 & 0.0118 & 0.0068 & 0.0101 \\
4  & 0.0363 & 0.0260 & 0.0260 & 0.0508 & 0.0321 & 0.0379 & 0.0084 & 0.0061 & 0.0061 & 0.0129 & 0.0074 & 0.0102 \\
5  & 0.0389 & 0.0282 & 0.0282 & 0.0575 & 0.0364 & 0.0392 & 0.0089 & 0.0066 & 0.0066 & 0.0143 & 0.0083 & 0.0106 \\
6  & 0.0416 & 0.0308 & 0.0308 & 0.0644 & 0.0389 & 0.0403 & 0.0095 & 0.0072 & 0.0072 & 0.0156 & 0.0089 & 0.0109 \\
7  & 0.0448 & 0.0328 & 0.0328 & 0.0743 & 0.0420 & 0.0399 & 0.0101 & 0.0077 & 0.0077 & 0.0172 & 0.0095 & 0.0108 \\
8  & 0.0490 & 0.0342 & 0.0342 & 0.0856 & 0.0446 & 0.0373 & 0.0109 & 0.0080 & 0.0080 & 0.0192 & 0.0101 & 0.0102 \\
9  & 0.0532 & 0.0346 & 0.0346 & 0.0998 & 0.0457 & 0.0329 & 0.0117 & 0.0082 & 0.0082 & 0.0211 & 0.0103 & 0.0090 \\
10 & 0.0589 & 0.0354 & 0.0354 & 0.1243 & 0.0480 & 0.0300 & 0.0128 & 0.0084 & 0.0084 & 0.0235 & 0.0107 & 0.0083 \\
11 & 0.0649 & 0.0354 & 0.0354 & 0.1481 & 0.0579 & 0.0311 & 0.0139 & 0.0084 & 0.0084 & 0.0266 & 0.0127 & 0.0086 \\
12 & 0.0745 & 0.0362 & 0.0362 & 0.1867 & 0.0712 & 0.0362 & 0.0156 & 0.0085 & 0.0085 & 0.0312 & 0.0155 & 0.0099 \\
13 & 0.0622 & 0.0316 & 0.0316 & 0.1338 & 0.0828 & 0.0469 & 0.0137 & 0.0074 & 0.0074 & 0.0343 & 0.0178 & 0.0128 \\
14 & 0.0673 & 0.0282 & 0.0282 & 0.1556 & 0.0980 & 0.0490 & 0.0147 & 0.0066 & 0.0066 & 0.0401 & 0.0209 & 0.0134 \\
15 & 0.0740 & 0.0301 & 0.0301 & 0.1680 & 0.1080 & 0.0394 & 0.0161 & 0.0071 & 0.0071 & 0.0430 & 0.0228 & 0.0108 \\
\midrule
Mean & 0.0512 & 0.0305 & 0.0305 & 0.0983 & 0.0526 & 0.0379 & 0.0113 & 0.0072 & 0.0072 & 0.0221 & 0.0116 & 0.0103 \\
\bottomrule
\end{longtable}
\end{center}
\end{small}

\vspace{-.4in}

In Table~\ref{tab:AP_interval_overall}, we present an evaluation of interval forecast accuracy by the age-period benchmark, averaged over 47 prefectures and 15 forecast horizons. From the averaged CPD and interval score, the age-period benchmark tends to over-estimate at the 80\% nominal coverage probability, and under-estimate at the 95\% nominal coverage probability. Between the functional time-series and age-period models, there is a marginal difference in their interval forecast accuracy. The horizon-specific results obtained by the age-period benchmark can be obtained upon request from the corresponding author.
\begin{table}[!htb]
\centering
\tabcolsep 0.21in
\caption{\small Averaged over 47 prefectures and 15 forecast horizons, we compute the overall interval forecast accuracy for the age--period benchmark, as measured by the ECP, CPD, and mean interval scores, between the ARIMA and ETS forecasting methods for two nominal coverage probabilities of 80\% ($\alpha=0.2$) and 95\% ($\alpha=0.05$).}\label{tab:AP_interval_overall}
\begin{tabular}{@{}lllrrrrrr@{}}
\toprule
        &           &       & \multicolumn{3}{c}{Female} & \multicolumn{3}{c}{Male} \\
\cmidrule(lr){4-6}\cmidrule(lr){7-9}
Method  & $\alpha$  & Gap   & ECP & CPD & $\overline{S}_{\alpha}$ & ECP & CPD & $\overline{S}_{\alpha}$ \\
\midrule
ARIMA   & 0.20      & Gender & 0.848 & 0.071 & 0.009 & 0.831 & 0.055 & 0.013 \\
        &           & Region & 0.834 & 0.049 & 0.009 & 0.856 & 0.079 & 0.007 \\
        &           & Double & 0.834 & 0.049 & 0.009 & 0.896 & 0.103 & 0.010 \\
\\
        & 0.05      & Gender & 0.935 & 0.038 & 0.014 & 0.898 & 0.078 & 0.019 \\
        &           & Region & 0.918 & 0.047 & 0.015 & 0.928 & 0.050 & 0.010 \\
        &           & Double & 0.918 & 0.047 & 0.015 & 0.971 & 0.045 & 0.013 \\
\midrule
ETS     & 0.20      & Gender & 0.844 & 0.075 & 0.008 & 0.817 & 0.053 & 0.009 \\
        &           & Region & 0.851 & 0.075 & 0.006 & 0.865 & 0.082 & 0.006 \\
        &           & Double & 0.851 & 0.075 & 0.006 & 0.801 & 0.034 & 0.009 \\
\\
        & 0.05      & Gender & 0.947 & 0.043 & 0.011 & 0.907 & 0.062 & 0.013 \\
        &           & Region & 0.962 & 0.036 & 0.008 & 0.955 & 0.042 & 0.008 \\
        &           & Double & 0.962 & 0.036 & 0.008 & 0.884 & 0.069 & 0.013 \\
\bottomrule
\end{tabular}
\end{table}

\newpage
\bibliographystyle{agsm}
\bibliography{Gap_modeling.bib}

\end{document}